\documentclass{scrartcl}
\usepackage{bbm,amsfonts,amssymb,epsfig,verbatim,mathbbol,graphics,graphicx}

\def\tr{{\rm tr}}
\def\p{\phi}
\def\P{\mathcal{P}}
\def\N{\mathcal{N}}

\def\la{\triangle}
\def\z{\zeta}
\def\e{\epsilon}
\def\a{\alpha}

\def\hm{\hat{\mu}}

\begin{document}
\title{\begin{flushright}
\normalsize
{\rm\large HU-EP-05/42}\\
\end{flushright}
\vspace{1cm}
Laplacian modes probing gauge fields}
\author{{\bf Falk Bruckmann}${}^{\,a}$ and
{\bf Ernst-Michael Ilgenfritz}${}^{\,b}$\\
\\
${}^{a}$ Instituut-Lorentz for Theoretical Physics, University of Leiden,\\
P.O.Box 9506, NL-2300 RA Leiden, The Netherlands\\
\\
${}^{b}$ Institute of Physics, Humboldt University Berlin,\\ 
Newtonstr.\ 15, D-12489 Berlin, Germany\\
}
\date{}
\maketitle

\begin{abstract}
We show that low-lying eigenmodes of the Laplace operator 
are suitable to represent
properties of the underlying $SU(2)$ lattice configurations. 
We study this for the case of finite temperature background fields,
yet in the confinement phase.
For calorons as classical solutions put on the lattice,
the lowest mode localizes one of the constituent monopoles
by a maximum and the other one by a minimum, respectively.
We introduce adjustable phase boundary conditions in the 
time direction,
under which the role of the monopoles in the mode localization is interchanged.
Similar hopping phenomena are observed for thermalized configurations.
We also investigate periodic and antiperiodic modes
of the adjoint Laplacian for comparison.

In the second part we introduce a new Fourier-like low-pass
filter method. It provides link variables by truncating 
a sum involving the Laplacian eigenmodes. The filter not only 
reproduces classical structures, but also preserves the confining 
potential for thermalized ensembles.
We give a first characterization of 
the structures emerging from this procedure.

\end{abstract}

\section{Introduction}

The question what drives confinement and other nonperturbative phenomena 
of QCD at strong coupling
is a long-standing one. In lattice gauge theory, the simulation of these 
effects at the observational level of correlation functions is well-established.
However, to extract the relevant degrees of freedom of the QCD vacuum 
remains a controversial problem which might not have a unique answer. 
The desire behind such attempts is to provide support for certain models 
like the instanton liquid or the dual Abelian Higgs model 
and to estimate their basic parameters.
 
Abelian and center projection have
been used to focus on objects 
like Abelian magnetic monopoles and center vortices,
respectively, which 
are thought of as localizing
certain embedded solutions that, strictly speaking, would exist
 only in the presence of Higgs fields.
Despite the ambiguity in their definitions, these
degrees of freedom  are
used especially for
confinement-related scenarios of the QCD vacuum.
Instantons as selfdual solutions, on the other hand, can be
conveniently
related to chiral symmetry breaking.
Field excitations 
resembling instantons have indeed been observed as the result of cooling.
It has been objected 
 that these methods modify the 
field configurations in an uncontrolled way such that the observed excitations
could actually be fake and do not represent the relevant nonperturbative fields.
Other smoothing techniques also reveal lumps of action which are usually
interpreted as instantons~\cite{degrand:98b,feurstein:96}.
In order to relate these
structures to confinement, some additional degrees of freedom 
seemed to be necessary.

Fermionic modes capture chiral and topological aspects of lattice gauge
theory provided a Dirac operator with good chiral properties is implemented.
Via the index theorem the number of
(left-handed minus right-handed) 
zero modes gives the total topological charge.
Using a definition of the 
topological charge density based on the overlap Dirac operator
\cite{niedermayer:98},
non-zero modes are contributing to the topological charge distribution, too.
Analyzing the latter, global coherent structures of lower dimension 
have been identified \cite{horvath:03a}.

It has become customary to use
 the localization of the fermionic modes 
as a means to probe  
the vacuum structure.
A specific scaling law of the inverse participation ratio
(IPR, see below) of the low-lying fermionic modes
w.r.t.\ lattice spacing and volume seems to be
able to recognize the codimension of the underlying gauge field structure
\cite{aubin:04,gubarev:05}.
However, to interpret the result,
the latter has to be subject to a model,
for instance one of those mentioned above.

The fact, that fermion zero modes are localized to classical objects in smooth
backgrounds, has lead to a thorough investigation of the local properties of 
zero and near-zero modes
also in equilibrium backgrounds.
Their small `energy' eigenvalue
should suppress contributions from large momenta and hence these modes are
less sensitive to UV fluctuations. Very recently such a programme
has been carried out using zero modes in the adjoint representation
\cite{gonzalez-arroyo:05}.

\begin{figure}
\centering
\begin{tabular}{cc}
\includegraphics[width=0.5\linewidth]
{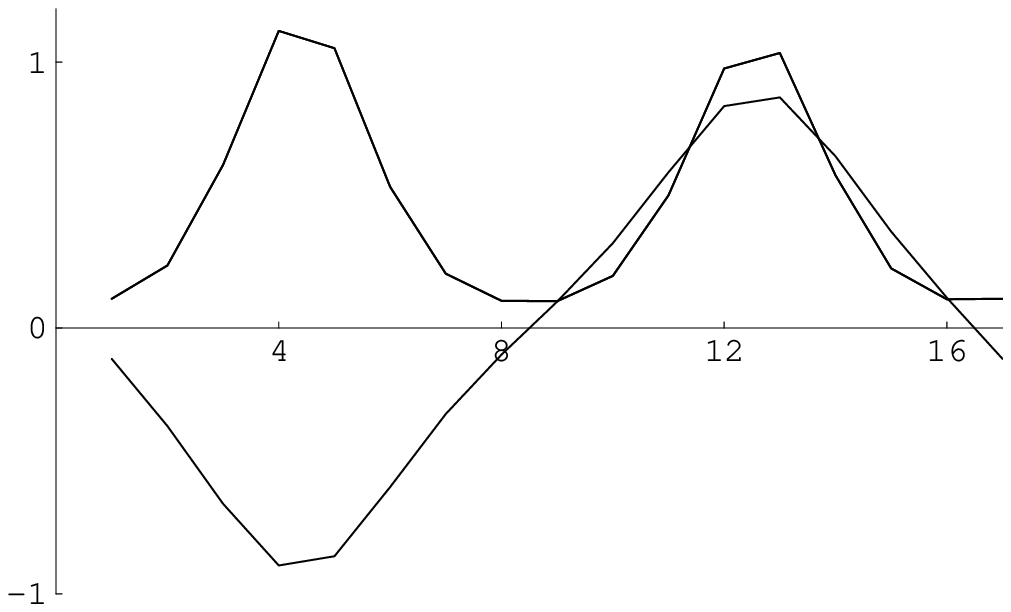}&
\includegraphics[width=0.48\linewidth]
{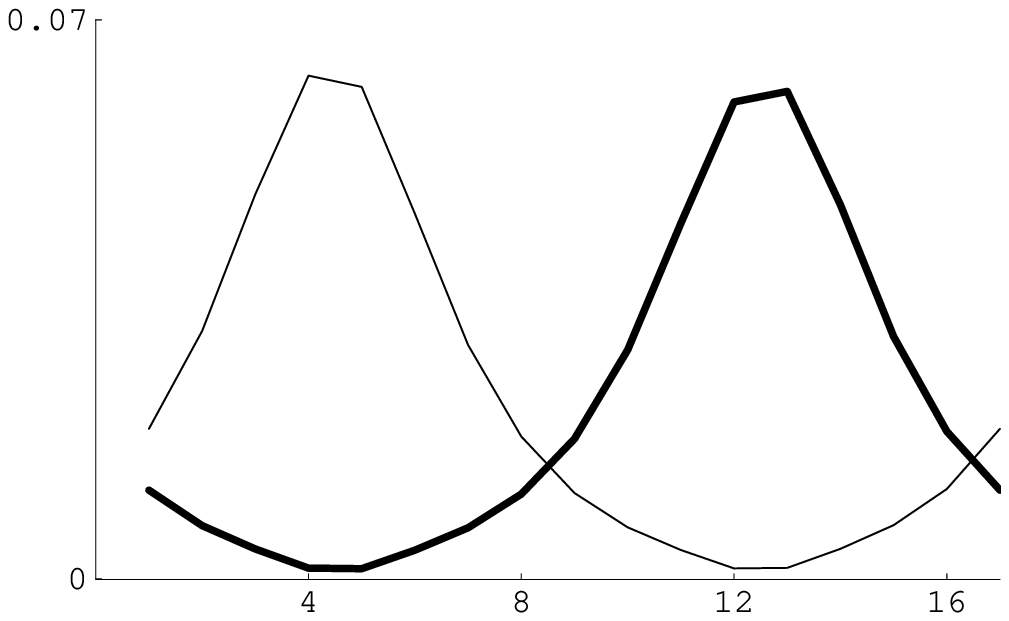}\\
(a)&(b)
\end{tabular}
\caption{
(a) The action and topological density (not distinguishable, both 
multiplied by 400) and the Polyakov loop shown along the line connecting 
the constituents in a large caloron on a $16^3\cdot 4$ lattice. 
(b) The modulus
of periodic (boldface line) and antiperiodic (thin 
line) fermion zero modes for the same caloron.}
\label{fig_bosonic}
\end{figure}

The trick of modifying the boundary conditions for the fundamental fermions by
a complex phase has
been introduced as a general tool in \cite{gattringer:02b} guided by the
knowledge of calorons.
These are instantons at finite temperature, i.e.\ on $R^3\times S^1$.
Calorons have become attractive over the recent years because -- when taken with
nontrivial holonomy \cite{kraan:98a,lee:98b} (see below) --
they account for a Polyakov loop not in
the center of the gauge group, as is the case on average in the confined
phase. Furthermore, calorons contain (gauge independent)
 magnetic monopoles, 
see Fig.\ \ref{fig_bosonic} (a), which realize the scenario of fractional
charge objects (also called instanton quarks \cite{belavin:79}).
For recent progress on calorons, both in the continuum and on the lattice,
see \cite{bruckmann:04c}.

The caloron zero mode with different boundary conditions is correlated to
different constituent monopoles \cite{garciaperez:99c},
see Fig.\ \ref{fig_bosonic} (b)\footnote{We thank Dirk Peschka for providing the
fermion zero mode for the (numerical) caloron.}. In a similar
way the zero modes on equilibrium configurations have been observed to localize
to different locations on the lattice
when scanning through the boundary conditions
\cite{gattringer:02b}.
This effect has been reported even for symmetric lattices representing `zero
temperature' \cite{gattringer:04a}. Given the intuition from calorons, one
expects these modes to detect carriers of topological charge, including such of
fractional charge. However, to support this one would need additional evidence
that the underlying structures actually have fractional or integer charge.
A mechanism analogous to Anderson localization in a random 
potential \cite{anderson:58}
has been proposed as an alternative explanation for the
localization and hopping of the fermionic modes.

In this paper we are going to study the analytic power of the low-lying 
eigenmodes of the gauge covariant Laplace operator. They have been first 
discussed with the aim to fix a particular gauge, the Laplacian 
gauge~\cite{vink:92}. 
Recently they have been investigated with respect to
their localization behaviour~\cite{greensite:05}
to shed additional light on the QCD vacuum.

Our interest concentrates on the local behaviour of these modes testing 
three ideas. The first idea is whether the Laplacian modes are sensitive 
to the constituents of the caloron. Unlike fermion zero modes, the 
Laplacian modes are not incorporated in the
ADHM-Nahm formalism\footnote{although the Greens function is}
that describes the caloron solutions analytically. So we investigate them 
numerically on the lattice, on which caloron configurations can be put with 
very good control. We find indeed that the Laplacian modes can detect the 
monopole constituents by their extrema.
Furthermore, we enforce adjustable
 phase boundary conditions on the Laplacian
modes  
as a function of which they are hopping in a similar way as the fermion 
modes do.

Most of our studies are concerned with eigenmodes
of the Laplacian in the fundamental representation.
For comparison, we also explore adjoint modes,
including those with antiperiodic boundary conditions.

At this point we want to emphasize that the Laplacian modes are advantageous
compared to the fermionic modes in that they have neither chirality nor doubler
problems.
Hence, the straightforward translation of the continuum Laplace operator
to the lattice can be used for the purposes we have in mind.
We will mostly study
 the lowest `energy' eigenmode which has no topological
origin (and therefore is not expected to give information about the topological
charge).

The other idea is how Laplacian modes could be used to 
reflect properties of equilibrium
configurations. As we will demonstrate, they do this in the 
`conventional way' by being pinned to some points on the lattice and hopping 
between several such locations under a change of the boundary conditions.

The relation between the localization mechanisms for classical and thermalized
backgrounds is not straightforward. As we will show, the 
lowest Laplacian eigenmode for the caloron
is similar to a modified wave 
(in contrast to the exponentially localized fermion zero mode). 
Its minimum, but even more the maximum for small
calorons, is not a perfect tool, whereas in the thermalized backgrounds
the maxima obviously determine the localization and
divide the set of 
boundary conditions into intervals.

The third main aspect of the present work is the introduction of a
general and hopefully powerful method to apply a low-pass filter
based on the Laplacian eigenmodes to generic equilibrium configurations.
We were inspired by Gattringer's earlier work who has constructed a smoothed 
field strength tensor based on fermionic modes~\cite{gattringer:02c}.
Our procedure is not limited to a particular observable,
but aims to reconstruct the link variables filtering out UV fluctuations.
It relies on the idea of
truncating a sum involving the Laplacian (`harmonic') modes which 
would give back the original links exactly. The only free parameter (besides
the number of modes used in the truncation)
is the phase angle in the 
boundary condition. We will show that in order to optimize the low-pass filter 
for our circumstances (confined phase, nontrivial holonomy calorons),
this angle needs to be chosen 
halfway between periodic and antiperiodic boundary conditions.

For the filtered configuration we measure the Polyakov loop, 
the topological charge and action density. Again, we present the calorons 
as a testing ground and find, indeed, that the selfdual monopole constituents 
are reproduced. Qualitative agreement with the original configuration
is found already with a surprisingly small number of modes.

What is even more interesting is that this method approximately
preserves the string tension 
when applied to Monte Carlo configurations representing some finite temperature
below the deconfinement phase transition.
Apparently, the Laplacian modes capture
enough of the long range disorder. 
This fact justifies to study the filter in more detail,
especially the emerging tomography of the
configurations in equilibrium.
The structure of the action density of the filtered
configurations includes narrow peaks, for which we do not have a final 
interpretation yet. 
They might be close to the gauge singularities found in the Fourier-filtered 
Landau gauge fields~\cite{gutbrod:00}.
Despite the appearance of the peaks
we will point out a similarity to 
smearing/cooling in an early stage.

The paper is organised as follows. In the next section we will give the
definition as well as some properties of the lattice covariant Laplacian
and briefly summarize the knowledge about calorons. Then we investigate 
low-lying Laplacian eigenmodes in caloron backgrounds. 
In Sect.\ \ref{sec_hopping} we show the typical hopping of the Laplacian 
eigenmodes in the background of thermalized configurations at finite 
temperature. The new filter method and its features are discussed 
in Sect.\ \ref{sec_filter}. We end with some discussion and an outlook. We will 
stick to the gauge group $SU(2)$ throughout this paper.

\section{Preliminaries}

\subsection{Definition and properties of the Laplace operator}

We consider the gauge covariant Laplace operator
\begin{eqnarray}
\la_{xy}^{ab}\equiv\sum_{\mu=1}^D\left[U_{\mu}^{ab}(x)\delta_{x+\hat{\mu},y}+
U_{\mu}^{\dagger ab}(y)\delta_{x-\hat{\mu},y}-2\delta^{ab}\delta_{xy}\right]\,,
\qquad a,b=1,2\,,
\label{eq_def_lapl}
\end{eqnarray}
in the background of a given configuration of lattice links $U_{\mu}(x)$ in
$D=4$ dimensions and in the
fundamental representation.
It is a hermitean (and non-positive) matrix of size
$\N\times\N$ where $\N$ is the number of lattice sites times the dimension of
the representation (here two).

We use the ARPACK package~\cite{arpack} to solve for up to 200
(out of $\N>10^5$) eigenvalues and eigenmodes in 
\begin{eqnarray}
-\la^{ab}_{xy}\p^b_{n,\z}(y)=\lambda_{n,\z}\p^b_{n,\z}(x)\,
\qquad\mbox{no sum in } n \mbox{ and } \z.
\label{eqn_eigen}
\end{eqnarray}
We allow for complex phase boundary conditions in the time-like direction
\begin{eqnarray}
\p_{\z}(x_4+N_4)=e^{2\pi i \z}\p_{\z}(x_4)\,.
\label{eqn_bc}
\end{eqnarray}
A way to implement this is by writing
\begin{eqnarray}
\varphi_{\z}(x)=e^{-2\pi i x_4 \z/N_4}\p_{\z}(x)\,,
\label{eqn_phiphi}
\end{eqnarray}
which is now fully periodic but solves the Laplace equation (\ref{eqn_eigen})
with the replacement $U_4\to U_4e^{2\pi i \z/N_4}$ in $\la$.
Effectively, this promotes the link to an element of $U(2)$, but
does not change any contractible Wilson loop.
Correspondingly, the gauge field receives a
constant identity component, which does not modify the field strength.
In the numerical computations we use $\varphi$ and transform back
to $\p$ by virtue of (\ref{eqn_phiphi}).

The spectrum of the Laplace operator is subject to two symmetries. The charge
conjugation (which is special for $SU(2)$)
\begin{eqnarray}
\p'^a_{n,-\z}=\e^{ab}\p_{n,\z}^{*b}\,,\qquad \lambda'_{n,-\z}=\lambda_{n,\z}
\label{eqn_symm_first}
\end{eqnarray}
relates two modes with the
{\em same eigenvalue} but opposite boundary conditions,
as indicated by the index $-\z$. In particular, the spectrum is two-fold
degenerate for periodic ($\z=0$) and antiperiodic ($\z=1/2$) boundary
conditions. Notice that $\p'$ is automatically orthogonal to $\p$.
We will therefore restrict ourselves to
$\z\in[0,1/2]$ without loss of generality.

The authors of Ref.~\cite{greensite:05} have found a relation between the lower and
upper end of the spectrum,
\begin{eqnarray}
\p''_{\N-n\,,\z}=(-1)^{\sum_\mu x_\mu}\p_{n,\z}\,,
\qquad\lambda''_{\N-n\,,\z}=4D-\lambda_{n,\z}\,.
\label{eqn_symm_second}
\end{eqnarray}
We will refer to this symmetry as the staggered symmetry.
It is only for even numbers $N_\mu$ of lattice points in all
directions that it (obviously) preserves both the boundary condition 
Eq. (\ref{eqn_bc}) and the periodicity in the space-like directions.
Thus, even numbers $N_{\mu}$ will be used throughout.
This symmetry, since it flips the sign of $\p$ at every other point, is
restricted to the discrete lattice (otherwise the spectrum of the continuum
Laplacian would be bounded from above as well as from below).

\subsection{Laplacian modes in vacuum backgrounds}

In order to illustrate the interplay of the boundary condition 
angle $\z$ and the Polyakov loop 
\begin{eqnarray}
\P(\vec{x})=\prod_{x_4=1}^{N_4} U_4(\vec{x},x_4)
\end{eqnarray}
with the spectrum of Laplacian modes, we now discuss the latter for vacuum
configurations.
We choose all time-like links to be identical
and to belong to the Abelian subgroup
consisting of diagonal matrices $U_4(x)\equiv \exp(2\pi i\a\sigma_3/N_4)$
which leads to a constant but adjustable Polyakov loop
\begin{eqnarray}
\frac{1}{2}\,\tr\,\P(\vec{x})=\cos(2\pi\a)\,,
\label{eqn_polloop_alpha}
\end{eqnarray}
whereas the space-like links are set equal to the identity.

\begin{figure}
\centering
\begin{tabular}{cc}
\hspace{-0.5cm}
\includegraphics[width=0.5\linewidth]
{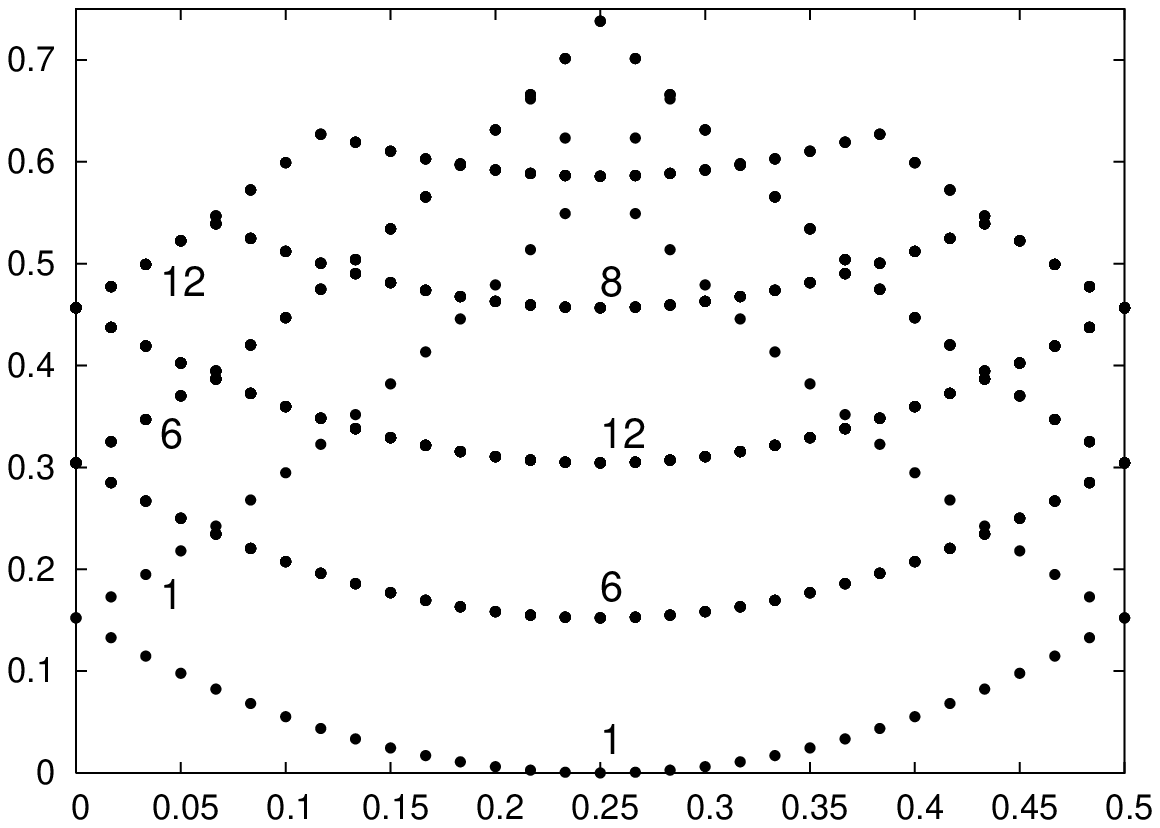}&
\hspace{-0.5cm}
\includegraphics[width=0.5\linewidth]
{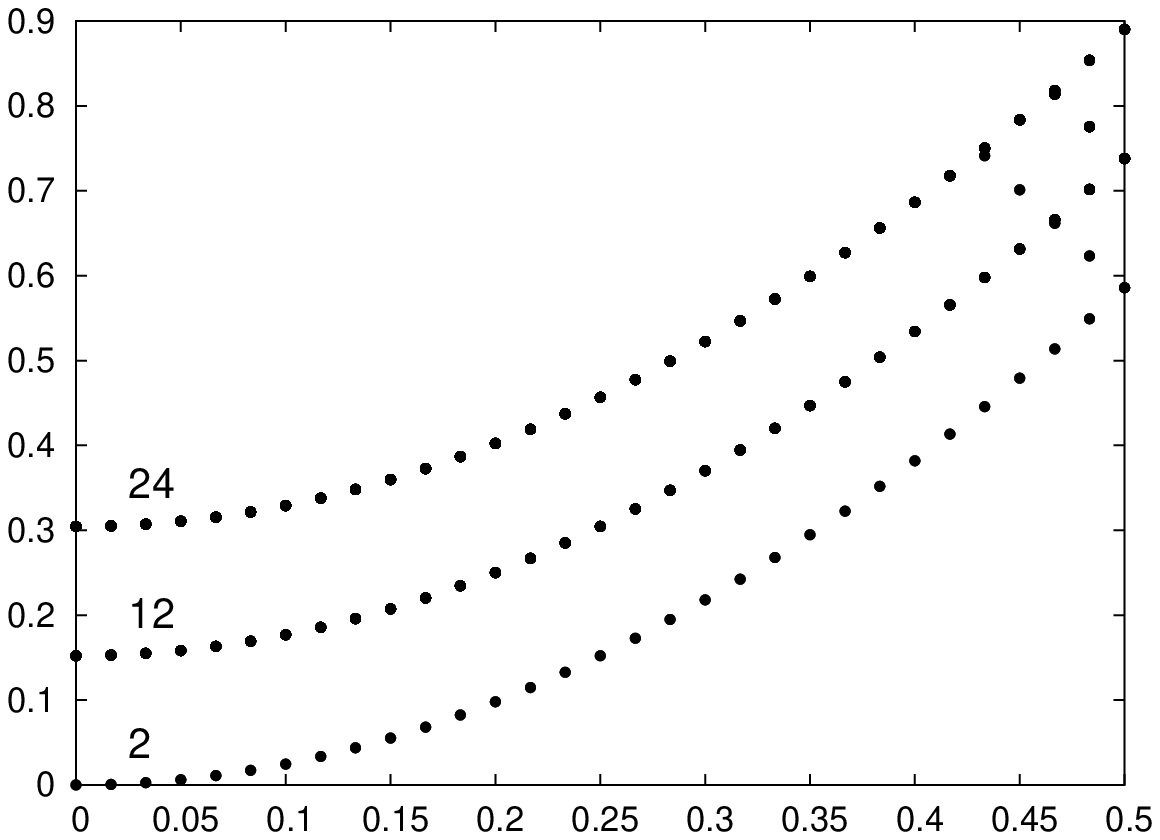}\\
(a)&(b)
\end{tabular}
\caption{
Spectrum of the Laplacian in vacuum backgrounds with
(a) traceless ($\a=1/4$) and (b) trivial Polyakov loop ($\a=0$) 
on a $16^3\cdot4$ lattice.
Plotted are the 38 lowest eigenvalues as a function of the angle 
$\z$ in the boundary condition. 
The numbers indicate the degeneracy of the bands.
The unit on the vertical scale is $0.152=2(1-\cos (2\pi/16))$, 
cf.\ Eq.\ (\ref{eqn_spec_vac}).}
\label{fig_spec_vac}
\end{figure}

The Laplace equation in these backgrounds can be solved
by considering the upper
and lower component separately and a simple product ansatz $\p(x)=\prod_\mu
\exp(2\pi i \beta_\mu x_\mu)$. The periodicity requirements give
$\beta_i=n_i/N_i,\,\beta_4=(n_4+\z)/N_4$ with integers $n_\mu\in[0,N_\mu-1]$. The eigenvalues are
\begin{eqnarray}
\lambda_{n_\mu,\z}=2\left[1-\cos2\pi(\frac{n_4}{N_4}+\frac{\z\pm\a}{N_4})+
\sum_{i=1}^3(1-\cos2\pi\frac{n_i}{N_i})\right]\,.
\label{eqn_spec_vac}
\end{eqnarray}
One can immediately read off 
that the dependence on $\z$ is trigonometric and shifted
by the Polyakov loop parameter $\a$, while the
spatial part shifts the spectrum and contributes to the degeneracy.

In Fig.\ \ref{fig_spec_vac} we give the vacuum spectra for
$\a=0$ and $\a=1/4$, which will be of particular interest below.
We plot the 38 lowest modes\footnote{
in order to completely 
fill the three lowest-lying bands at $\z=0$} 
and indicate the degeneracy of the bands.
They are reflected at the boundary of the `Brillouin zone' (where the
eigenmodes are at least two-fold degenerate as they should) and then 
cross each other.

Aspects of these spectra, especially the symmetry around $\z=1/4$ for $\alpha=1/4$,
will be reproduced by Laplacian modes in calorons and, to some extent,
in thermalized backgrounds.

\subsection{Calorons -- in the continuum and on the lattice}

Calorons are (anti-)selfdual instanton solutions on
$R^3\times S^1$. We will be particularly interested in caloron solutions with
the so-called holonomy being maximally nontrivial~\cite{kraan:98a,lee:98b}.
That is $\tr\,\P_\infty/2\equiv\cos(2\pi\alpha_\infty)=0$,
where $\P_\infty$ is the Polyakov loop at spatial 
infinity.  These, rather than the old Harrington-Shephard 
solutions~\cite{harrington:78} (with $\P_\infty=\pm \Eins_2$), shall be of 
relevance for the confined phase, where the trace of the Polyakov loop vanishes
on average. The authors of~\cite{diakonov:04a} have computed the contribution 
of calorons to the effective potential driving the Polyakov loop to that value. 

The nontrivial holonomy gives rise to a symmetry breaking $SU(2) \rightarrow
U(1)$. Therefore, it is plausible that calorons can be described by a
monopole and an antimonopole (calorons of charge $k$ consist of $|k|$ monopoles
and $|k|$ antimonopoles). Their masses are given by the eigenvalues of the
holonomy and equal for maximally nontrivial holonomy.

The moduli space of these solutions contains both small and large calorons, 
where the size is compared to the extension of the compact direction. 
Large calorons have two almost static action density lumps, which are 
identical for our choice of holonomy.
They merge for small calorons and this results
 in a strong time-dependence of the
action density (just like for conventional instantons).

The (untraced) Polyakov loop acts like a Higgs field (exponentiated to give
an element of the gauge group). It passes through $\Eins_2$ and $-\Eins_2$ 
near the core of the monopoles, where the symmetry is restored,
see Fig.\ \ref{fig_bosonic} (a).
This dipole persists even for small calorons within the single action density 
lump.

The monopoles make up the topological charge with the help of the so-called
Taubes winding \cite{taubes:84a}: one of the monopoles performs a full rotation
in the unbroken $U(1)$-subgroup relative to the other monopole
when completing a full period in the
time-like direction. For a gauge invariant statement one has to connect
the field strength at the different monopole cores by a Schwinger 
line~\cite{ilgenfritz:02a}.
In the periodic gauge that we use for the calorons \cite{kraan:98a}
put on the lattice, the link
variables $U_\mu(x)$ are static at the $\P=\Eins_2$ monopole, while they
rotate around the holonomy direction $\tau_3$ at the $\P=-\Eins_2$ monopole. 

The fermion zero mode in the caloron background follows the action density 
in that it has a maximum at one constituent monopole \cite{garciaperez:99c} 
plus a zero near the other one \cite{bruckmann:05a}. The first fact can be 
understood from the Callias index theorem \cite{callias:77}, which also 
explains why the zero mode hops with the angle $\z$, namely to the other 
monopole when the boundary condition is changed from periodic to antiperiodic.
When the phase $\z$ equals the holonomy parameter 
$\alpha_\infty$, the zero mode 
`sees' both monopoles, decaying with a power-law instead of exponentially.
The zero of the zero mode is connected to the nontrivial caloron topology.

On the lattice, calorons were first obtained by cooling in 
Ref.~\cite{ilgenfritz:02a}, where the typical behaviour of the action density, 
the Polyakov loop and the fermion zero mode was confirmed. Furthermore, the 
Polyakov loop averaged over the low-action region of the lattice was proposed 
to play the role of the asymptotic holonomy in the infinite continuum.
As for generic equilibrium configurations,
recent smearing studies revealed clusters of topological charge, that --
according to their content of Abelian monopoles and the corresponding behaviour
of the Polyakov loop -- 
have charges close to either $\pm 1$ or $\pm 1/2$,
resembling calorons and their constituents, respectively~\cite{ilgenfritz:04a}.

An alternative possibility to analyze calorons on the lattice is
to evaluate the discrete parallel transporters
from the continuum gauge field
and to cool the emerging lattice configuration by a few steps
in order to adapt it to periodic boundary conditions in space. 
This was first done in \cite{gattringer:03a}
for the case of gauge group $SU(3)$.
Because of lattice artefacts, the use of improved or even 
overimproved cooling \cite{garciaperez:93}
is advantageous when aiming at large calorons.
We will mainly use this approach to generate caloron backgrounds since it 
permits to control the locations of the constituent monopoles.

\section{Laplacian modes in caloron backgrounds}

\subsection{Eigenvalue spectra}

\begin{figure}[t]
\centering
\begin{tabular}{cc}
\hspace{-0.5cm}
\includegraphics[width=0.5\linewidth]
{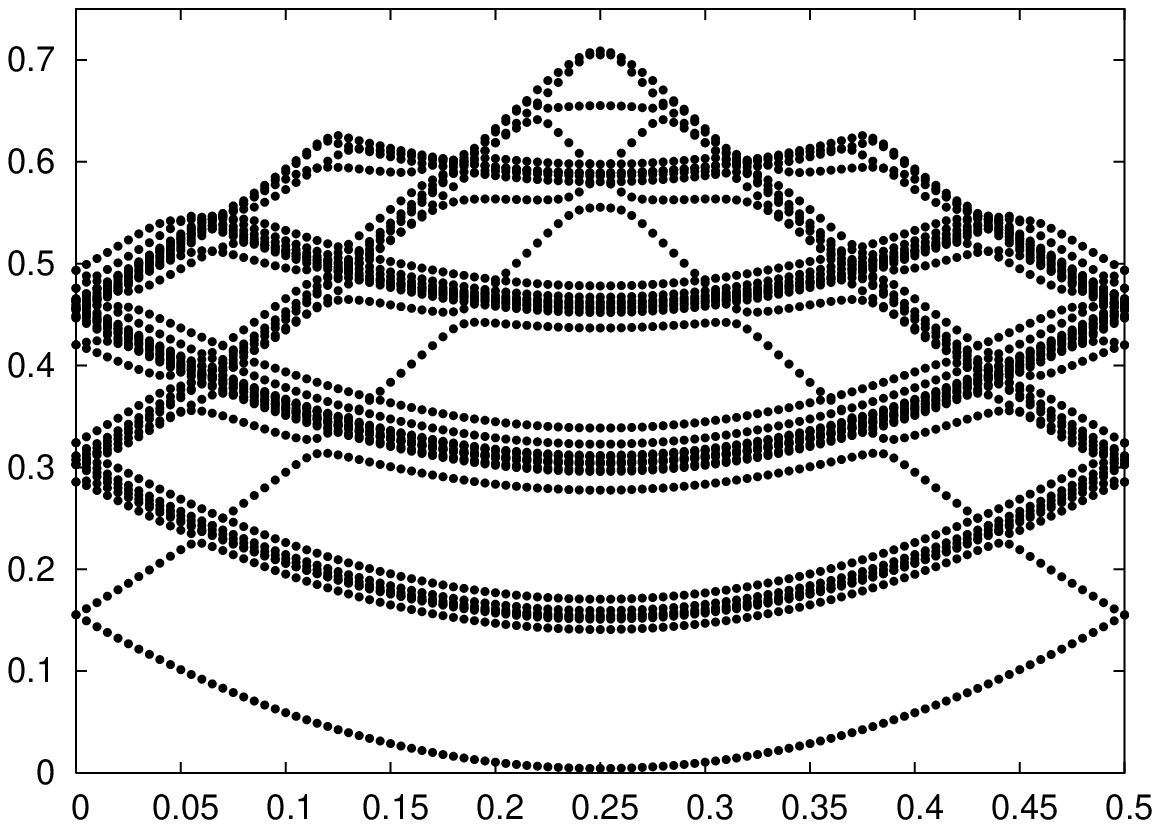}&
\hspace{-0.5cm}
\includegraphics[width=0.5\linewidth]
{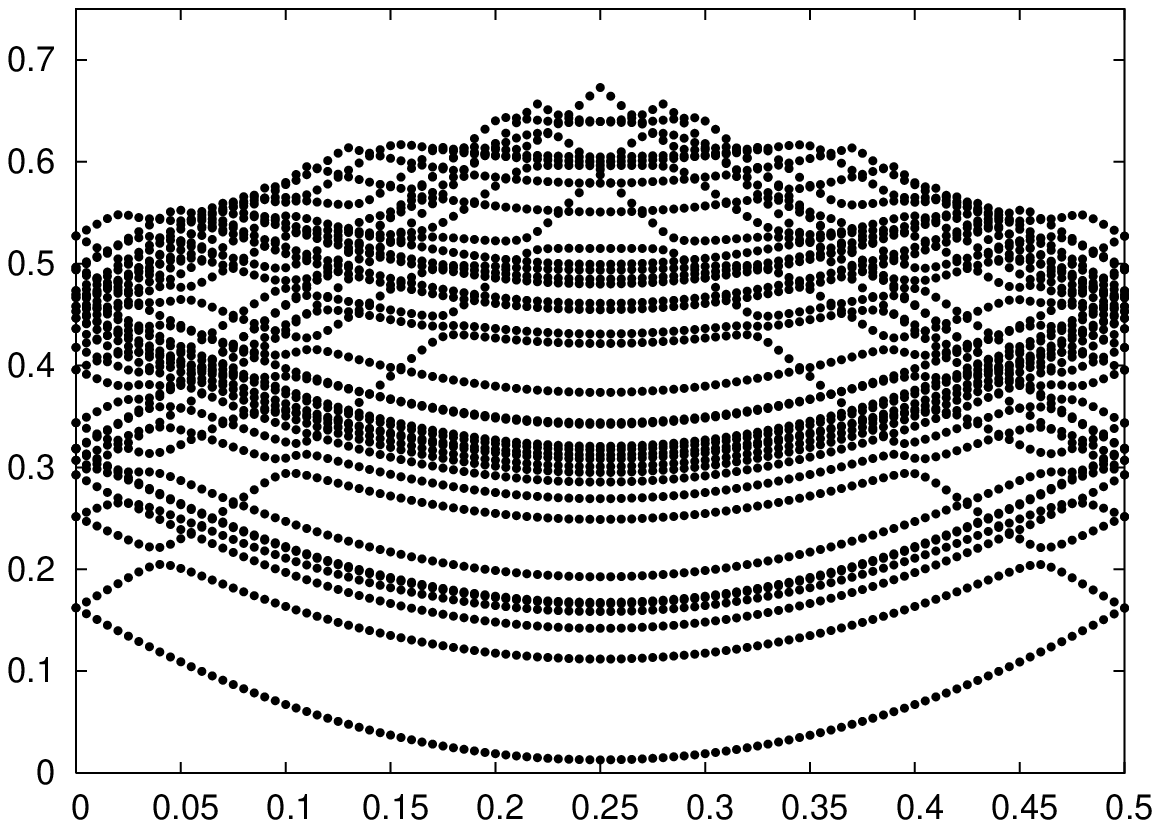}\\
(a)&(b)
\end{tabular}
\caption{
Dependence of the 38 lowest-lying eigenvalues on the boundary condition 
for (a) a small caloron and (b) a large caloron, both with maximally 
nontrivial holonomy on a $16^3\cdot4$ lattice.}
\label{fig_spec_cal}
\end{figure}

The lowest-lying Laplacian eigenvalues
in a background of a small and
a large caloron, both with maximally nontrivial holonomy, 
are displayed in Fig.\ \ref{fig_spec_cal} (a) and (b), respectively.
By construction, the trace of the Polyakov loop of these
lattice configurations vanishes outside of the
action density lumps of the constituent monopoles.
Accordingly, the spectra of the Laplacian are similar to those of a vacuum
background with the corresponding Polyakov loop with $\a=1/4$
in Eq.\ (\ref{eqn_polloop_alpha})
(see Fig.\ \ref{fig_spec_vac} (a)).
However, the zero in the vacuum spectrum at $\z=1/4$ is lifted,
the degenerate bands are splitted and some of the level crossings are 
inevitably avoided. Such effects are known form quantum systems at weak 
coupling (with the Laplacian playing the role of a Schr\"odinger operator). 
In this respect, the caloron backgrounds act like a small perturbation.

\begin{figure}[!b]
\centering
\begin{tabular}{cc}
\hspace{-0.5cm}
\includegraphics[width=0.5\linewidth]
{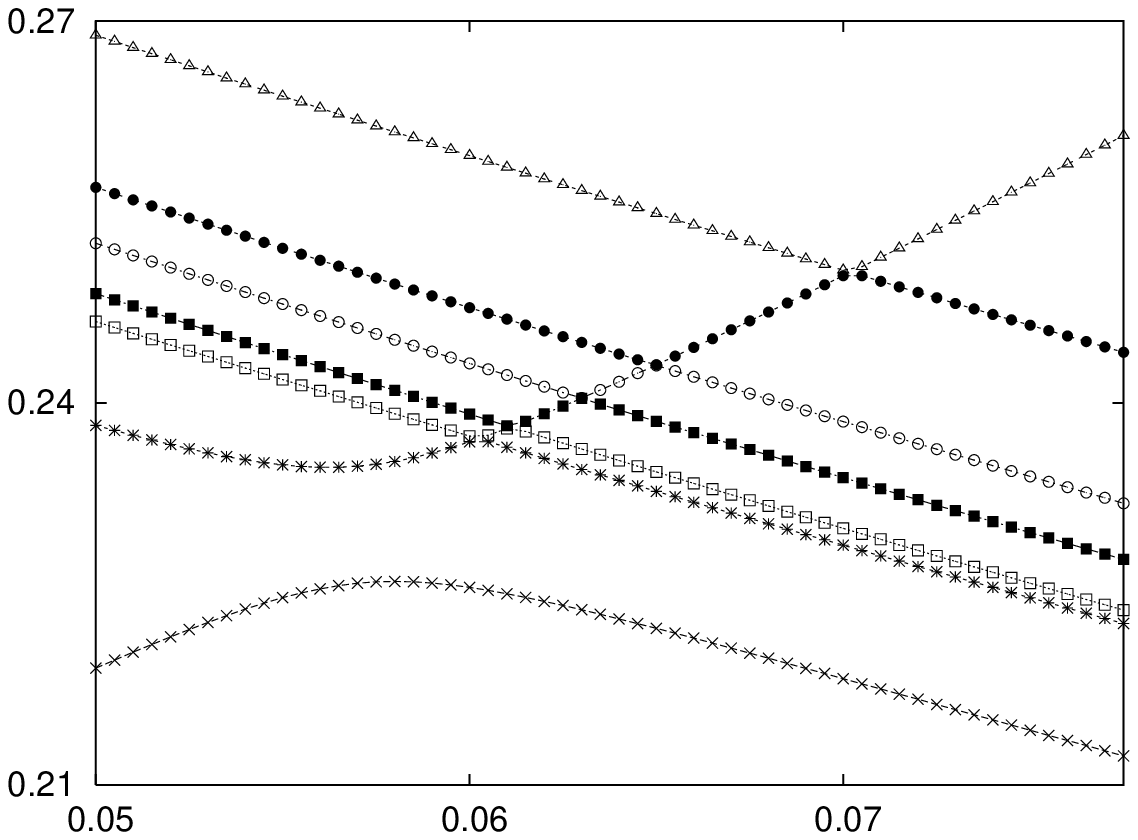}&
\hspace{-0.5cm}
\includegraphics[width=0.5\linewidth]
{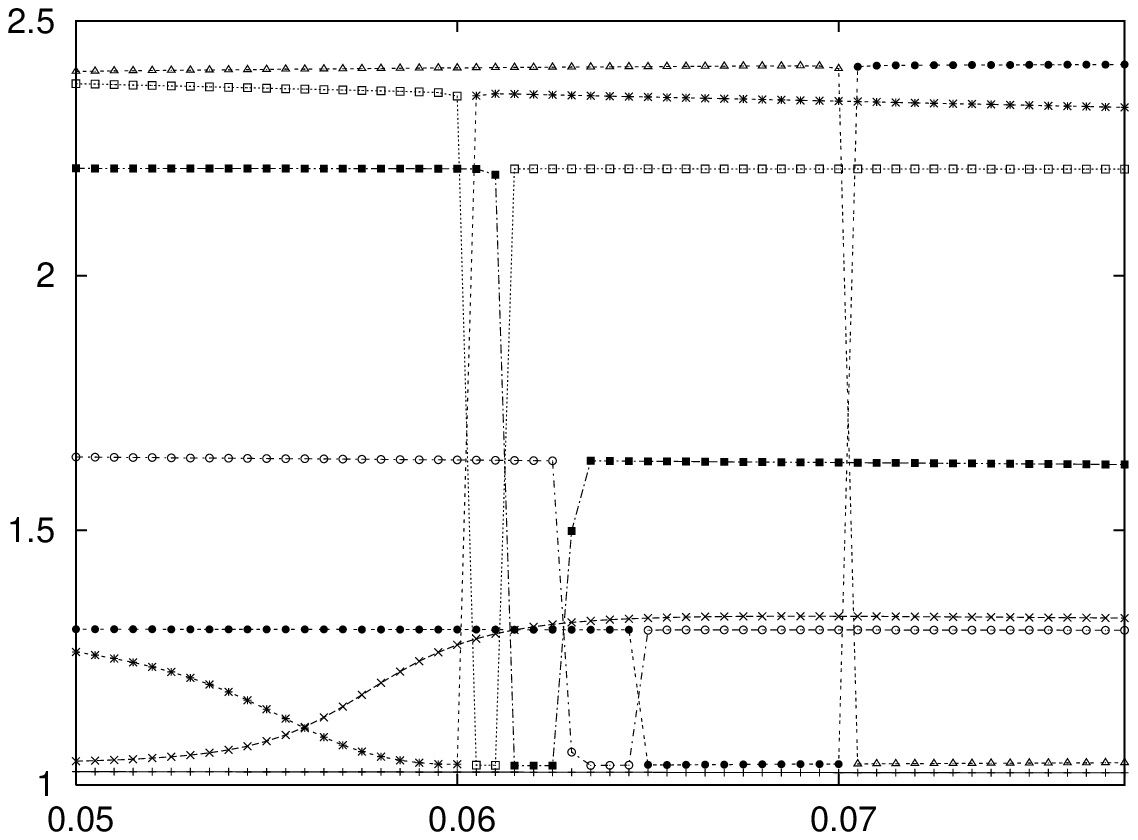}\\
(a)&(b)
\end{tabular}
\caption{Zooming in to
(a) the eigenvalues number 2 to 8 and (b) the corresponding inverse participation
ratios for the small caloron shown in the previous 
Fig.\ \ref{fig_spec_cal} (a).
For comparison also the IPR of the lowest mode (closest to $I=1$) is shown.}
\label{fig_spec_zoom}
\end{figure}

In order to characterise the localization of the eigenmodes by a global 
quantity, we will use the inverse participation ratio (IPR)
\begin{eqnarray}
I(\phi)={\rm Vol}\cdot\sum_x\rho^2(x)\,,
\qquad \rho(x)=|\p|^2(x)\,,
\qquad {\rm Vol}=N_1N_2N_3N_4\,.
\end{eqnarray}
A constant profile has the minimal $I=1$, whereas large IPR's signal
strong localization (with the lattice $\delta$-function saturating the 
upper bound $I={\rm Vol}$). For the IPR's of fermionic modes in caloron 
backgrounds we refer the reader to Ref.~\cite{ilgenfritz:05}.

\begin{figure}[t]
\centering
\begin{tabular}{cc}
\hspace{-0.5cm}
\includegraphics[width=0.5\linewidth]
{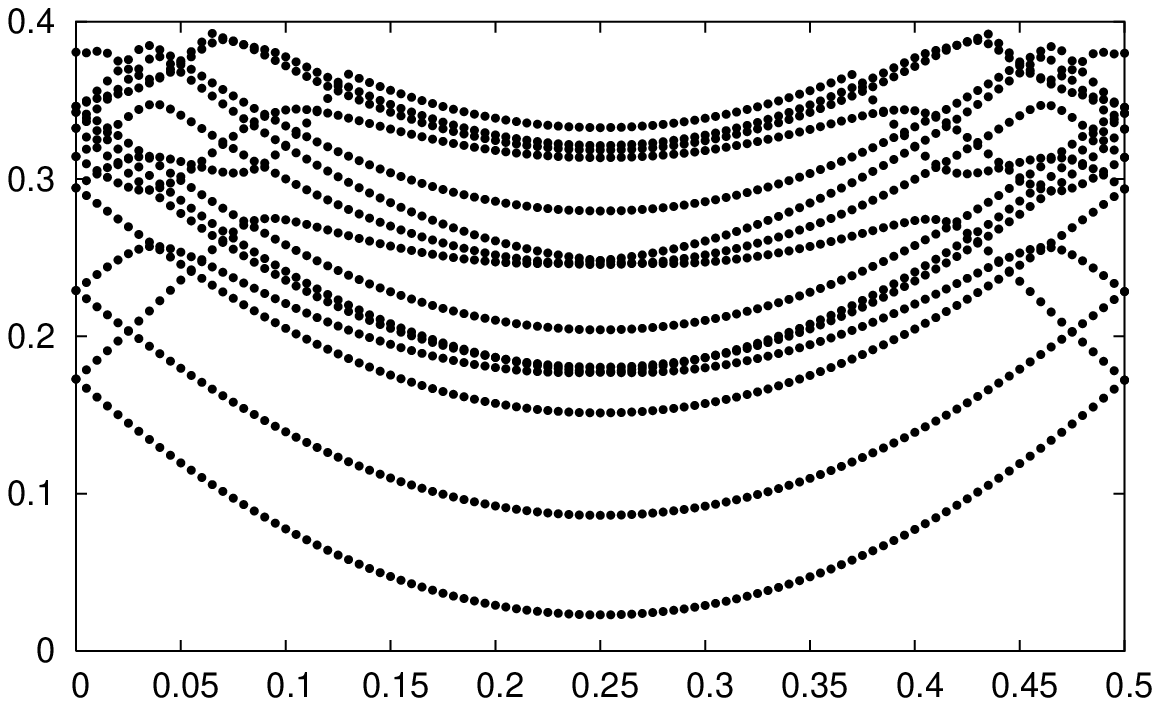}&
\hspace{-0.5cm}
\includegraphics[width=0.5\linewidth]
{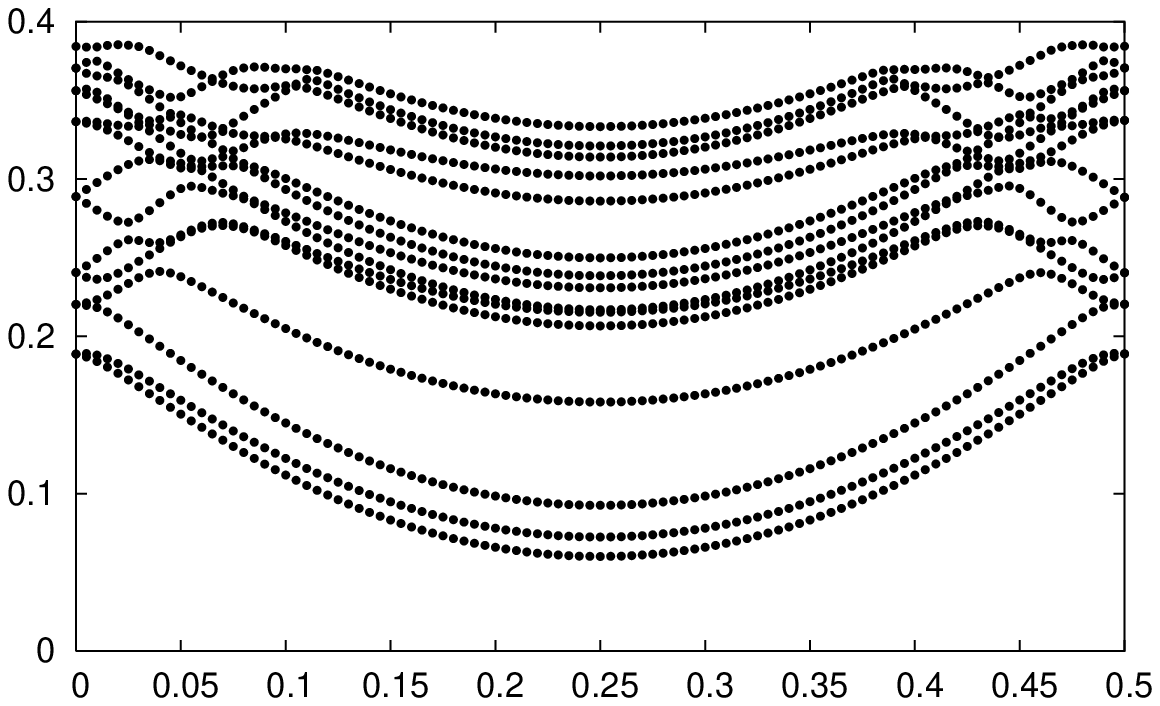}\\
(a)&(b)
\end{tabular}
\caption{Low-lying spectra (lowest 15 modes) for  charge 2 calorons,
(a) the case of pairwise well-separated constituents, (b) the case of like-charge 
constituent monopoles merged forming double-monopoles (rings, see the
text and Ref.\ \cite{bruckmann:04b}).}
\label{fig_cal_two}
\end{figure}

In Fig.\ \ref{fig_spec_zoom} we zoom in to a region of (almost) crossing
eigenvalues. 
In contrast to the vacuum spectrum, the eigenvalues of the second and 
third state are repelled from each other. As a remnant of the crossing 
these modes exchange their IPR's around that $\z$, which indicates that 
in the next $\z$-region each mode is similar to the complementary mode
in the previous region.
Outside of the crossing regions the IPR's are hardly changing.

\enlargethispage{\baselineskip}

At some other crossing points the eigenvalues come very close and the 
switch to the other eigenvalue branch is an instantaneous one 
even within the enlarged resolution in $\z$. 
The drastic changes in the IPR's illustrate this again.
One might speculate whether some of the level crossings
are exact ones in the continuum limit.
They may be governed by the 
quantum number corresponding to the axial symmetry of the caloron.
However, as the spatial volume increases, $N_i\rightarrow\infty$,
many bands will approach each other 
to form a continuous spectrum and to investigate the existence of 
discrete eigenstate would require a more detailed study.

In Fig.\ \ref{fig_cal_two} the spectra of two extreme cases of
charge-2 calorons 
\cite{bruckmann:04a} are shown. The four constituent 
monopoles are maximally separated in the first example, whereas in the second 
case the like-charge monopoles sit on top of each other forming rings 
(see Fig.\ 7 and Fig.\ 5 of Ref.\ \cite{bruckmann:04b}).

In an overall view,
the spectra are similar to those of charge-1 calorons. 
However, the eigenvalues are shifted upwards,
especially for the second example 
(Fig.\ \ref{fig_cal_two} (b)). The band structure is different as well,
with three close eigenvalues at the bottom of that spectrum.
The other case 
(of maximally separated constituents in the charge-2 caloron) seems to have 
curves with different curvature (seen in the middle of 
Fig.\ \ref{fig_cal_two} (a))
which upon a closer look turn out to have their minima
slightly away from $\z=1/4$. Although these details are interesting,
it is not clear how one can infer, for instance, the topological charge 
of the background configuration from these features.

In all the gauge field backgrounds studied so far, the IPR of the 
lowest eigenmode is close to $I=1$.
It means that this mode is quite spread out. 
For the excited states the IPR takes values up to $I=3.5$.

\subsection{Eigenmode profiles}

In this subsection we study
the local properties of the Laplacian modes 
for a caloron background
and show how they can reveal the underlying monopole constituents.

\begin{figure}[b]
\centering
\includegraphics[width=0.5\linewidth]
{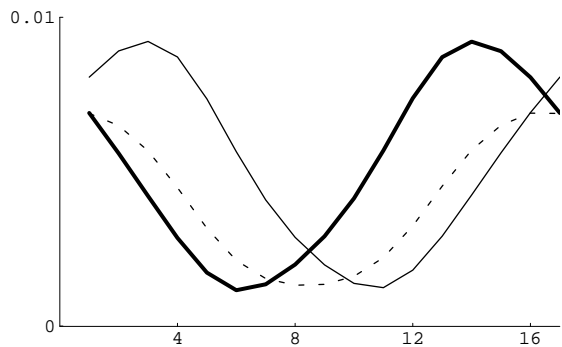}
\caption{The modulus $|\p|$
 of the lowest Laplacian mode with 
periodic (boldface line), antiperiodic (thin line) and intermediate 
(dashed line) boundary conditions 
along the line connecting the constituents of a large caloron on $16^3\cdot4$,
to be compared with Fig.\ \ref{fig_bosonic}.}
\label{fig_modulus_line}
\end{figure}

The modulus $|\p|$ of the lowest mode
 for a large caloron is plotted in
Fig.\ \ref{fig_modulus_line} for different boundary 
conditions. This figure should be compared to
Fig.\ \ref{fig_bosonic} (a) (for the background field) 
and 
Fig.\ \ref{fig_bosonic} (b) (for the fermion zero modes). 
One can see that the 
lowest Laplacian eigenmode with periodic boundary conditions 
has a maximum (located at $x_3=14$) near the monopole with a positive
Polyakov loop (at $x_3=12$).
Furthermore, it has a minimum (at $x_3=6$) near the 
monopole with negative Polyakov loop (at $x_3=4$).
Thus, it approximately
{\em localizes the two constituents
by virtue of a minimum and a maximum}.  
However, both are not very pronounced and occur 
with a shift of up to 2 lattice spacings (compared to a time-extent of
$N_4=4$ lattice spacings)
relative to the locations of the action density lumps.

We have also investigated excited eigenstates of the Laplacian and found 
that the first excited one has a higher maximum, but both minimum
and maximum are now shifted in the opposite direction.
Even higher states 
seem to be rather sensitive to the finite spatial volume.

What can also be read off from Fig.\ \ref{fig_modulus_line} is that 
the minimum and the maximum of the lowest mode
{\em move to the other constituent 
upon changing the boundary condition 
from periodic to antiperiodic}. 
This symmetry can be understood in the continuum. The caloron action  
density is invariant under an antiperiodic gauge transformation that  
exchanges the locations and masses of the constituent monopoles and  
inverts the holonomy. The latter two replacements are uneffective for  
a caloron at maximally nontrivial holonomy. 
Hence, for the Laplacian mode this gauge transformation reflects the  
profile function at the caloron's center of mass (around $x_3=8$).  
Furthermore, it replaces $\z$ by $\z+1/2$ (due to the antiperiodicity)  
which by charge conjugation is equivalent to $1/2-\z$. Thus the periodic  
mode turns into the antiperiodic one. 
This particular caloron symmetry also explains the mirror symmetry of the 
spectrum in Fig.\ \ref{fig_spec_cal} at $\z=1/4$.

In this respect the lowest Laplacian eigenmode
resembles the fermion zero mode.
However, the interpolation between the two extreme boundary
conditions is different: The minimum moves through the center of mass, 
whereas the maximum decreases and goes through the `boundary' at `infinity'.
To demonstrate this we have included the intermediate boundary condition
$\z=0.25$ in the figure.

\begin{figure}
\centering
\begin{tabular}{cc}
\includegraphics[width=0.45\linewidth]
{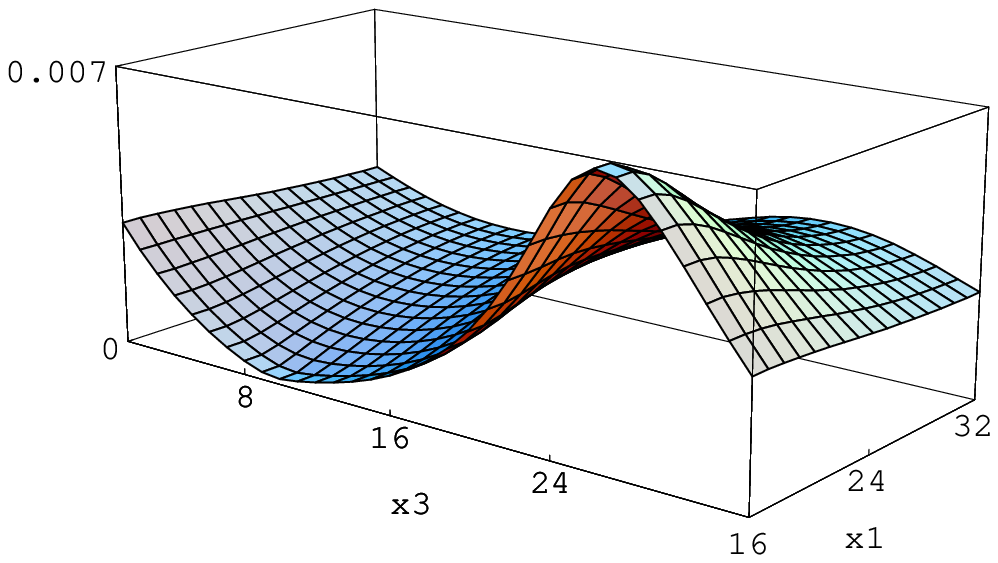}&
\includegraphics[width=0.45\linewidth]
{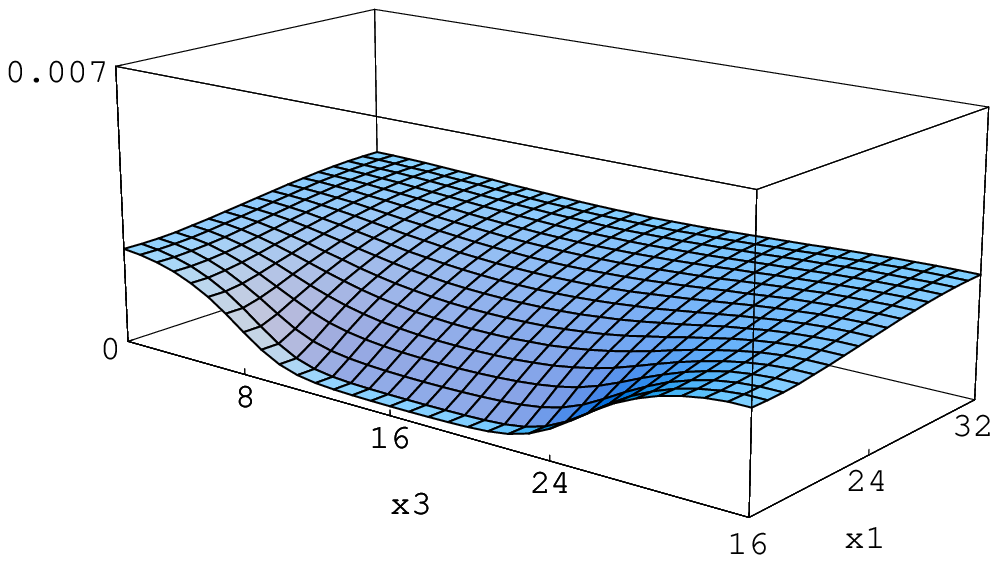}\\
(a)&(b)
\end{tabular}
\caption{Space-space plots of the modulus of the lowest eigenstate
for a large caloron on a $32^3\cdot4$ lattic, with (a) periodic and (b) 
intermediate boundary condition.}
\label{fig_modulus_large}
\end{figure}

In order to clarify how finite volume effects influence these findings, we
have doubled both the spatial extension of the lattice and the size of the
caloron, which now has its constituents at $x_1=x_2=16$, $x_3=8$ and $25$. 
The lowest eigenstate with periodic boundary condition, shown in 
Fig.\ \ref{fig_modulus_large} (a), has a slightly more pronounced maximum,
still shifted, to $x_3=26$. Around the maximum the modulus is actually close to 
spherically symmetric as expected since the constituents are quite well 
separated compared to the typical scale $N_4=4$. The minimum is shifted 
as well, namely to $x_3=10$.

This configuration also clarifies the behaviour of the minimum under the change
of boundary conditions. Fig.\ \ref{fig_modulus_large} (b) shows that 
for an intermediate boundary condition the minimum extends to a {\em valley} 
between the minimum locations corresponding to the periodic and antiperiodic 
modes, respectively. 

\begin{figure}
\centering
\begin{tabular}{cc}
\includegraphics[width=0.45\linewidth]
{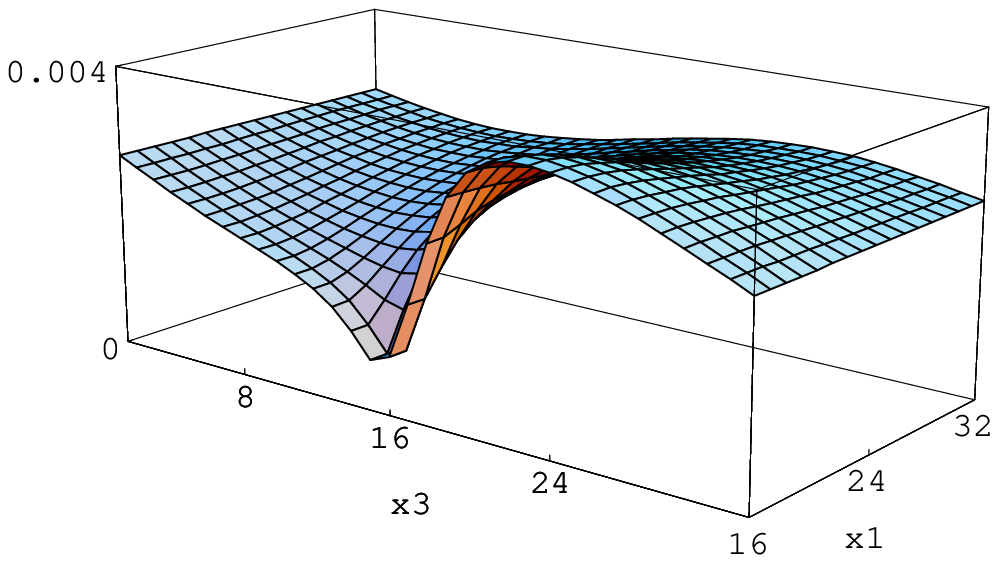}&
\includegraphics[width=0.45\linewidth]
{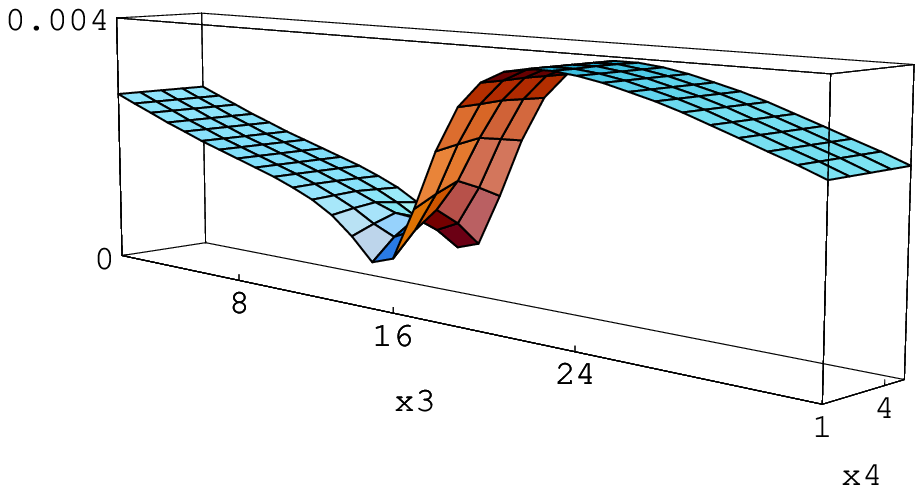}\\
(a)&(b)
\end{tabular}
\caption{(a) Space-space and (b) space-time plot of the modulus of the 
lowest eigenstate with periodic boundary conditions on the background 
of a small caloron on a $32^3\cdot4$ lattice.}
\label{fig_modulus_small}
\end{figure}

It is instructive to see how the lowest Laplacian eigenmode looks like
in the background of a smaller caloron.
The configuration that we employ for this demonstration 
has only one lump of action density, but the typical dipole behaviour of 
the Polyakov loop,
the latter being close to $-\Eins_2$ at $x_3=15$ and close to $\Eins_2$ 
at $x_3=18$.
The modulus of the lowest Laplacian eigenmode has a strong gradient 
at the location of the lump, too, see Fig.\ \ref{fig_modulus_small} (a).
As a matter of fact, the minimum ($x_3=15$) reflects the Polyakov loop
minimum quite well, while the maximum ($x_3=23$) is shifted far outwards. 

The maximum is also less pronounced compared to the one of the large caloron.
Actually all the presented 
profiles of the lowest Laplacian eigenmodes away from the monopoles become
close to the value an entirely constant mode would have on the corresponding
lattice
(0.0078 for $16^3\cdot4$, 0.0028 for $32^3\cdot4$). Therefore, the Laplacian 
modes should be viewed as a wave with local modifications rather than a 
localized state.
The rationale behind that might be the absence of an effective mass that
would localize the modes exponentially. For the fermion zero modes this role is
played by (the difference of the boundary condition and) the eigenvalues of the holonomy.

Moreover, the maximum of the Laplacian mode is almost static,
even for the small `instanton-like' 
caloron, as can be seen in Fig.\ \ref{fig_modulus_small} (b). The minimum, on 
the other hand, shows up in a particular time-slice, exactly
where the action density is maximal.
This is another sign that, from a practical point of view, the use of the 
minimum is superior compared to the maximum when looking for small calorons 
(or instantons). The time dependence of the minimum becomes much weaker for 
large calorons as we have observed (but not shown here), reflecting the
more static character of the solution.  
In the continuum, Laplacian eigenmodes in nontrivial backgrounds will have a 
zero of topological origin, like fermions do \cite{bruckmann:05a}. 

To summarize, the assignment of a maximum and a minimum in
the modulus of the lowest Laplacian eigenmode
to the constituents in a caloron is similar to the
behaviour of the fermion zero mode. Yet the profile is less localized and
shifted. For small calorons 
it is recommendable to use the minimum, as it was done in 
Ref.~\cite{deforcrand:01a}, to find instantons on the lattice
(in the continuum, the coincidence of the minimum with the instanton center 
was shown analytically in \cite{bruckmann:01a}). The shift observed in that 
reference nicely agrees with our findings.

\begin{figure}
\centering
\begin{tabular}{cc}
\includegraphics[width=0.48\linewidth]
{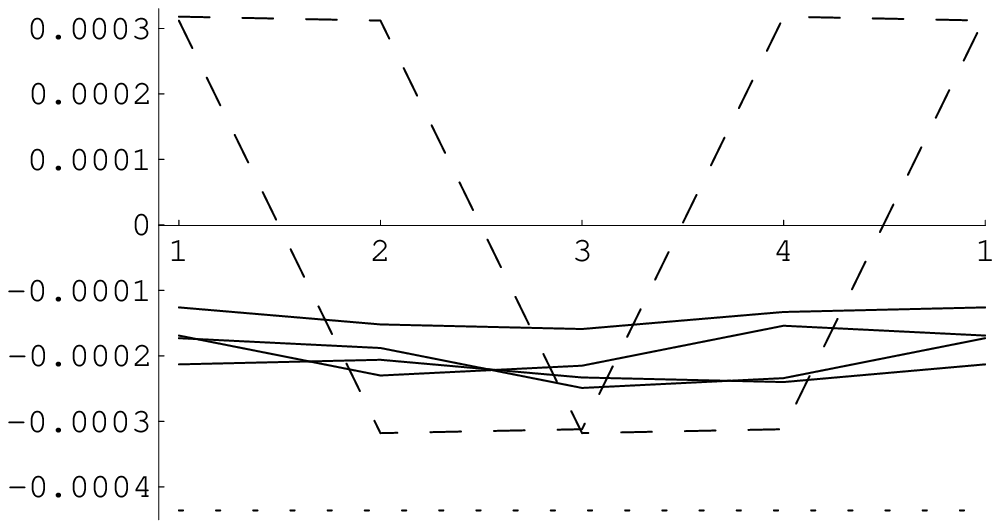}&
\includegraphics[width=0.48\linewidth]
{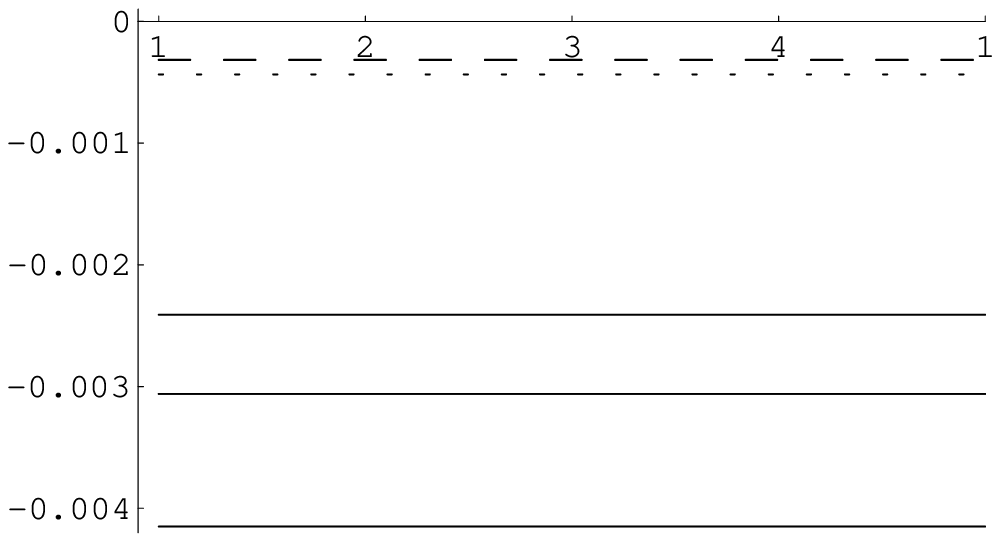}\\
(a)&(b)
\end{tabular}
\caption{Components of the lowest eigenstate with periodic 
boundary conditions, in both the fundamental (full lines, real and
imaginary part of both components) and the
adjoint representation (dashed line, the $\tau_3$-component as dotted line),
as function of $x_4$ at (a) the point $x_1=x_2=16$, $x_3=8$ showing the 
Taubes winding and (b) the point $x_1=x_2=16$, $x_3=25$ (no winding).
Two pairs of lines fall on top of each other in plot (b): real and
imaginary part of the second fundamental component and the first two 
adjoint components.}
\label{fig_taubes_winding}
\end{figure}

By inspecting the components of the Laplacian modes one finds signatures 
of the Taubes winding. We plot the real and imaginary part of
$\p^{a=1,2}_{\z=0}$ 
in Fig.\ \ref{fig_taubes_winding}, showing the time dependence at both 
monopoles. At one monopole core  the components are fully static,
Fig.\ \ref{fig_taubes_winding} (b). 
At the other monopole, carrying the Taubes winding, they oscillate over 
one period, however, they seem to contain an additional constant part,
Fig.\ \ref{fig_taubes_winding} (a).

The general picture of Laplacian modes described so far passes over to the
caloron examples of charge $Q=2$. The periodic and antiperiodic modes 
possess maxima and minima near the corresponding constituents.
However, the ring structure in the action density of two overlapping like-charge
monopoles is not resolved by the Laplacian modes,
in contrast to the fermion zero modes. 

\subsection{Adjoint representation}

In this subsection we discuss (in short) the properties of
the lowest Laplacian eigenmode in the adjoint representation. 
In the Laplacian, the fundamental links $U_\mu^{ab}(x)$ are simply replaced 
by the adjoint ones $\tr (\sigma^A U_\mu(x)\sigma^B U_\mu^\dagger(x))/2$
with indices $A,\,B$ running from 1 to 3. 
The eigenfunctions $\p_n^A(x)$ are real in this representation and so there 
is no continuous phase available to modify the boundary conditions as this
was possible in the fundamental representation,
Eq.\ (\ref{eqn_bc}) and (\ref{eqn_phiphi}). Nevertheless, we have 
included the possibility of antiperiodic boundary conditions in the computations.

\begin{figure}
\centering
\includegraphics[width=0.5\linewidth]
{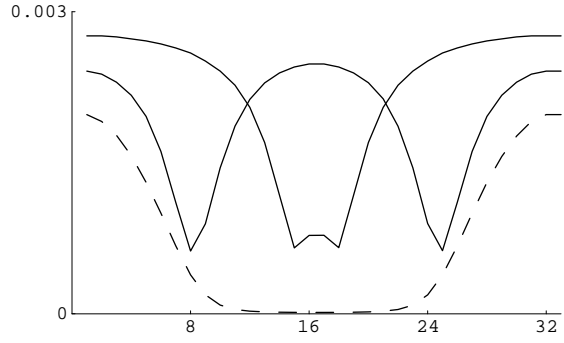}
\caption{The lowest adjoint eigenmodes (with periodic boundary condition) 
for the large and the small caloron on a $32^3\cdot4$ lattice.
The dashed line shows the lowest adjoint eigenmode with antiperiodic 
boundary condition for the large caloron.} 
\label{fig_adj}
\end{figure}

In caloron backgrounds we basically reproduce the findings of
\cite{alexandrou:00a}, namely that the constituent monopoles are localized 
by {\em minima} in the modulus of the lowest-lying adjoint Laplacian eigenmode 
with periodic boundary condition.
In other words, the single adjoint mode
`sees' monopoles of both kinds simultaneously.
As Fig.\ \ref{fig_adj} shows, these minima have no shift problems and 
start to join for a small caloron. That the lowest adjoint Laplacian
eigenmode
on an instanton background has a zero of second order at the instanton 
location has been shown in the continuum in \cite{bruckmann:01a}. 
Zeroes in this mode
underly the construction of Abelian 
projected monopoles in the Laplacian Abelian gauge \cite{vandersijs:97}. 
Thus, for calorons these monopoles coincide with the (gauge independent) 
constituent monopoles.

\begin{figure}[b]
\centering
\begin{tabular}{cc}
\includegraphics[width=0.48\linewidth]
{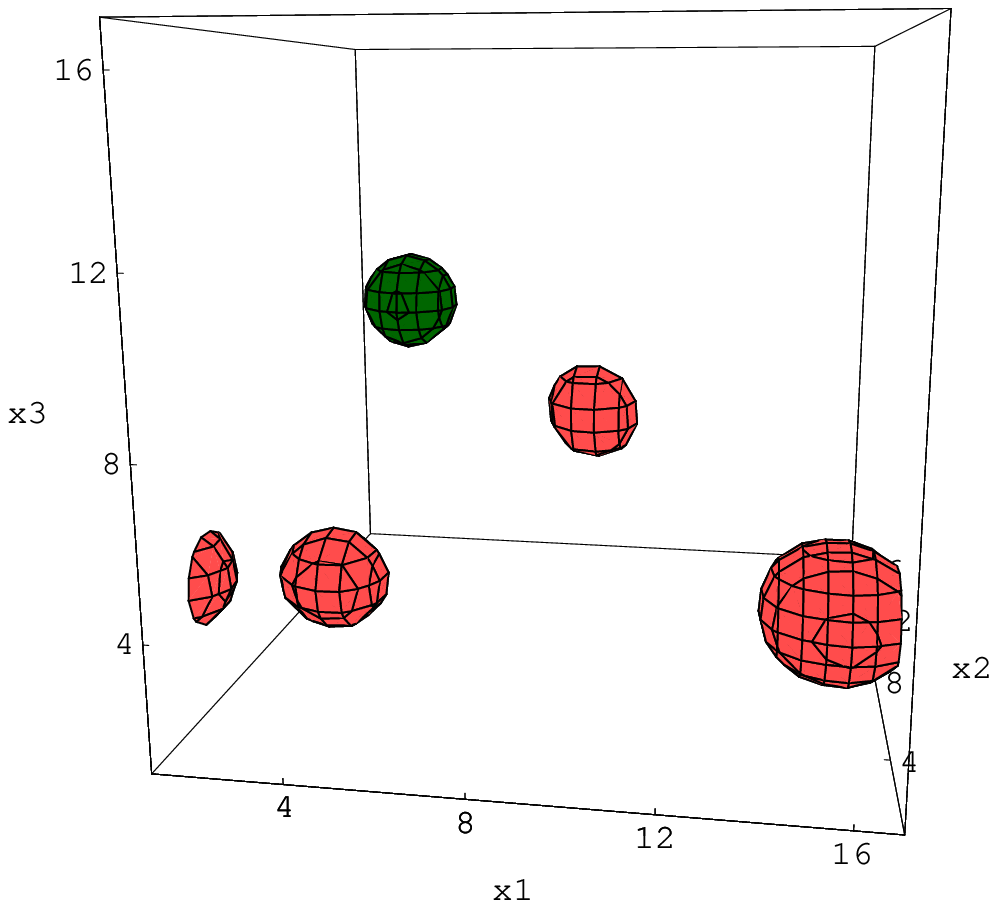}&
\includegraphics[width=0.48\linewidth]
{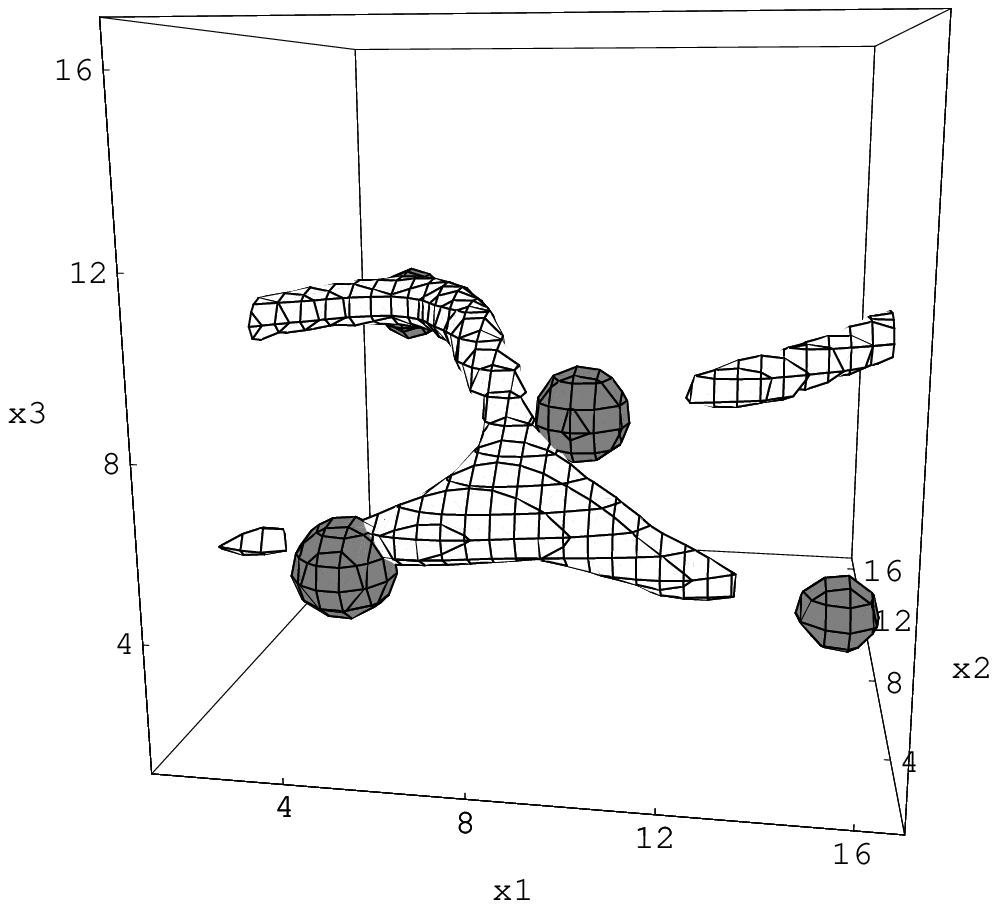}\\
(a)&(b)
\end{tabular}
\caption{The semiclassical configuration described in the text:
(a) isosurfaces of the
topological density (positive as dark green, negative as light red), 
(b) isosurfaces locating the minima of the modulus of the lowest
adjoint modes, both for
periodic (pointlike, balls corresponding to the constituents seen in (a)) 
and antiperiodic boundary conditions (one-dimensional network).} 
\label{fig_adj_extended}
\end{figure}

The lowest adjoint Laplacian eigenmode with {\em antiperiodic} boundary 
condition develops an extended static zero sheet between the monopoles 
(similar to the lowest fundamental eigenmode with intermediate
boundary conditions). The corresponding profile is shown in 
Fig.\ \ref{fig_adj} as a dashed line.

This feature can be of interest for detection purposes, too.
In order to illustrate this, we have investigated a semiclassical case.
The gauge field
configuration has been obtained by cooling down to a plateau with an action 
of 1.92 instanton units and a topological charge of $-1$ \cite{bruckmann:04b}. 
As has been described in that reference,
there are one selfdual and three antiselfdual monopoles, distinguished by 
color in Fig.\ \ref{fig_adj_extended} (a).
The Polyakov loop at one of the cores is found close to $\Eins_2$
and close to $-\Eins_2$ at the three others.
Isosurfaces (at small value) of the modulus of the lowest adjoint Laplacian 
eigenmode with either boundary condition are shown in 
Fig.\ \ref{fig_adj_extended} (b).
As for the simple (dissociated) caloron, the periodic 
mode has static minima at the monopole locations, whereas the regions of
small modulus of the lowest antiperiodic mode form a network connecting them.

The lowest adjoint Laplacian eigenmode with periodic boundary condition
actually reveals the Taubes winding in a very clear manner, 
cf.\ Fig.\ \ref{fig_taubes_winding}. At the `rotating' monopole (a) the 
$\tau_1$ and $\tau_2$ components rotate around the holonomy subgroup
generated by $\tau_3$, whereas at the other monopole (b) all components 
are static. For the antiperiodic lowest mode all components perform half a
rotation 
everywhere (to account for the antiperiodicity) with the $\tau_3$-component
being suppressed.

\section{Hopping of the lowest Laplacian mode for thermalized configurations}
\label{sec_hopping}

In the last section we have shown that the 
lowest Laplacian eigenmodes (fundamental and adjoint)
reflect certain properties of smooth classical gauge field backgrounds.
Now we explore these modes as an analyzing tool
for thermalized gauge field configurations
and concentrate on how the dependence on boundary conditions
gives additional information.

As a set of thermalized background configurations we take an
ensemble of 50 configurations on a $16^3\cdot 4$ lattice, generated
by Monte Carlo heat bath sampling with the Wilson action
at $\beta=2.2$. It represents the
confining phase at finite temperature, namely at $T \simeq 0.75\, T_c$ 
(deduced from $\sigma(0)a^2=0.22$ for our $\beta$ \cite{bloch:03}
and $T_c/\sqrt{\sigma(0)}=0.709$ for $SU(2)$ YM theory \cite{lucini:05}).

\begin{figure}[t]
\centering
\begin{tabular}{cc}
\hspace{-0.5cm}
\includegraphics[width=0.5\linewidth]
{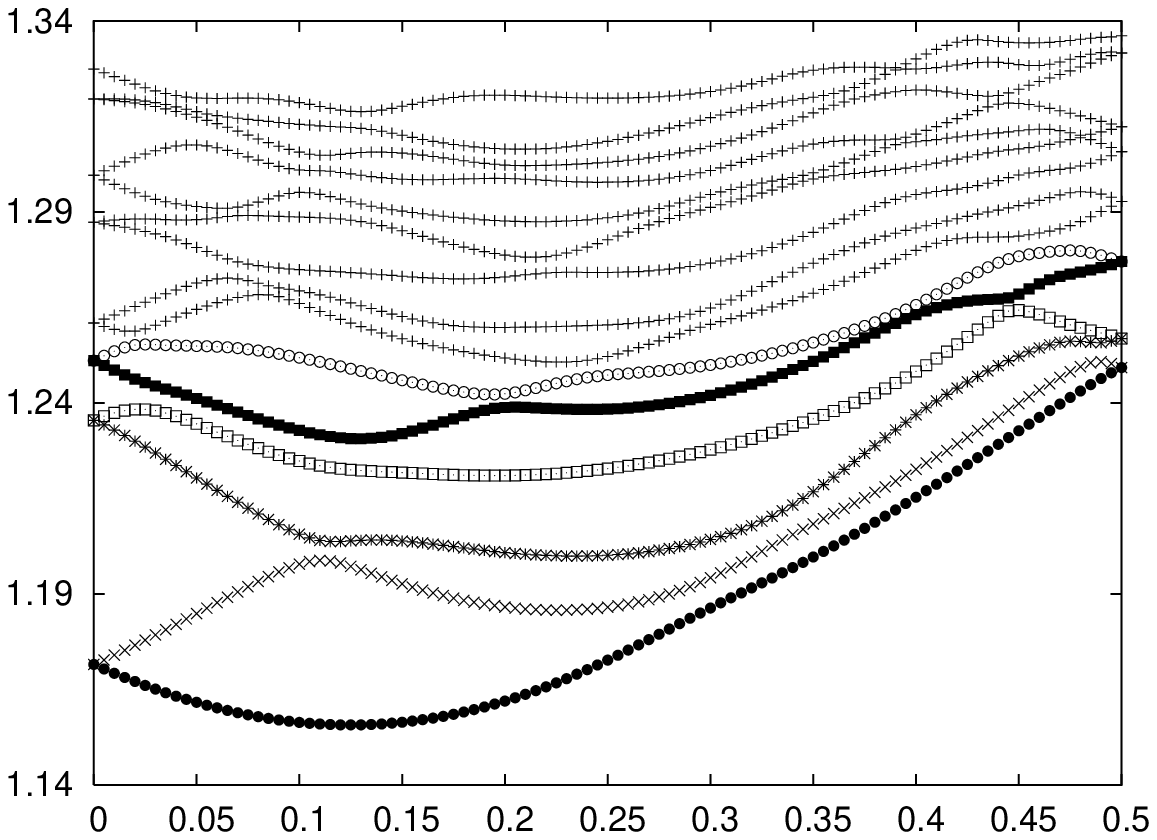}&
\hspace{-0.5cm}
\includegraphics[width=0.5\linewidth]
{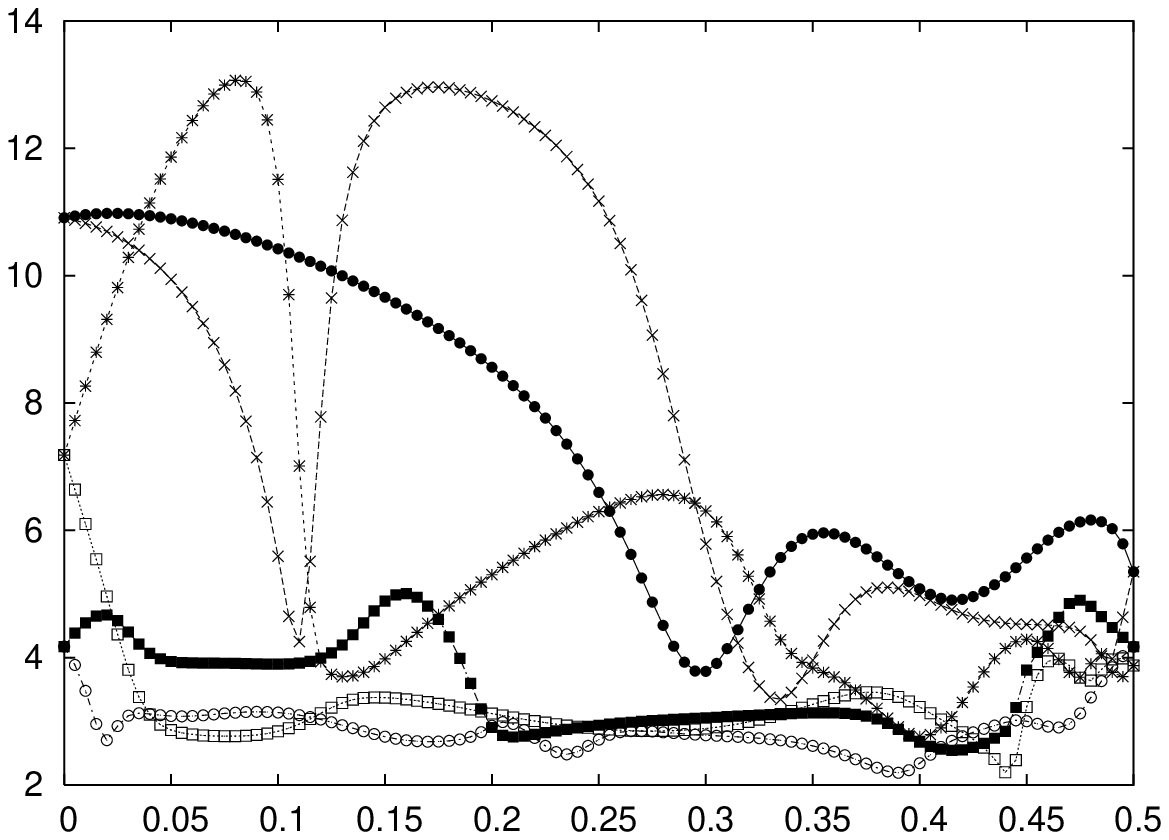}\\
(a)&(b)\\
\hspace{-0.5cm}
\includegraphics[width=0.5\linewidth]
{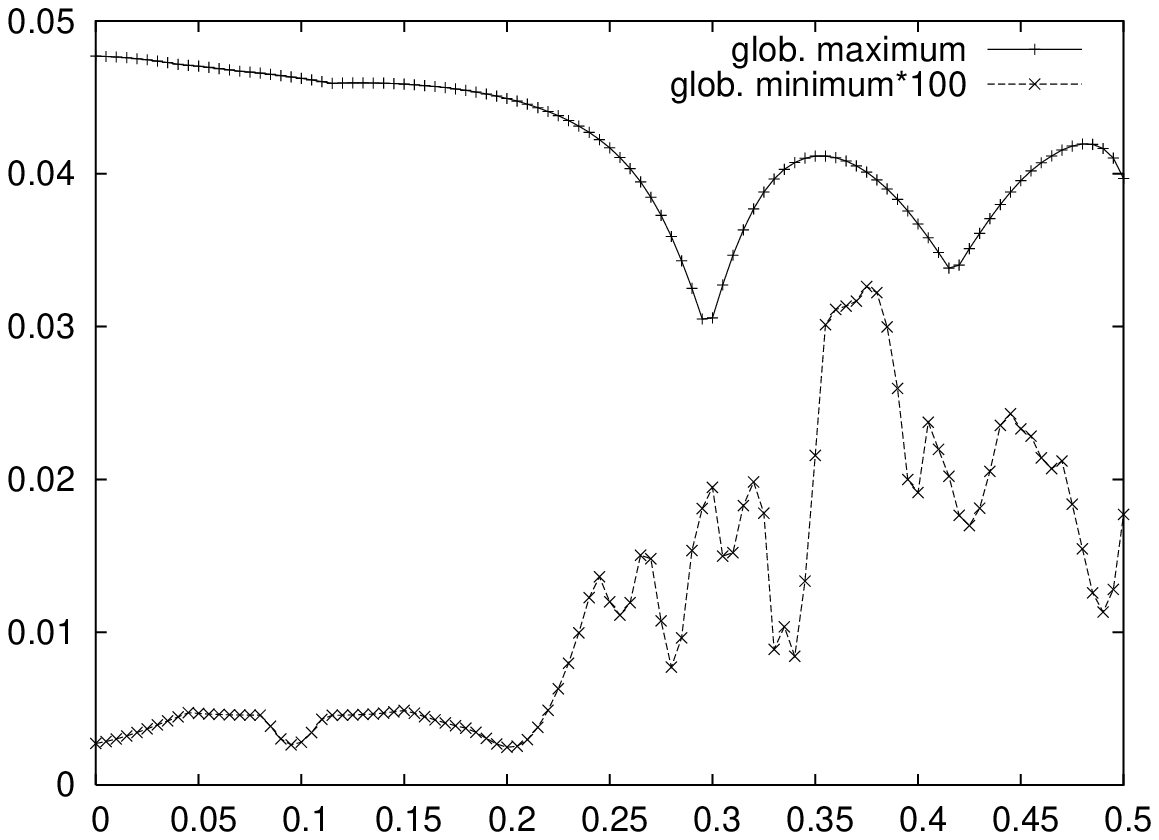}&
\hspace{-0.5cm}
\includegraphics[width=0.5\linewidth]
{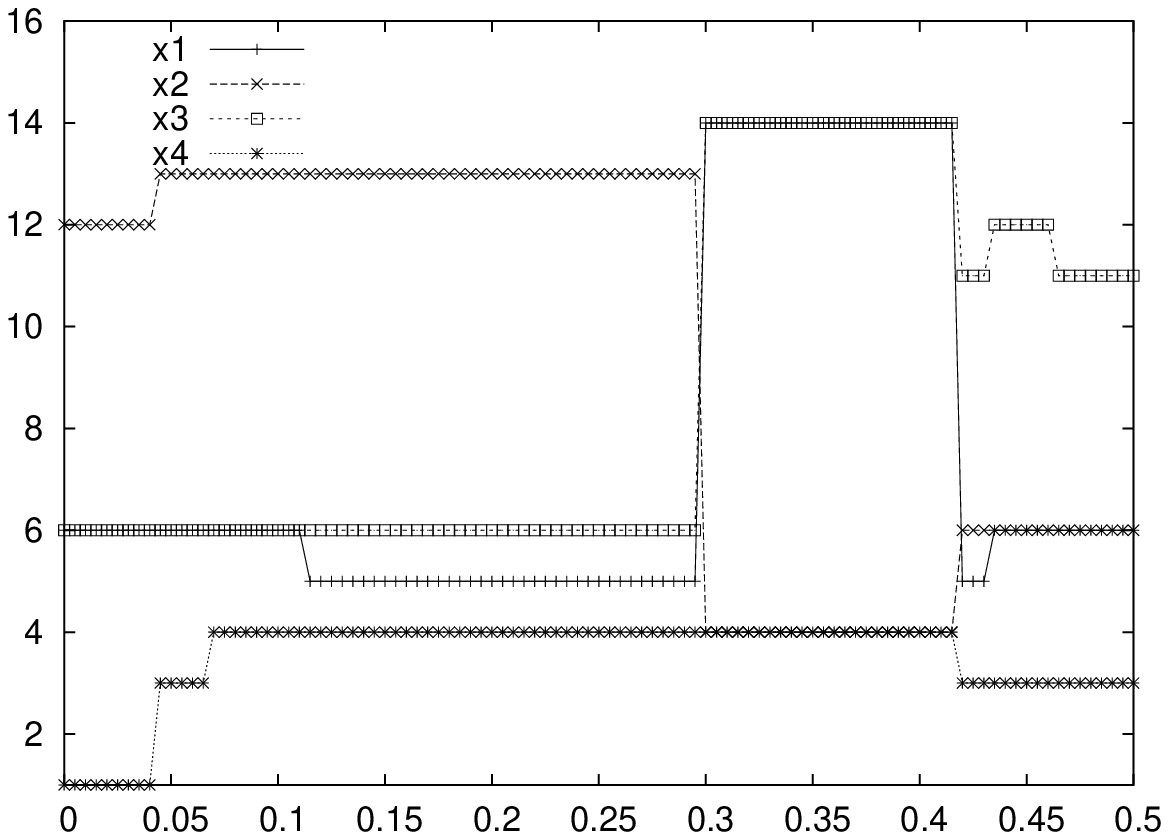}\\
(c)&(d)
\end{tabular}
\caption{Laplacian modes for a configuration in the confined phase.
Plotted over the boundary condition $\zeta$ are
(a) the 15 lowest-lying eigenvalues in lattice units,
(b) the IPR of the 6 lowest-lying modes (with corresponding symbols);
(c) the value of the modulus of the lowest eigenmode at its global maximum 
and minimum (multiplied by a factor 100),
(d) coordinates of the global maximum of that mode,
displaying the typical $\zeta$-intervals.}
\label{fig_spec_finite001}
\end{figure}

Fig.\ \ref{fig_spec_finite001} (a) shows a typical spectral flow
of the 15 lowest-lying Laplacian
modes with the boundary condition angle $\z$.
The first outstanding feature
to notice is that the eigenvalues themselves are much bigger\footnote{
With our lattice spacing
the eigenvalue unit in Fig.\ \ref{fig_spec_finite001} (a) is
$1/a^2=0.87\,GeV^2$.}
 than for the 
smooth backgrounds considered so far.
This is most naturally ascribed to the ultraviolet noise present in the 
background. The latter has also removed any remnants of the vacuum (or 
caloron) band structure in this plot. 
Still, the typical near-crossing points are present,
where again the IPR signals big rearrangements in 
Fig.\ \ref{fig_spec_finite001} (b),
see e.g.\ the behaviour of the second and third mode around $\z=0.12$.

Inspecting several independent configurations in the confined phase, 
the lowest eigenvalue takes on its smallest value around $\z=1/4$
(in this respect the configuration in Fig. 12
is not a typical one). 
This reflects the fact that the average Polyakov loop is close to traceless.
On the contrary, in the deconfined phase
all low-lying eigenvalues become minimal 
very close to $\z=0$ (and are much bigger at $\z=1/2$)
with the spectrum {grossly similar to Fig.\ \ref{fig_spec_vac} (b),
in accordance with the asymmetric Polyakov loop distribution in that phase.

The main observation in the context of Laplacian modes for thermalized
backgrounds is 
the effect of `hopping' of these modes with changing $\z$.
The IPR of the lowest mode already gives a hint on the existence of what 
we will call the `$\z$-intervals'. In the example of 
Fig.\ \ref{fig_spec_finite001} (b) there are three such intervals,
inbetween which the IPR has a minimum.
It means that the lowest mode 
delocalizes there in order to rearrange itself.
This is confirmed by the inspection of the global maximum 
in Fig.\ \ref{fig_spec_finite001} (c): the value of the modulus at the
maximum is minimal at the two transition points $\z=0.300$ and $\z=0.415$.
The main feature of the $\z$-intervals is
the pinning down of the lowest mode to particular locations.
That means that the coordinates of the global maximum
have constant values within the $\z$-intervals and
jump at the transition points,
as it is clearly visible in Fig.\ \ref{fig_spec_finite001} (d). 

For some configurations we have observed two distinct minima 
in the lowest eigenvalue as a function of $\z$,
which turned out to be another signal for the existence of 
the $\z$-intervals.\\

\begin{figure}[b]
\centering
\includegraphics[width=0.5\linewidth]
{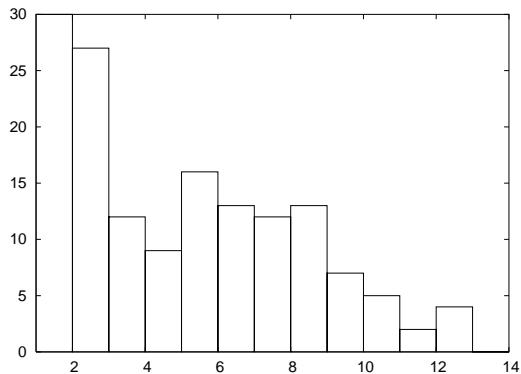}
\caption{
Statistics of the spatial distance of
jumps from 50 thermalized configurations
(160 jumps occured with spatial distance between 1 and 2 lattice spacings
but were not considered further).}
\label{fig_jump_statistics}
\end{figure}

In the configuration discussed
above the global maximum jumps over spatial distances of
8.5 and 12.7. We will restrict ourselves for the moment to a spatial analysis,
since the Laplacian modes are almost static
(see below and Fig.\ \ref{fig_hop_finite001_time}). We have inspected 50
configurations with 31 boundary conditions and recorded the spatial jumps. Most
of them (160) are smaller than two lattice spacings, which can be viewed as a
discretization error, where coordinates of the maximum change by one lattice
spacing.
We have not considered these minimal jumps further.

There remained 120 jumps over
at least two lattice spacings.
Their distribution is plotted in Fig.\ \ref{fig_jump_statistics}.
It is still dominated by small jumps, however, the jump distance also reaches
values almost as high as the
maximally possible distance $13.9=(16/2\cdot\sqrt{3})$. There seems to be no
correlation of this distance to the $\z$-value at which the jump
occurs. Instead, the fact that the jump distribution is rather flat around half
the linear extension of the lattice (here 8), may point to a random distribution
of pinning centres,
as suggested for the hopping of fermions in \cite{gattringer:04a}.
In order to make such a statement more precise, one would need a model also for
the small jumps.

Having underlying calorons in mind, small jumps would be related to small
calorons (and larger calorons might actually be suppressed in
the same way as large instantons).
However, calorons of holonomy close to maximally nontrivial (as expected for the
confined phase) would prefer one jump at $\z\simeq1/4$.
On the other hand, out of the 50 thermalized configurations
we find only 11 configurations with only one jump,
but 16 with two, 15 with three and 8 with even four jumps, respectively,
and the jumps are scattered 
between $\z=0$ and $\z=1/2$.

In the last section we have argued that actually the minima of the modulus
of the lowest Laplacian mode are good markers
for classical topological objects. In the thermalized background, however, the
global minimum is not as stable as the global maximum. 
Fig.\ \ref{fig_spec_finite001} (c) shows that for our example
its value is changing smoothly
only in a certain fraction of the first $\z$-interval. 
For most values of $\z$ the value is highly fluctuating and so is the location
(not shown).
Furthermore, the number of local minima is typically an order of magnitude 
bigger than the number of local maxima.
Therefore, it is difficult to employ those minima 
to eventually localize background instantons or calorons
within a thermalized gauge field.\\

\begin{figure}
\centering
\begin{tabular}{ccc}
\includegraphics[width=0.32\linewidth]
{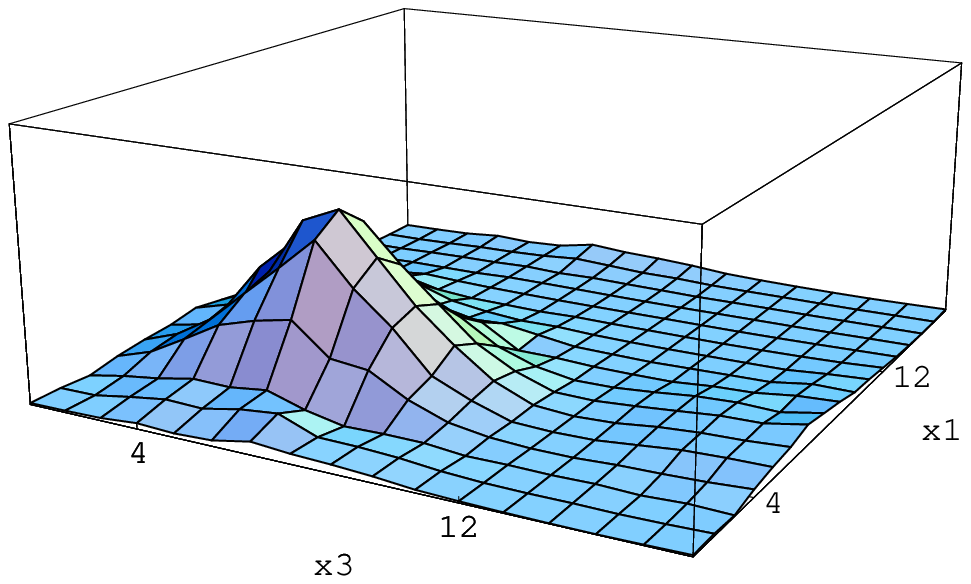}&
\includegraphics[width=0.32\linewidth]
{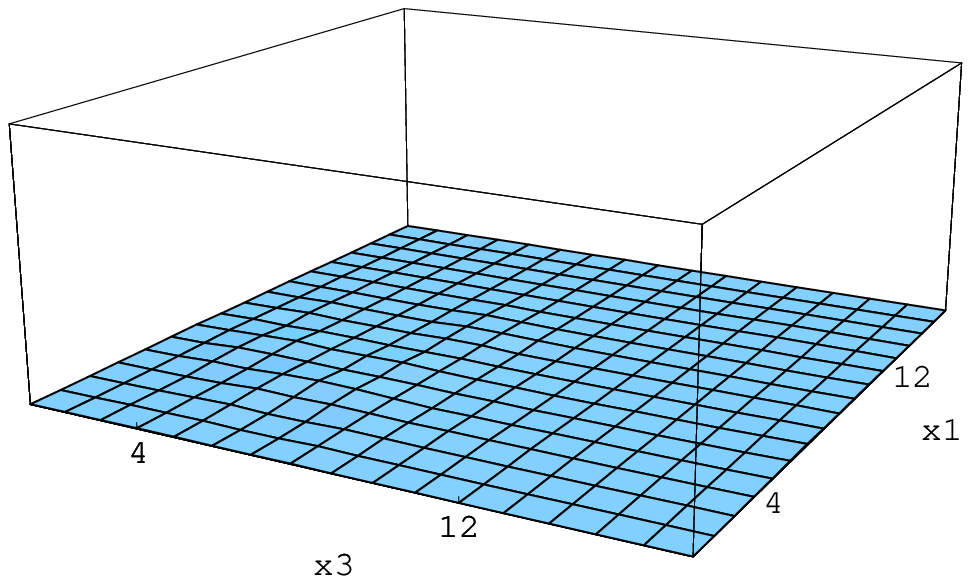}&
\includegraphics[width=0.32\linewidth]
{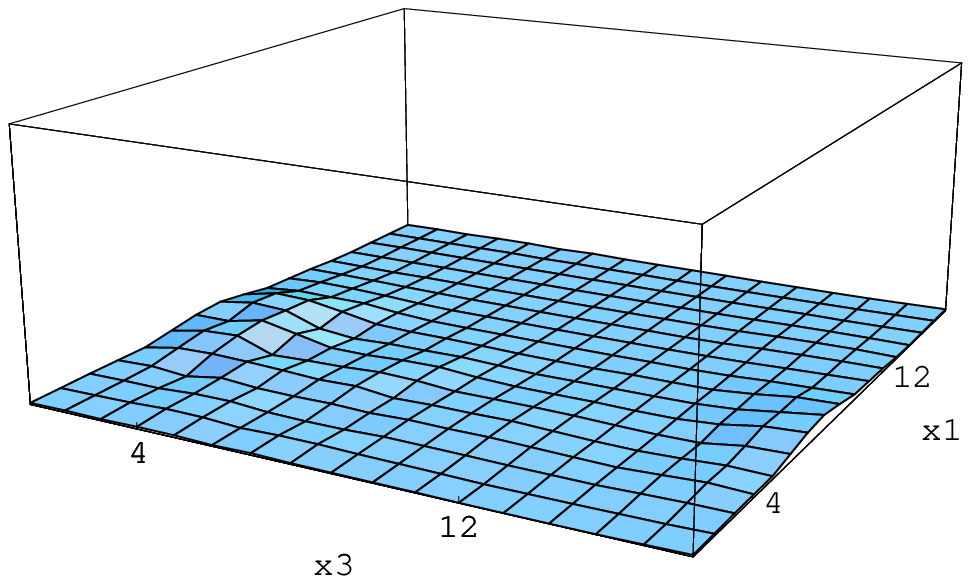}\\
\includegraphics[width=0.32\linewidth]
{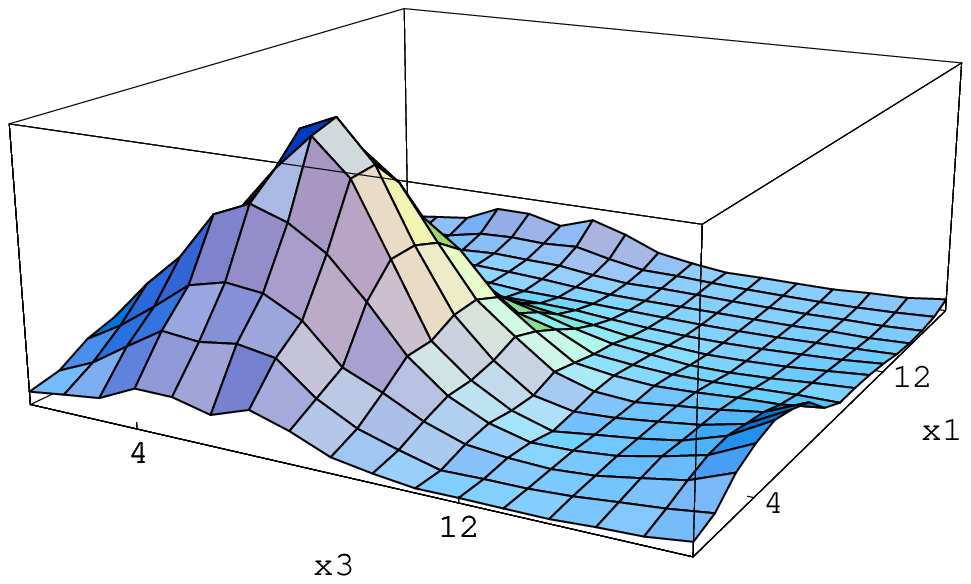}&
\includegraphics[width=0.32\linewidth]
{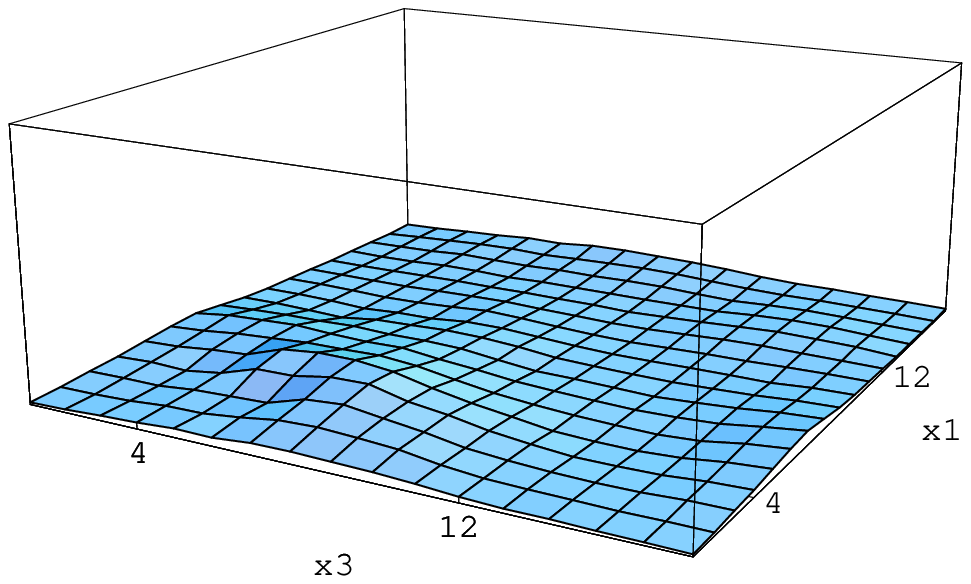}&
\includegraphics[width=0.32\linewidth]
{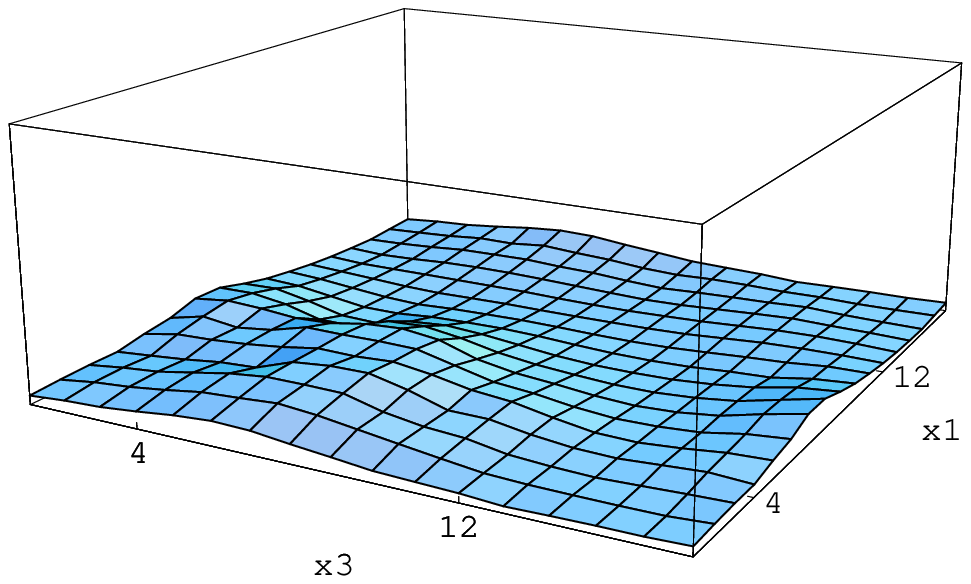}\\
\includegraphics[width=0.32\linewidth]
{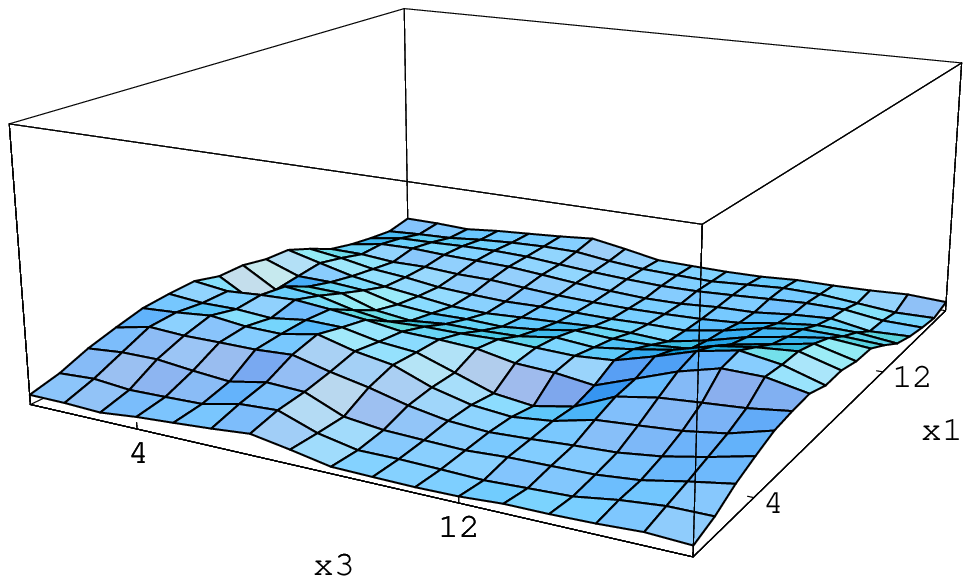}&
\includegraphics[width=0.32\linewidth]
{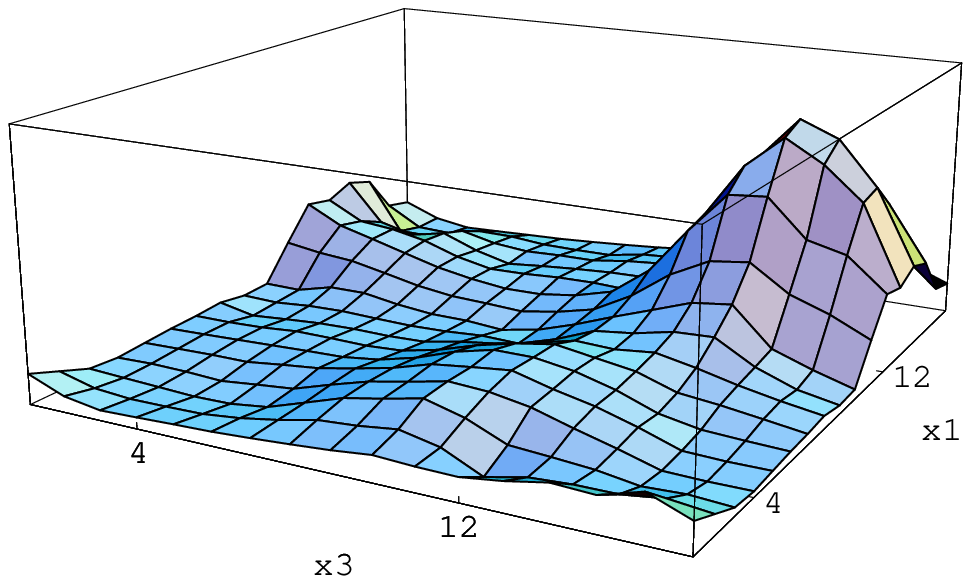}&
\includegraphics[width=0.32\linewidth]
{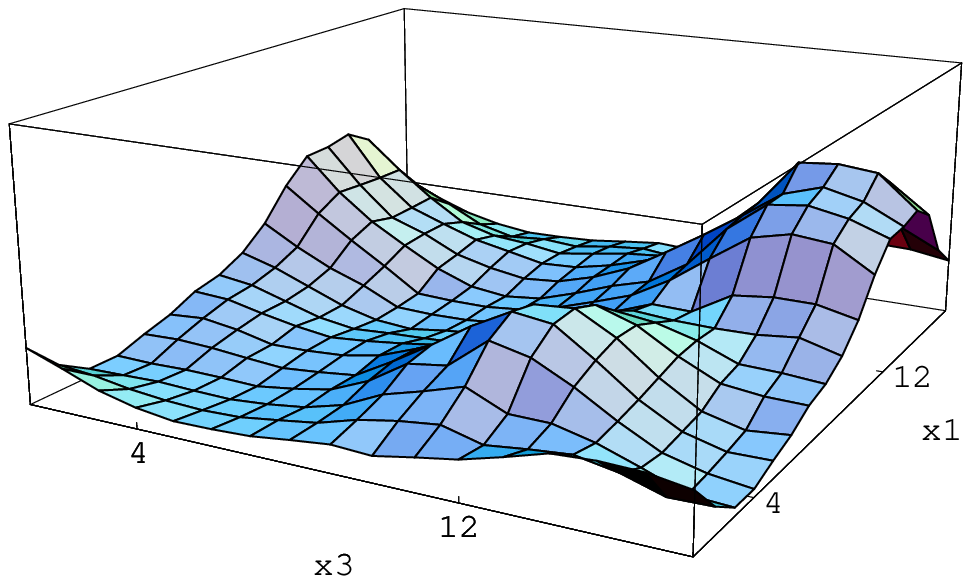}\\
\includegraphics[width=0.32\linewidth]
{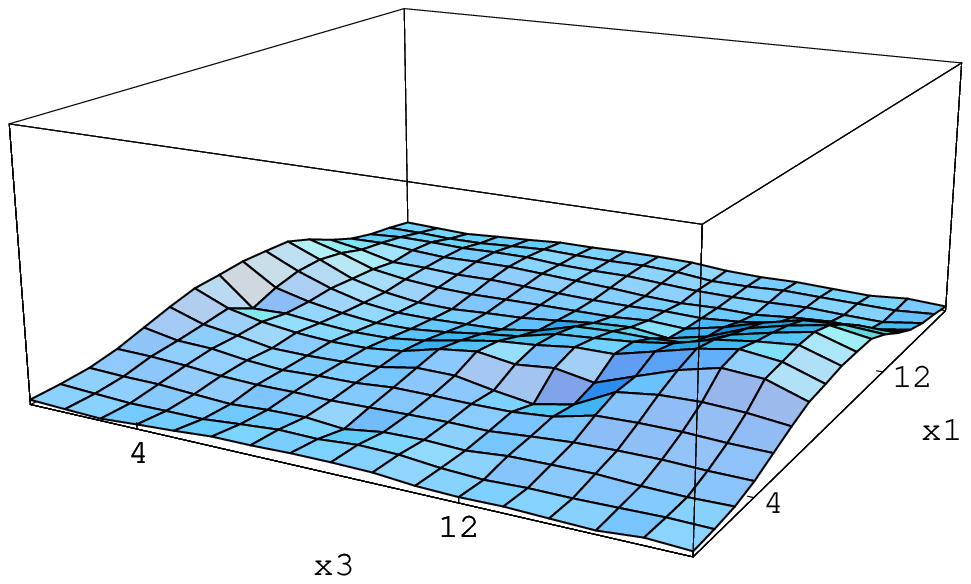}&
\includegraphics[width=0.32\linewidth]
{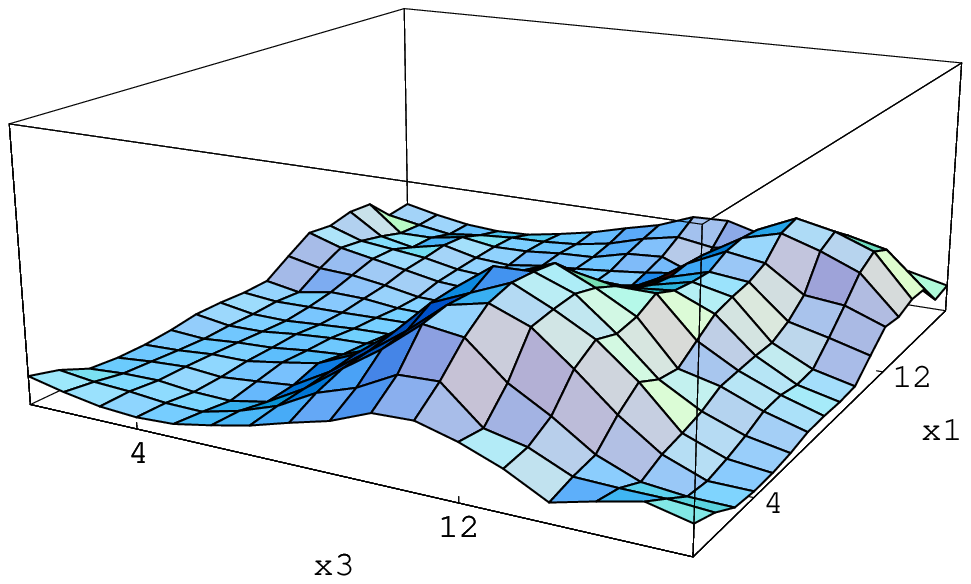}&
\includegraphics[width=0.32\linewidth]
{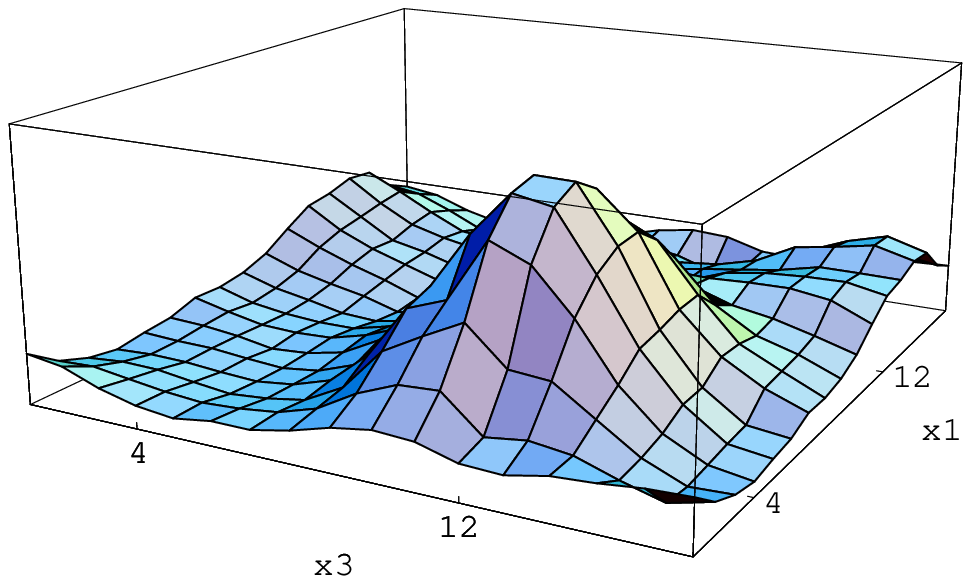}\\
\includegraphics[width=0.32\linewidth]
{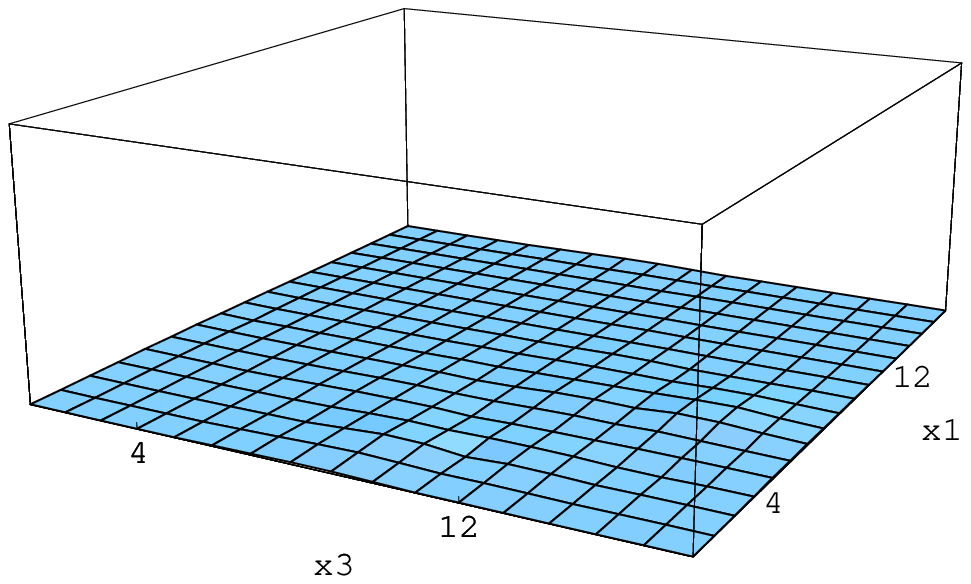}&
\includegraphics[width=0.32\linewidth]
{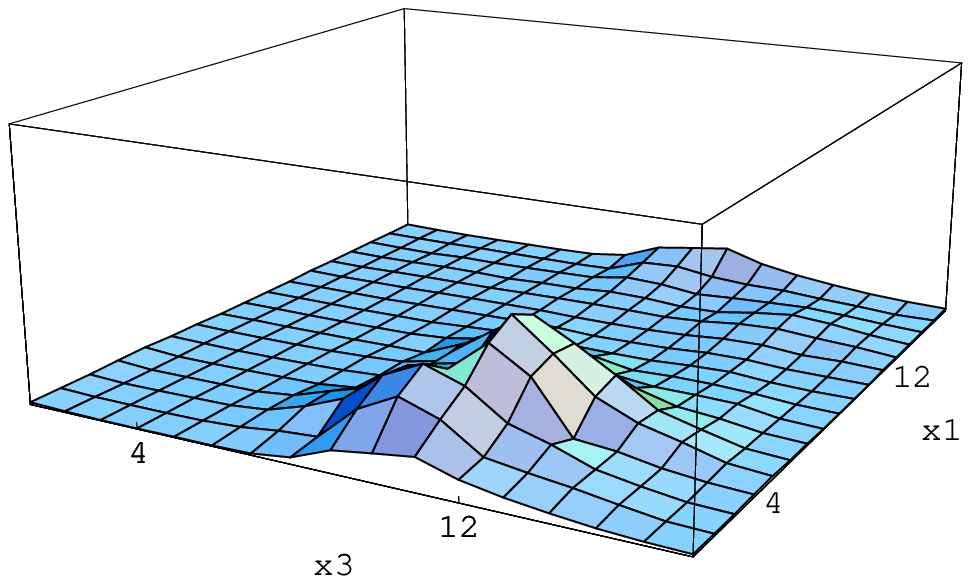}&
\includegraphics[width=0.32\linewidth]
{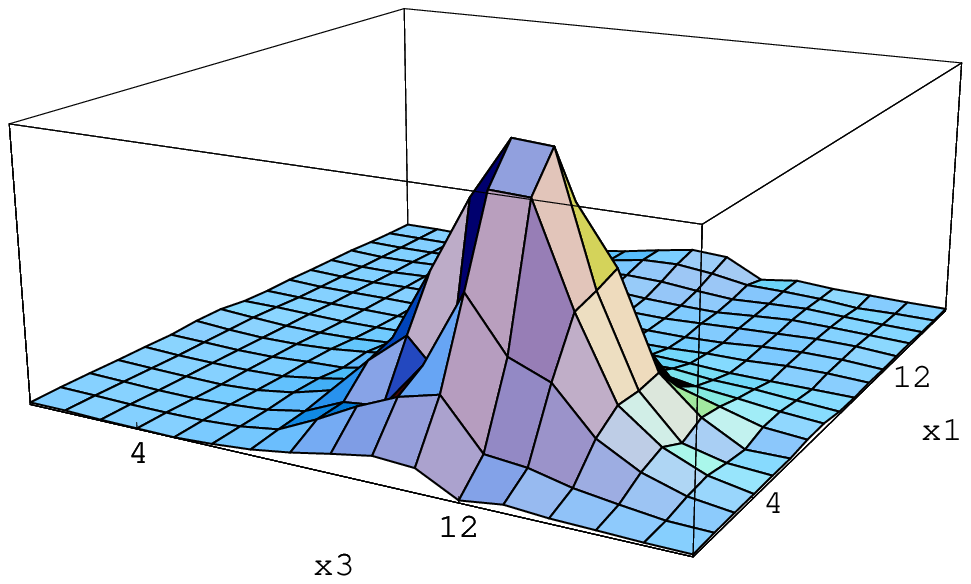}
\end{tabular}
\caption{`Hopping' of the modulus of the lowest Laplacian eigenmode.
The rows represent different modes.
First row: periodic adjoint mode; 
second to fourth row:
fundamental mode with $\z=0.020$, $0.355$
and $\z=0.480$ representing the three $\z$-intervals; 
fifth row: antiperiodic adjoint mode.
The columns show different planes through the lattice.
Left:   plane through $(x_2,x_4)=(12, 1)$ with the maximum at 
$(x_1,x_3)=( 6, 6)$ occurring in the first interval as seen in the second row;
middle: plane through $(x_2,x_4)=( 4, 4)$ with the maximum at 
$(x_1,x_3)=(14,14)$ occurring in the second interval as seen in the third row;
right:  plane through $(x_2,x_4)=( 6, 3)$ with the maximum at
$(x_1,x_3)=( 6,11)$ occurring in the third interval as seen in the fourth row;
such that the plots contain the respective global maximum.  
The vertical scale is 0.1 for the adjoint and 0.05 for the fundamental plots,
respectively.}
\label{fig_hop_finite001_space}
\end{figure}

In Fig.\ \ref{fig_hop_finite001_space} we inspect the lowest mode
locally.
As can be seen immediately, the modulus of the lowest mode
is quite smooth which justifies the expectation
that these modes do not possess UV fluctuations. 
Horizontally in the figure we have plotted
three lattice planes that contain the locations
of the global maximum taken in the different $\z$ intervals.
The precise $\z$-values representing the intervals are chosen by the demand
for a maximal IPR.
Following the maxima one can see that they either remain as local 
maxima or disappear in the other $\z$-intervals.
In other words, the hopping effect comes about by local maxima
turning into global ones in particular
$\z$-intervals, as visualized in Fig.\ \ref{fig_local_max}.

For comparison we show in the first and fifth row
of Fig.\ \ref{fig_hop_finite001_space} 
the lowest adjoint mode with periodic and antiperiodic boundary
condition, respectively. These modes are correlated to the
fundamental ones with the closest boundary conditions. Hence the adjoint
Laplacian modes are also subject to hopping
(when one allows for antiperiodic boundary conditions).

\begin{figure}[t]
\centering
\hspace{-0.5cm}
\includegraphics[width=0.6\linewidth]
{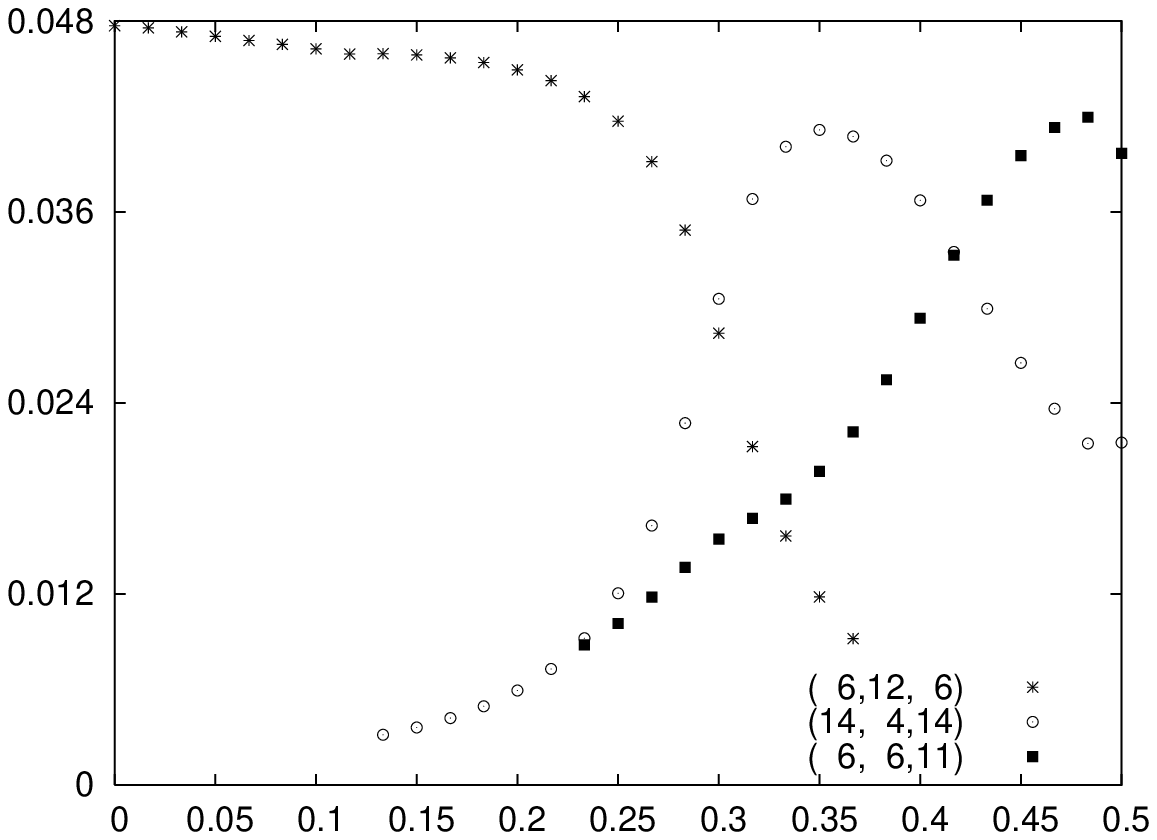}
\caption{Value and location ($\pm$ one lattice
spacing) of all local maxima that become  global ones in
some $\z$-intervals.}
\label{fig_local_max}
\end{figure}

\begin{figure}[b]
\centering
\begin{tabular}{ccc}
\includegraphics[width=0.32\linewidth]
{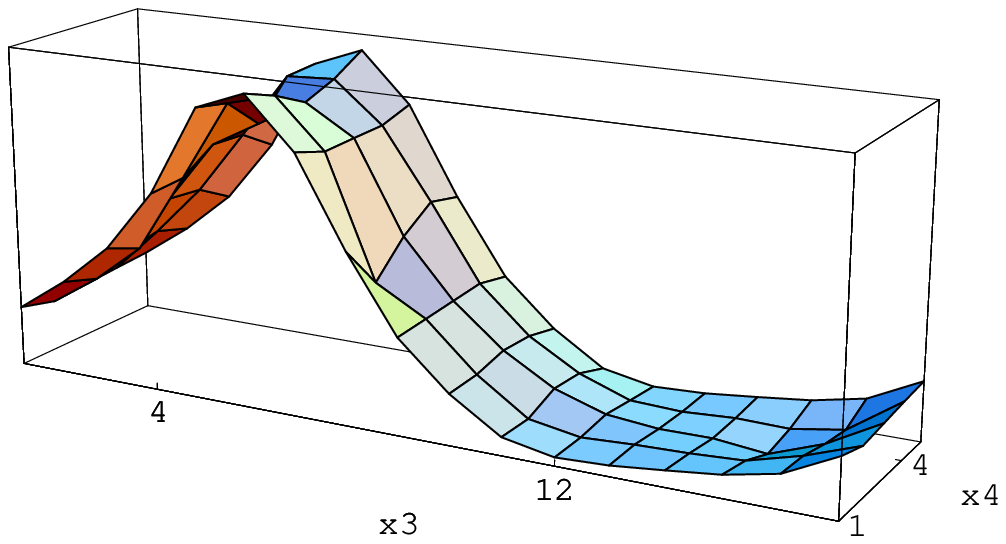}&
\includegraphics[width=0.32\linewidth]
{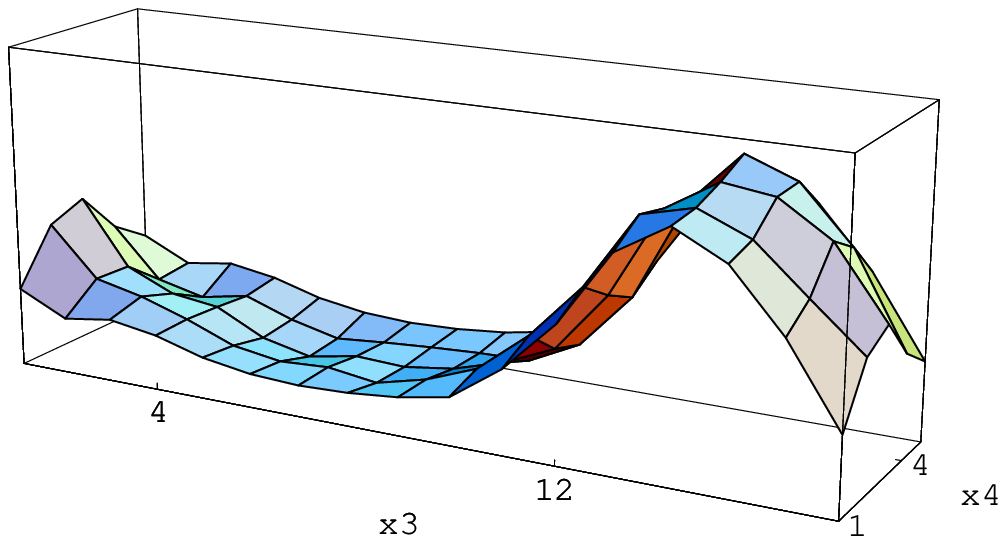}&
\includegraphics[width=0.32\linewidth]
{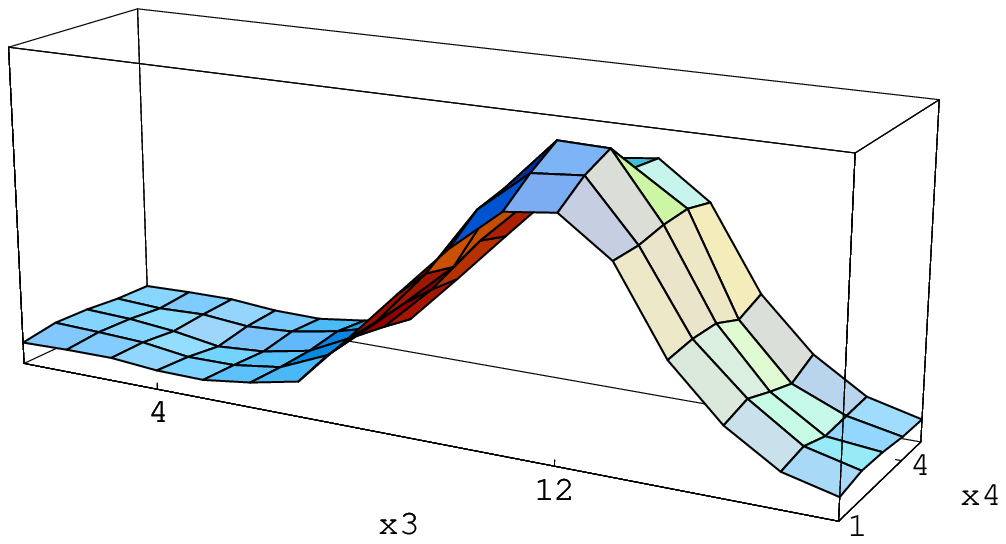}
\end{tabular}
\caption{
Time dependence of the maxima in the three intervals, 
i.e.\ at $\z=0.020$, $\z=0.355$ and $\z=0.480$ with 
$(x_1,x_2)=(6,12)$, $(x_1,x_2)=(14,4)$ and $(x_1,x_2)=(6,6)$,
from left to right, respectively. The vertical scale is 0.05 as in the
corresponding space-space plots of Fig.\ \ref{fig_hop_finite001_space}.}
\label{fig_hop_finite001_time}
\end{figure}

When compared to the classical backgrounds the maxima
in thermalized backgrounds are more pronounced,
i.e.\ more localized (compare the scales in 
Fig.\ \ref{fig_spec_finite001} (c) and 
Fig.\ \ref{fig_modulus_line}).
Accordingly, the IPR's are higher,
varying between $I=4$ and $I=11$ for the lowest fundamental mode
(cf. Fig.\ \ref{fig_spec_finite001} (b)).
Moreover, the regions outside the lumps have now a lower modulus. 
This can be quantified by the average of $|\phi|$, which for the thermalized
background is lower than the classical one (and rises at the points of 
transition).
For the lowest adjoint mode the IPR is even higher, $I=21$ and $I=53$ for
periodic and antiperiodic, respectively, an effect which has already been
observed in \cite{greensite:05}.

Thus, the overall behaviour of the lowest Laplacian eigenmodes in equilibrium 
background fields is again
analogous to the behaviour of fermionic zero 
modes \cite{gattringer:02b}\footnote{Ref.~\cite{gattringer:02b} deals 
with $SU(3)$ gauge fields (on a $20^3\cdot 6$ lattice) which should
not differ in the general picture.}. 
Still the lowest Laplacian mode seems to be broader
than the fermionic counterpart,
which is probably also the reason
why the maxima are quite static, 
see Fig.\ \ref{fig_hop_finite001_time}.\\

\begin{figure}[t]
\centering
\begin{tabular}{ccc}
\hspace{-0.5cm}
\includegraphics[width=0.3\linewidth]
{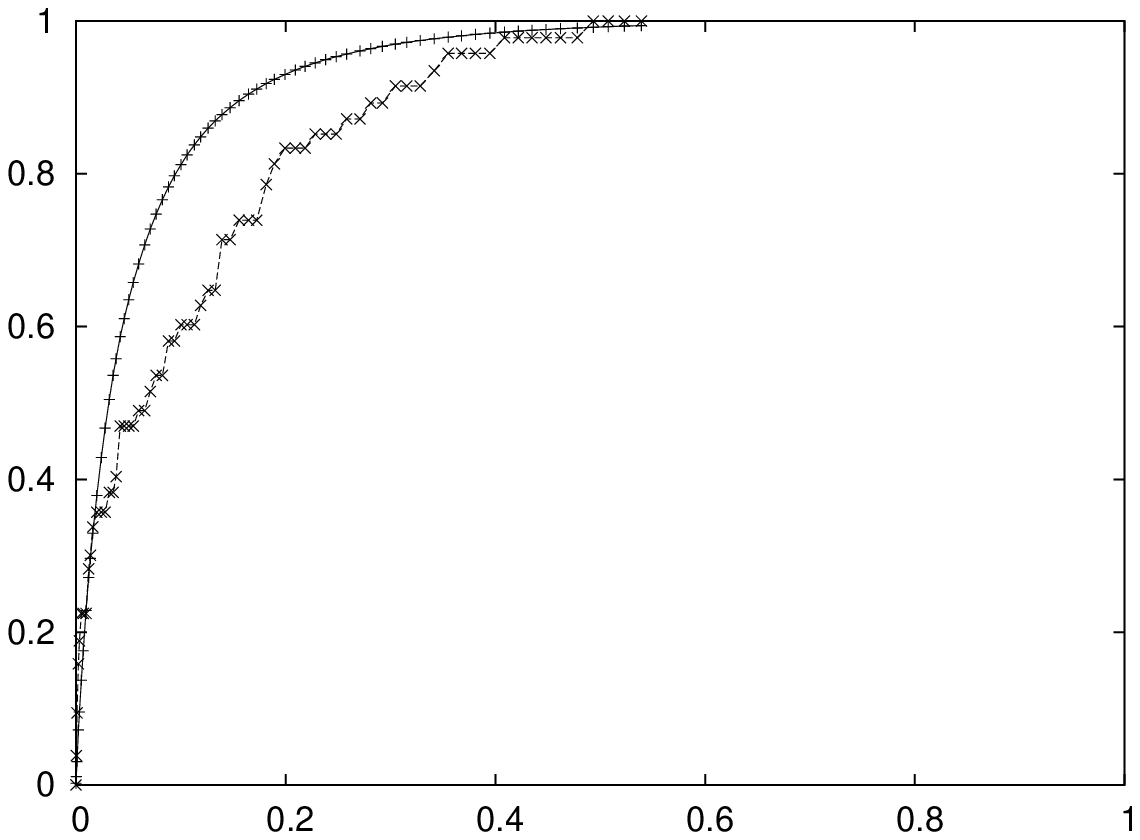}&
\hspace{-0.5cm}
\includegraphics[width=0.3\linewidth]
{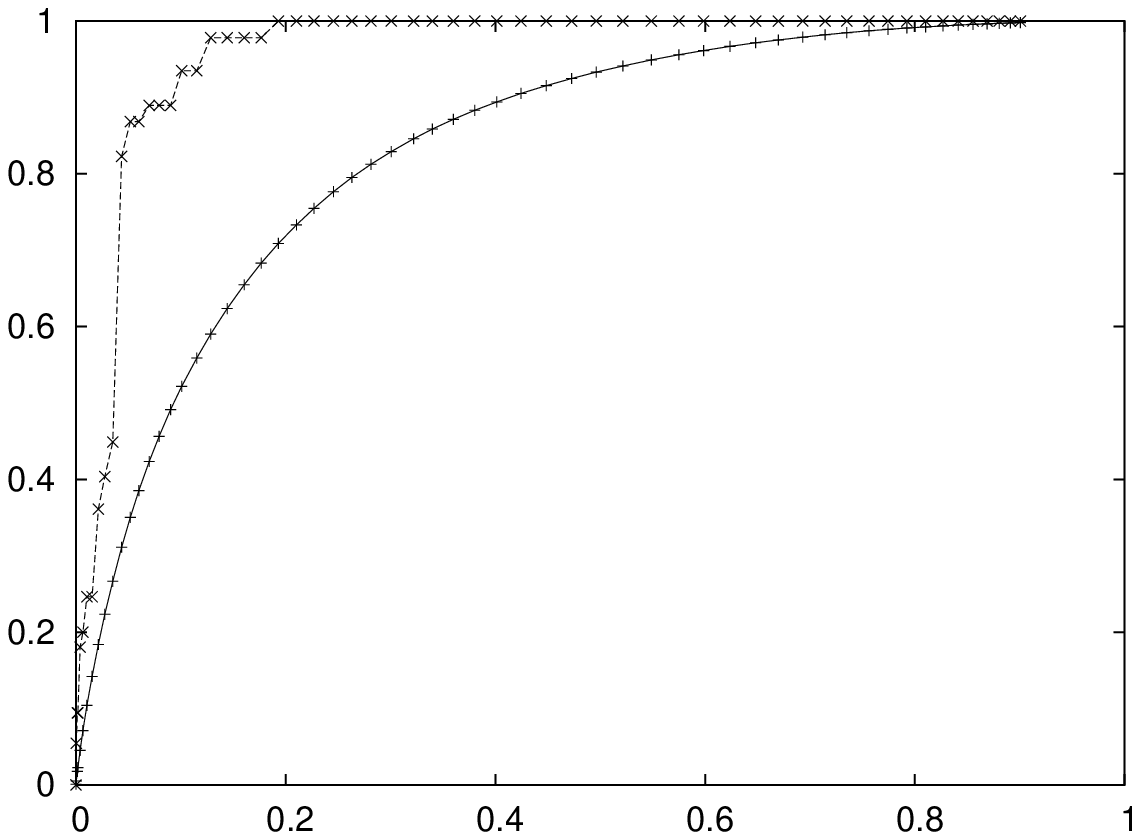}&
\hspace{-0.5cm}
\includegraphics[width=0.3\linewidth]
{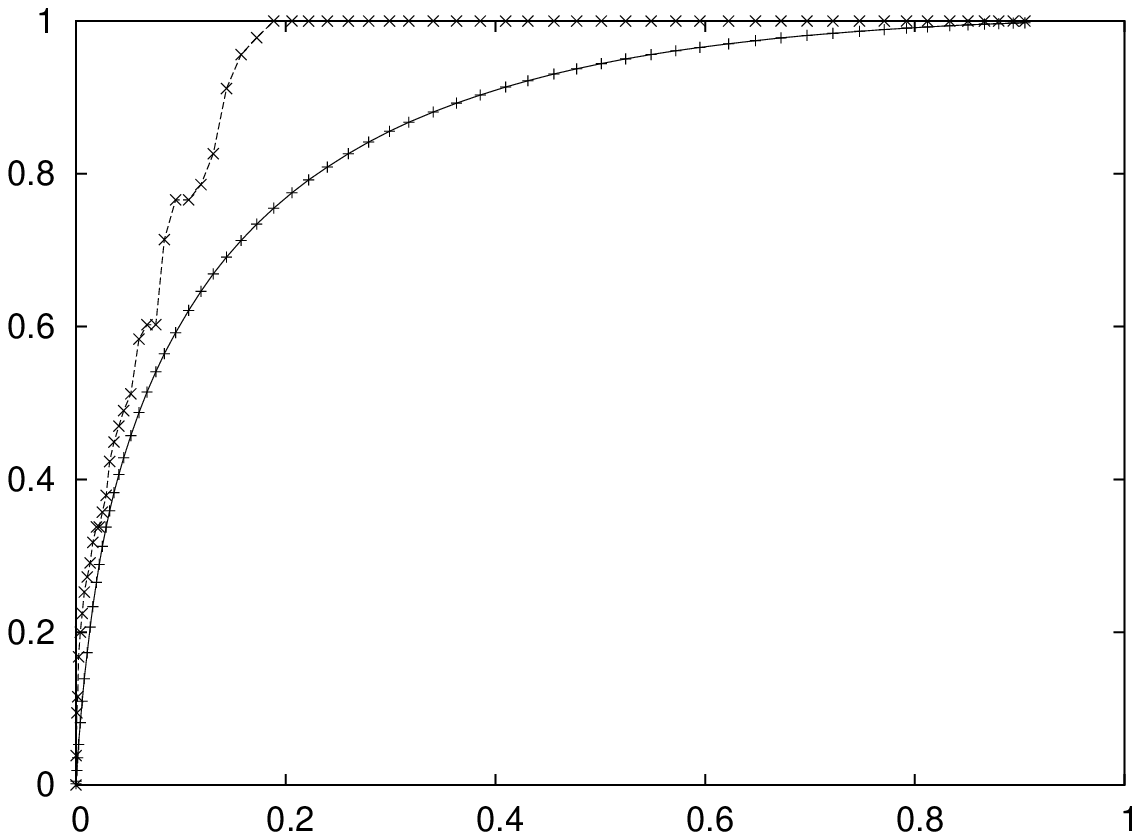}\\
(a)&(b)&(c)\\
\hspace{-0.5cm}
\includegraphics[width=0.3\linewidth]
{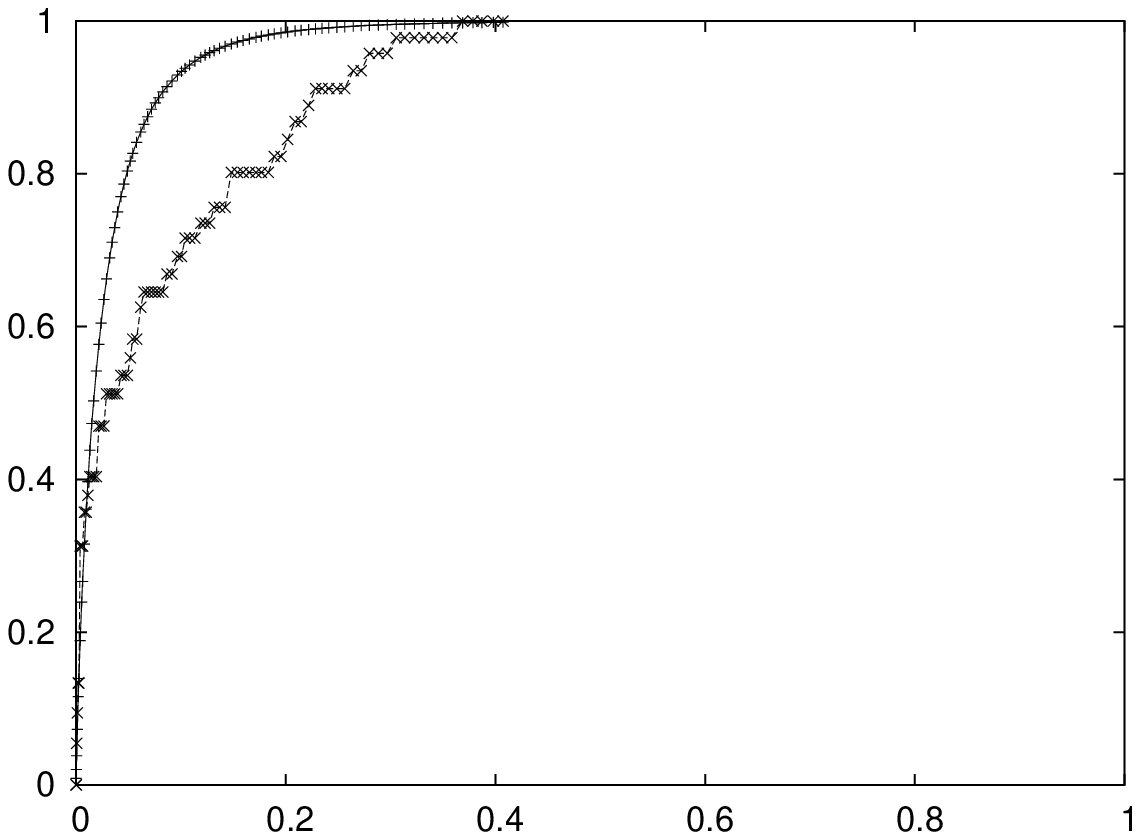}&
\hspace{-0.5cm}
\includegraphics[width=0.3\linewidth]
{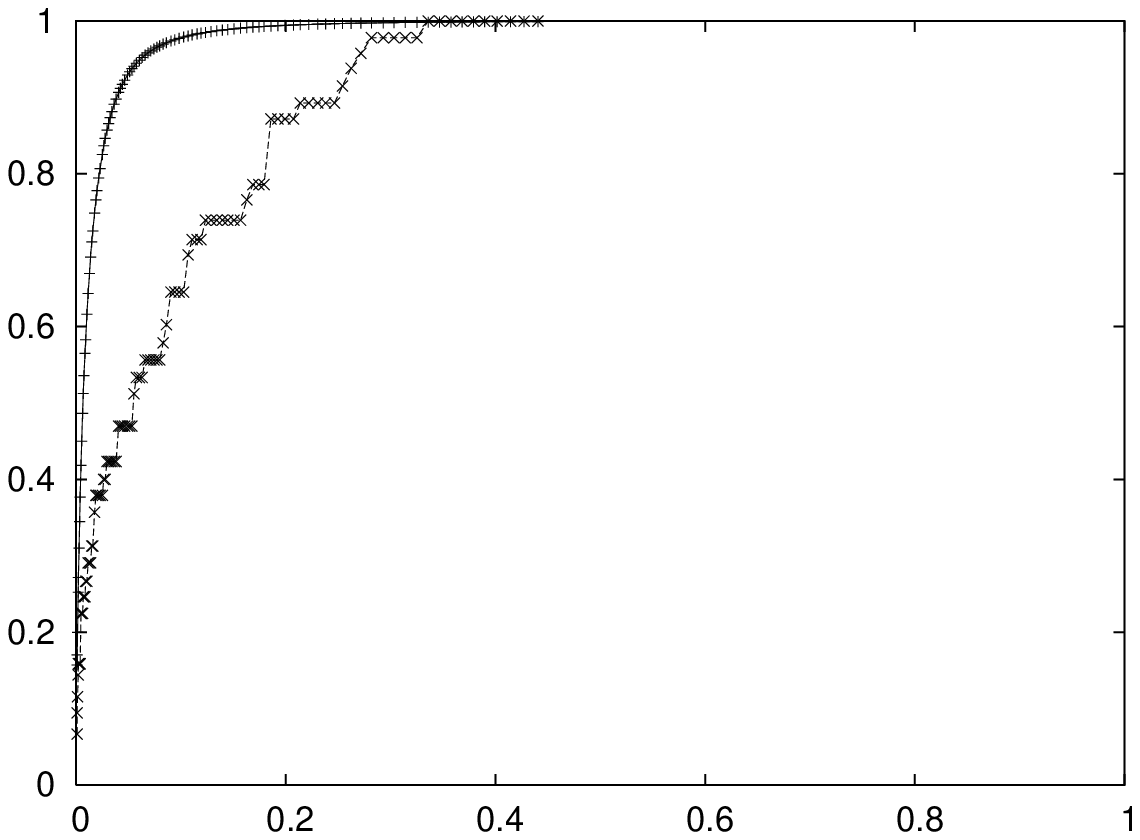}&\\
(d)&(e)&
\end{tabular}
\caption{Characterization of the cluster(s) occupied by Laplacian modes:
accumulated norm (smooth  curve) together with the relative size (step curve) 
as a function of the space-time filling fraction of the mode. 
The upper row shows
(a) the lowest fundamental Laplacian mode with $\z$ inside the first interval 
  (the mode is localized), 
(b) at the first transition (where it has turned to global) 
and (c) inside the second interval (less global). 
The lower row shows   
the lowest adjoint Laplacian mode with (d) periodic boundary condition 
and (e) antiperiodic boundary condition (both are highly localized).}
\label{fig_globalvslocal}
\end{figure}

At this point it is desirable to
discuss more quantitatively the feature of localization of the Laplacian modes.
 We have found it useful to
add to the IPR measurements done so far a description suggested by Horvath
\cite{horvath:04}. It compares how quickly clusters
(in our case of large modulus of the Laplacian mode)
grow in size and how much --
at the same time -- they gain in norm, when their total volume increases by
lowering the lower cutoff in the cluster definition. The size of the
cluster has been defined here as the diagonal of the enclosing 3D cube (since the
modes are almost static) and then divided by the diagonal of the 
3D lattice.
According to this description, a profile is called localized when
small clusters already accumulate a large norm fraction.

For the Laplacian modes we find mostly one big cluster.
As Fig.\ \ref{fig_globalvslocal} shows, the adjoint modes are always local,
but the fundamental modes are sometimes rather global, depending on the boundary
condition $\z$. This is the case for instance for $\z=0.300$ (first transition
point) and $\z=0.355$ (second $\z$-interval),
cf.\ Fig.\ \ref{fig_globalvslocal} (b) and (c).
There the cluster has already its maximal size,
i.e.\ has percolated through the lattice,
although the accumulated norm is around three quarters only.

Thus this analysis supports the view of Laplacian modes being close to wave-like
(as has already been argued in the caloron background).
Hence they should be able to
reflect the background gauge field quite well everywhere on the lattice.
This is an advantage for Laplacian gauge fixing
and the filter method proposed in the next section.\\ 

To end this section about Laplacian modes in equilibrium backgrounds, we want to
mention a possibility to connect the behaviour of Laplacian modes in thermalized
and classical backgrounds. It involves
the smoothing of the gauge field background by applying
smearing (for instance 5 steps) to it.
Then the maximum decreases in modulus
(and can move or join with local maxima), whereas the minimum value
increases.
More importantly, the number of local minima decreases drastically
and from some stage on it should be possible to use the minima as markers 
of the (gradually emerging) topological structure. However, such an approach
mixes two techniques to suppress UV fluctuations and it would be cleaner to
interpret directly the Laplacian modes of the original configuration.

\begin{figure}[t]
\centering
\begin{tabular}{ccc}
\hspace{-0.5cm}
\includegraphics[width=0.5\linewidth]
{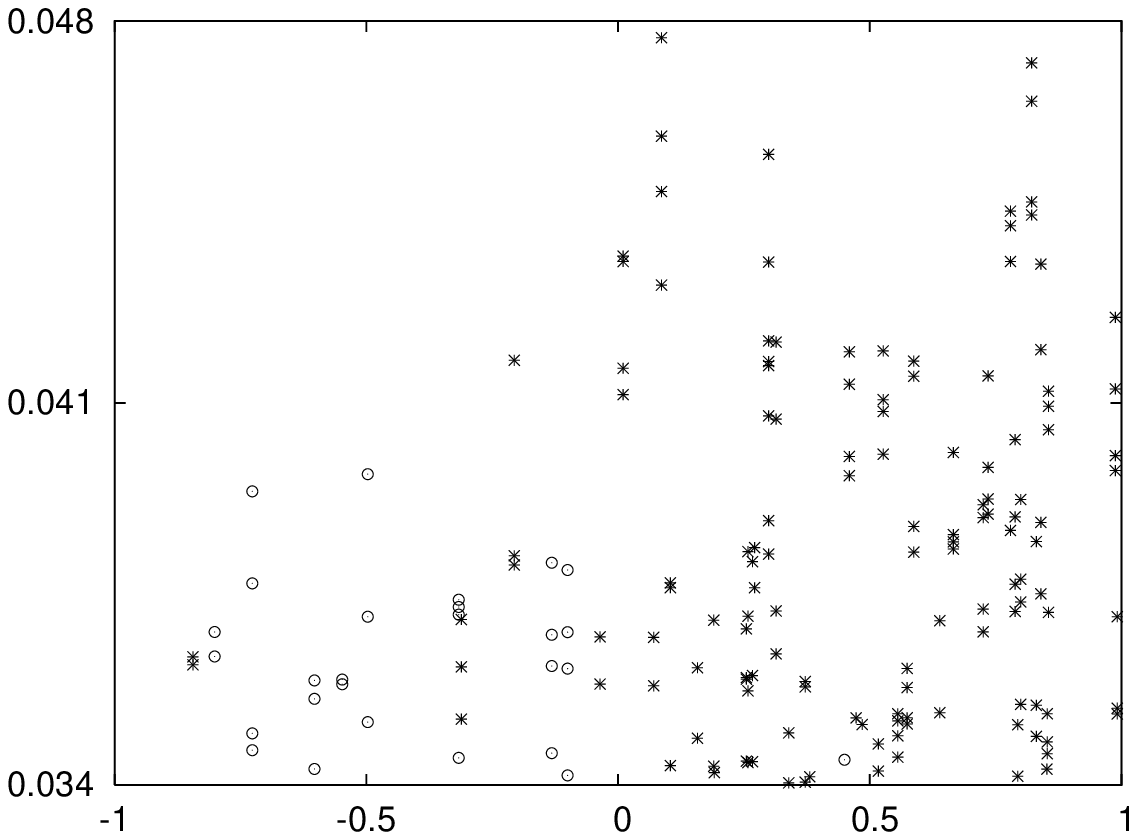}&
\hspace{-0.5cm}
\includegraphics[width=0.5\linewidth]
{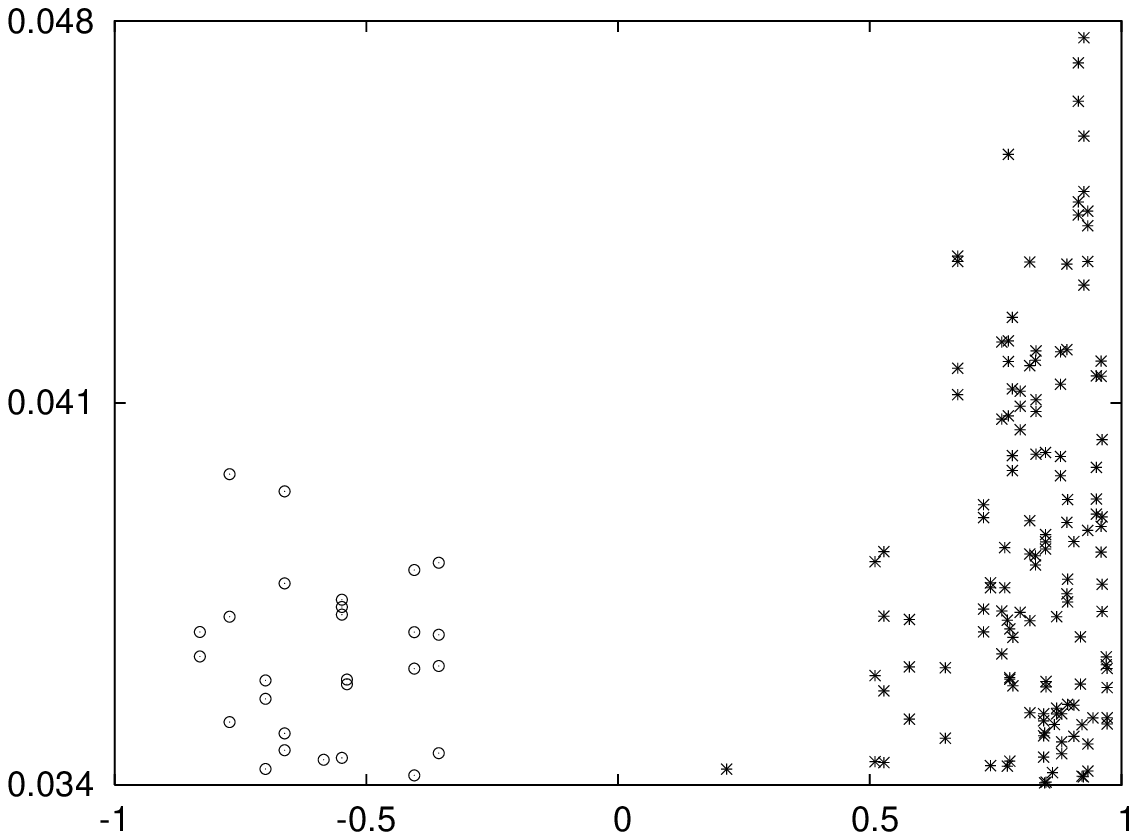}\\
(a)&(b)
\end{tabular}
\caption{Scatterplots of the Polyakov loop (horizontal) at lattice
sites where the modulus of the Laplacian modes (vertical) with
$\z=0.02$ (close to periodic, crosses)
and $\z=0.48$ (close to antiperiodic, circles) is large: (a)
for the original configuration and (b) for the Polyakov loop measured on the
configuration with 5 smearing steps applied.}
\label{fig_scatter_mode_polloop}
\end{figure}

\begin{figure}[b]
\centering
\begin{tabular}{cc}
\includegraphics[width=0.48\linewidth]
{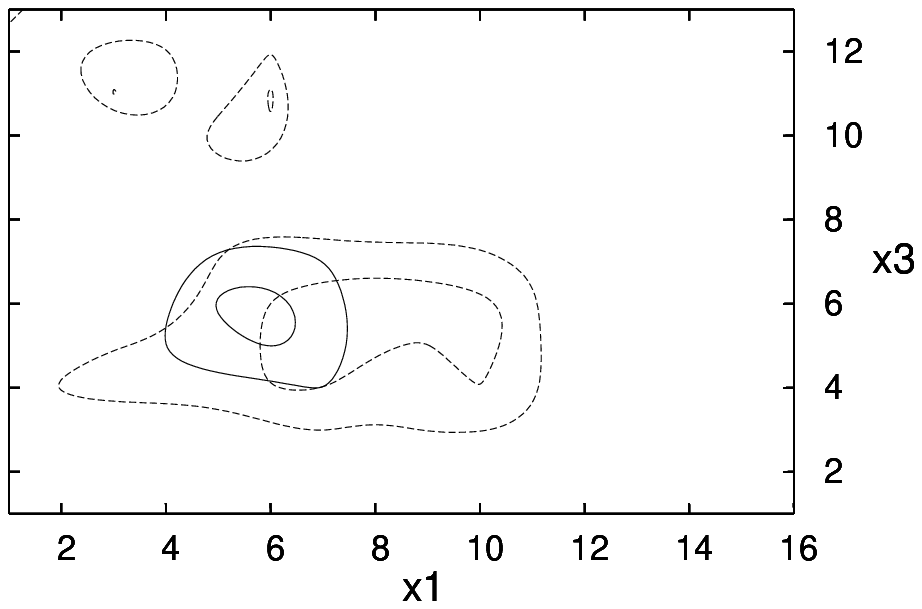}&
\includegraphics[width=0.48\linewidth]
{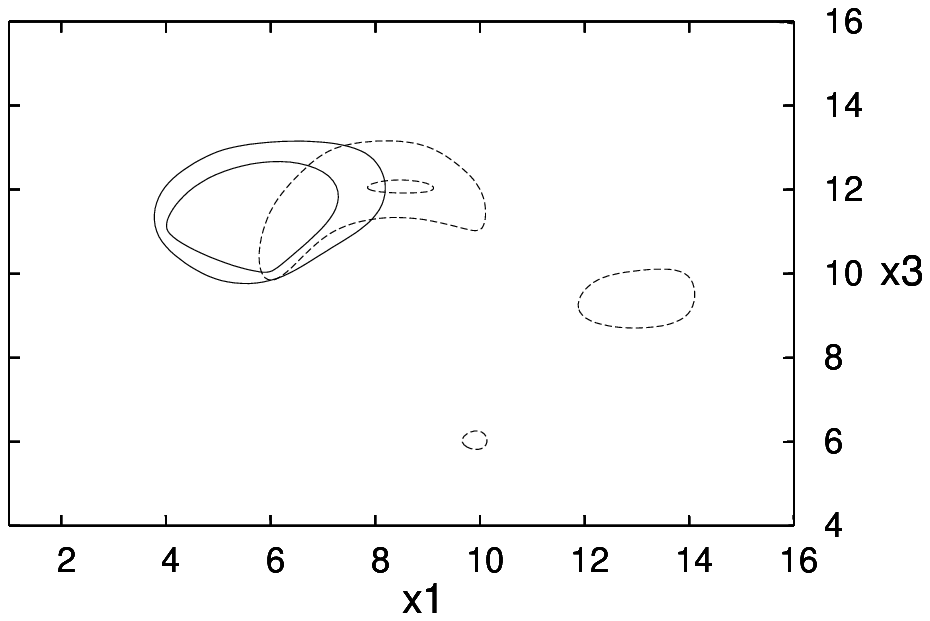}\\
(a)&(b)
\end{tabular}
\caption{Comparison of contours of constant Polyakov loop
(dashed curves) after smearing and
of the unsmeared Laplacian modes at large modulus (70\% and 90\%
of the maximum, full curves),
in lattice planes where the latter become maximal.
(a) the periodic mode vs.\ positive Polyakov
loop (0.7 and 0.9) in $(x_2,x_4)=(12, 1)$,
(b) the antiperiodic mode vs.\
negative Polyakov loop ($-0.7$ and $-0.9$) in $(x_2,x_4)=( 6, 3)$.}
\label{fig_pinning}
\end{figure}

Interestingly, we have found an interrelation of the pinning of the Laplacian
modes to a gluonic observable, namely the Polyakov loop.
The maximum of the modulus of the 
periodic mode is correlated to a positive Polyakov loop, whereas the maximum of
the antiperiodic mode prefers negative values,
see Fig.\ \ref{fig_scatter_mode_polloop} (a).
Of course, this expresses only an overall tendency,
because there is a mismatch between the smoothness of the
Laplacian mode and the roughness of the Polyakov loop.
To make this statement quantitative one can measure the sum over the
Polyakov loop weighted with $|\phi|^2$ of the periodic and antiperiodic mode (the
horizontal coordinate of the `center of mass' of that figure).
This gives weighted averages of 0.206 and $-0.1154$ for the two cases,
respectively, while the ordinary Polyakov loop average is 0.005.

We have smeared the gauge field
which results also in smoothing the Polyakov loop, then we
compared the latter to the Laplacian modes of the original configuration.
The mentioned tendency becomes more pronounced,
see Fig.\ \ref{fig_scatter_mode_polloop} (b)
and is now locally visible. 
As Fig.\ \ref{fig_pinning} shows, the smeared Polyakov loop provides a
collection of pinning centres for the modes to settle down at Polyakov loops near
$\Eins_2$ and $-\Eins_2$, repectively. Hence the Laplacian modes on the
unsmeared configuration know about the Polyakov loop landscape,
which only emerges after smearing.

The direction of this correlation agrees with the findings in classical
solutions (compare Fig.\ \ref{fig_bosonic} (a) and Fig.\ \ref{fig_modulus_line})
which seems to suggest calorons as `underlying' the thermalized gauge
fields. However, it can also be interpreted within the Anderson localization
scenario, with the Polakov loop localizing the candidate `minima 
of a random potential' to be occupied by the scalar or fermionic modes.

\section{A new filter method}
\label{sec_filter}

In the following we will introduce a new low-pass filter based on the Laplacian
modes. It uses an exact representation of the links in the form of a sum of the
latter, which, by truncation, should remove UV noise and find the
`underlying IR structures'. Hence, the spirit of our method is close to a
Fourier decomposition or a (gauge invariant) high momentum cut-off.

Technically, our approach is similar to the one in \cite{gattringer:02c},
where the field strength was given in terms of a sum over 
fermionic modes.
A similar mode truncation for the overlap-based topological
charge has been used in \cite{horvath:02} and studied in more detail in
\cite{koma:05,weinberg:05}.
However, we will obtain directly the link variables
and in this `reconstructed' configuration any observable can be measured.

\subsection{Derivation and properties}

We combine the definition of the gauge covariant Laplace
operator, Eq.\ (\ref{eq_def_lapl}), with a decomposition into its eigenmodes:
\begin{eqnarray}
-\la^{ab}_{xy}=\sum_{n=1}^{\N}\lambda_{n,\z}\p^a_{n,\z}(x)\p^{*b}_{n,\z}(y)
\label{eqn_decomp}
\end{eqnarray}
At $y=x+\hm$ one immediately obtains
\begin{eqnarray}
U_\mu^{ab}(x)=-\sum_{n=1}^{\N}\lambda_{n,\z}\p^a_{n,\z}(x)\p^{*b}_{n,\z}(x+\hm)
\label{eqn_link_filtered_first}
\end{eqnarray}
The corresponding formula for $U^\dagger_\mu$ at $y=x-\hm$ is fully equivalent to
Eq.\ (\ref{eqn_link_filtered_first}), while
\begin{eqnarray}
\delta^{ab}&\!=&\!\frac{1}{2}\sum_{n=1}^{\N}\lambda_{n,\z}\p^a_{n,\z}(x)\p^{*b}_{n,\z}(x)
\\
0&\!=&\!\sum_{n=1}^{\N}\lambda_{n,\z}\p^a_{n,\z}(x)\p^{*b}_{n,\z}(y)\qquad
\forall y\neq x,x\pm\hm\label{eqn_relations}
\end{eqnarray}
could be used as a check
for the approximation to be discussed now.
Notice that the eigenfunctions enter these expressions in a combination that is
invariant under a multiplication of the eigenfunctions with a complex
phase. More general, at a level with degeneracy $k$ the
expressions are invariant under a change of the basis
(a global $U(k)$ rotation
acting on the index $n$).

The idea of the filter is to {\em truncate the sum in Eq.\ 
(\ref{eqn_link_filtered_first}) at a rather small number $N$ 
of eigenmodes}.
When doing so, the question arises, how to relate the r.h.s.\ of
Eq.\ (\ref{eqn_link_filtered_first}) to a {\em unitary} link variable.
We remind the reader that in the cooling method staples are added. This gives a
quaternion, which becomes an
element of $SU(2)$ upon multiplying by a real number.
We will use the same projection.
However, the non-trivial task here will be to arrive at a quaternion
in the first place.
We will show now that the charge conjugation symmetry of the Laplacian
helps to resolve this issue.

For the $\phi$-terms in the sum we use the abbreviation
\begin{eqnarray}
u_\mu^{ab}(x)_{n,\z}=\p^a_{n,\z}(x)\p^{*b}_{n,\z}(x+\hm)\,.
\label{eqn_link_abbrev}
\end{eqnarray}
The charge conjugation symmetry (\ref{eqn_symm_first}) implies
\begin{eqnarray}
u_\mu'(x)_{n,-\z}=\e\, u_\mu(x)_{n,\z}^*\, \e^{-1}\,,\qquad\e=i\sigma_2\,.
\label{eqn_u_cc}
\end{eqnarray}
For $\z=0$ (periodic) and $\z=1/2$ (antiperiodic) these contributions are
automatically included in the sum in Eq.\ (\ref{eqn_link_filtered_first})
(as every level is two-fold degenerate with eigenmodes
$\p$ and $\p'$).
To make use of this relation for general $\z$, we take the
average of the link formula (\ref{eqn_link_filtered_first})
over $\z$ and $-\z$
\begin{eqnarray}
U_\mu(x)&=&-\frac{1}{2}\,\sum_{n=1}^{\N}\left[
\lambda_{n,\z} u_\mu(x)_{n,\z}+\lambda'_{n,-\z} u'_\mu(x)_{n,-\z}\right]
\nonumber\\
&=&-\frac{1}{2}\,\sum_{n=1}^{\N}\lambda_{n,\z}\left[
u_\mu(x)_{n,\z}+\e\, u_\mu(x)_{n,\z}^*\, \e^{-1}\right]\,.
\label{eqn_link_filtered_third}
\end{eqnarray}
So far this manipulation is exact, but seems artificial.
However, it will help in the truncation of the sum.
Parametrising the matrix in the bracket we have
\begin{eqnarray}
\left(\begin{array}{cc}
a&b\\c&d
\end{array}\right)+
\left(\begin{array}{cc}
d^*&-c^*\\-b^*&a^*
\end{array}\right)=
\left(\begin{array}{cc}
u&v\\-v^*&u^*
\end{array}\right)
\label{eqn_quaternion}
\end{eqnarray}
For a matrix of this form\footnote{Eq.\
(\ref{eqn_quaternion}) is actually the most general quaternion parametrised by
${\rm Re}\, u \cdot\Eins_2+{\rm Im}\, u \cdot i \sigma_3+{\rm Re}\, v \cdot
i\sigma_2+{\rm Im} v\,
\cdot i\sigma_1$.}
the inverse is the same as the hermitean conjugate up to a factor, which
is the determinant $|u|^2+|v|^2$
and positive for all practical purposes.
Hence, the matrix obtained by truncating the sum is an element of $SU(2)$ up to
the square root of the determinant
by which we will divide.

An alternative way to arrive at a unitary link is to use the unitary
matrix in the singular value or polar decomposition.
It agrees with our procedure under special circumstances,
for instance if the matrix to start with
has a real determinant as is the case for $\z=0$ and $\z=1/2$.
In our point of view the inclusion of both
$\z$ and $-\z$ is very natural since these boundary conditions contain the same
information of the Laplacian mode
(same eigenvalue and charge conjugated eigenmode). 

To summarize, there is an exact formula for the link variables in terms of the
Laplacian modes, namely Eq.\ 
(\ref{eqn_link_filtered_third}).
It is this sum we truncate to obtain the filtered links
\begin{eqnarray}
\tilde{U}_\mu(x)_{N,\z}=\left(-\sum_{n=1}^{N}\lambda_{n,\z}\left[
u_\mu(x)_{n,\z}+\e\, u_\mu(x)_{n,\z}^*\, \e^{-1}\right]\right)_{{\rm det}=1}
\quad M_{{\rm det}=1}\equiv M/\sqrt{{\rm det}\,M}\,,
\label{eqn_link_filtered_final}
\end{eqnarray}
where the operation $(\ldots)_{{\rm det}=1}$ forces the link variable to have
unit determinant\footnote{
We have used the positive root in Eq.\ (\ref{eqn_link_filtered_final}) for all
links. 
One can in principle also work with a local choice of the positive
or the negative root, but we do not see a reason to introduce a $Z_2$-freedom
here.} (and we have dropped the factor $1/2$).

Eq. (\ref{eqn_link_filtered_final}) 
is our final proposal for a {\em low-pass
filter acting on lattice configurations}. 
It is a mapping from the original links
$U_\mu(x)$ to the filtered ones $\tilde{U}_\mu(x)_{N,\z}$ via the Laplacian
eigenmodes
with the phase $\z$ in the boundary condition as a free parameter.
The quality of the filter is
controlled by $N$, where $N=\N$ reproduces
the original configuration exactly
(with no determinant correction necessary).
Formally, the filtered links $\tilde{U}_\mu(x)$ look like composite
fields, as they are produced by bilinears in $\phi$, Eq.\ (\ref{eqn_link_abbrev}).

Our approximation keeps the full gauge covariance, because if a gauge
transformation $g(x)$ acts on $\p(x)$ from the left, then $g^\dagger(x+\hm)$
acts on $\p^*(x+\hm)$ from the right (and the determinant is gauge invariant)
resulting in $g(x)\tilde{U}_\mu(x)g^\dagger(x+\hm)$ as it should be.

It is instructive to investigate the extreme case of $N=1$, i.e. the
contribution from the lowest mode only.
Then the filter gives -- independent of the starting configuration -- a pure
gauge $\tilde{U}_\mu(x)$, but with a Polyakov loop given by $\z$.

The best way to understand this is going into Laplacian gauge
\cite{vink:92}.
That is to rotate the lowest mode to have only an upper nonvanishing
component, which is real. Then the matrices $u_\mu(x)$ would have
only a real 11-component and adding the charge conjugate and normalising by the
square root of the determinant makes all $\tilde{U}_\mu(x)$ equal to the
identity.
However, for the nontrivial boundary conditions of Eq.\ (\ref{eqn_bc}) one has to
slightly rethink this procedure. The simplest way to define the gauge fixing
then is to demand the form $({\rm
real},\,0)^T$ for the (always periodic) $\varphi_\z(x)$.
By transforming back to $ \phi_\zeta(x)$, Eq.\ (\ref{eqn_phiphi}),
and plugging this into Eq.\ 
(\ref{eqn_link_abbrev}) the same considerations hold, with the only exception that
$u_{\mu=4}^{a=b=1}(x)$ receives an additional factor $\exp(-2\pi i
\z/N_4)$. After charge conjugation and normalising the filtered links become 
$\tilde{U}_4(x)=\exp(-2\pi i \z\tau_3/N_4)$. Hence the filtered Polyakov loop is  
\begin{eqnarray}
\frac{\tr}{2}\,\tilde{\P}(\vec{x})_{N=1,\z}=\cos(2\pi\z)\quad \forall\vec{x}\,.
\label{eqn_polloop_N=1}
\end{eqnarray}

These considerations did not use that $\p(x)$ is the ground state. Thus every
eigenmode alone gives vanishing action and a Polyakov loop as above.
Since the superposition in Eq.\ (\ref{eqn_link_filtered_final}) is nonlinear,
 we expect $N=2$ to be the first nontrivial filter with
nonvanishing action\footnote{This is true unless $\z=0$ or $\z=1/2$ for which
the two-fold degeneracy makes $N=3$ (equivalent to $N=4$) the first nontrivial
case. }.
As we will see in the next section, the trace of the Polyakov loop will
start to fluctuate in this case, but on average is still close to
the value of Eq.\ (\ref{eqn_polloop_N=1}).

Let us add a few remarks on the possible
stability of the filtered links $\tilde{U}_\mu(x)_{N,\z}$
under variations of $N$.
From the normalization of $\p_n(x)$ follows that the entries of $u_\mu(x)$ are
roughly $O(1/\N)$. The prefactor $\lambda_n$ is rising slightly with $n$, such that one
can expect the terms in the sum to be of the same order of magnitude.

The only exception to this argument is the caloron background at $\z=1/4$, where
the lowest Laplacian eigenvalue is strongly suppressed w.r.t.\
the other eigenvalues. As a result, the lowest mode practically does not
contribute to the filtered links. Here the situation is analogous to the
fermionic filter in \cite{gattringer:02c},
where the zero mode does not contribute.

Actually, the staggered symmetry of the Laplacian can be used to improve the
convergence. Relation (\ref{eqn_symm_second}) between low and high modes
immediately gives
\begin{eqnarray}
u_\mu''(x)_{\N-n,\z}=-u_\mu(x)_{n,\z}\,.
\end{eqnarray}
Therefore, the inclusion of the upper end of the spectrum does not change the
contributions to the filtered links locally. It only reweights the terms in the
sum (\ref{eqn_link_filtered_final})
\begin{eqnarray}
-\lambda_n\rightarrow 4D-2\lambda_n\,.
\end{eqnarray}
With this insertion one has to sum in Eq.\ (\ref{eqn_link_filtered_final})
only over half the spectrum to obtain the
original link. Now each subsequent term
$u_\mu(x)_n$ has a smaller weight.
In particular, the weight of the ground state is the biggest.

The price to pay is to include the upper end of the spectrum of the Laplacian,
which a priori contradicts the meaning of a low-pass filter. We decided not to
include the highest modes and to stick to Eq.\ (\ref{eqn_link_filtered_final}).
Nevertheless, we have checked the consequences of such a modification for the
results presented below. The changes are quite small apart from the case
of reconstructing
the caloron background from $N=2$ modes at $\z=1/4$, the case discussed
above. Therefore, it seems that the local information $u_\mu(x)_n$ is more
important for the filtered links than the relative weights $\lambda_n$ of the
eigenfunctions.
One might speculate whether eigenmodes other than the lowest ones
could be used and whether the weights could be chosen arbitrarily
(constant or random).

\subsection{Classical objects seen through the filter}

In this subsection we test the filter in a controlled environment, namely for
calorons as smooth configurations.

First we want to study a technical detail, namely how big the determinant is
that appears in (\ref{eqn_link_filtered_final})
to scale up the sum to an $SU(2)$-valued link.
In other words, how much of the link is
captured by the superposition of a finite number of modes (in a naive sense
without inspecting any structure in the links).

\begin{table}[h]
\centering
\begin{tabular}{cccccc}
$N$&1&4&10&50&200\\[0.5ex]\hline\\[-1ex]
min$_x\,\log_{10}$ det&$-15.3$&$-12.1$&$-8.5$&$-6.1$&$-4.1$\\
max$_x\,\log_{10}$ det&$-11.9$&$-8.6$&$-7.6$&$-5.5$&$-3.9$
\end{tabular}
\caption{Behaviour of the determinant used to
project to an $SU(2)$-element in the truncated sum
(\ref{eqn_link_filtered_final}), depending on the
number $N$ of modes. The background is the large caloron on $16^3\cdot 4$ with
$\z=1/4$.}
\label{tab_det}
\end{table}

From Tab.\ \ref{tab_det} one can read off
that this determinant is small (the correction
factor is large) when only a few modes are taken.
This is to be expected from the smallness of the $\phi_n(x)$
that are square-normalized on the entire lattice.
The determinant grows with $N$.
Of more relevance is that
 the minimum and the maximum of the determinant taken
over the lattice approach each
other, such that all links appear `equally well filtered'.

The most important question in the caloron context is
whether the filtered links produce the action density
lumps of the two constituent monopoles including the typical structure in the
Polyakov loop.
In Fig.\ \ref{fig_filter_bosonic_caloron}
we give the observables corresponding to Fig.\
\ref{fig_bosonic} (a) in the filtered configuration with
different $\z$ and $N$.

For the chosen $N\ll\N$, the boundary condition parameter governs the average
Polyakov loop.
In the intermediate case $\z=1/4$
the average Polyakov loop trace
is 0 with extrema of nearly $\pm 1$
at the monopole cores, see Fig.\ \ref{fig_filter_bosonic_caloron} (b).
In the periodic case $\z=0$ the Polyakov loop is almost
everywhere $\Eins_2$ with a dip resembling $\P=-\Eins_2$ at the corresponding
monopole, see Fig.\ \ref{fig_filter_bosonic_caloron} (a).
The antiperiodic case (not shown) is complementary, $\P\simeq -\Eins_2$ with
a signal at the other monopole.

\begin{figure}[h]
\centering
\begin{tabular}{ccc}
\includegraphics[width=0.32\linewidth]
{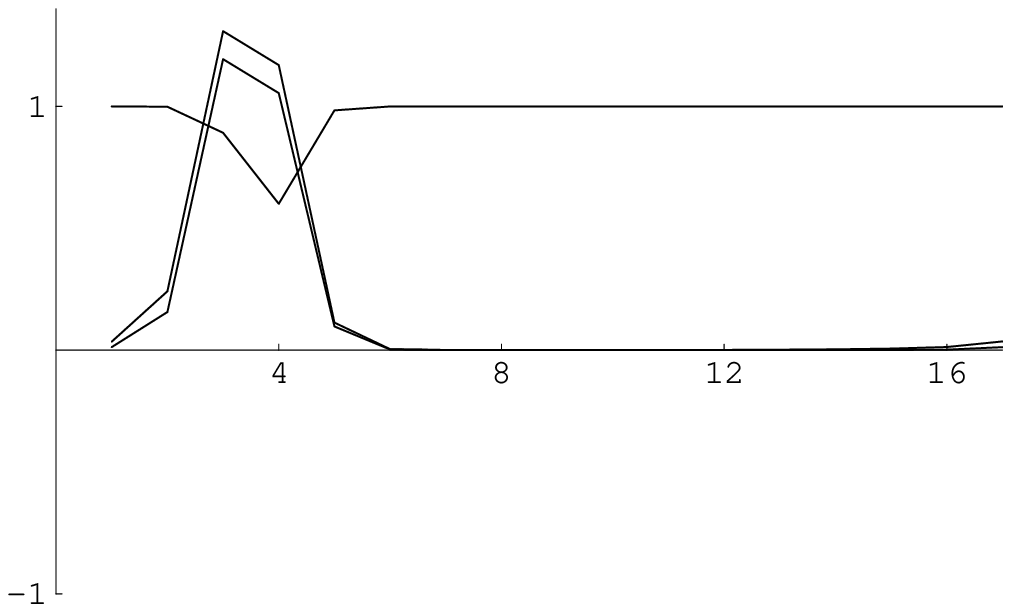}&
\includegraphics[width=0.32\linewidth]
{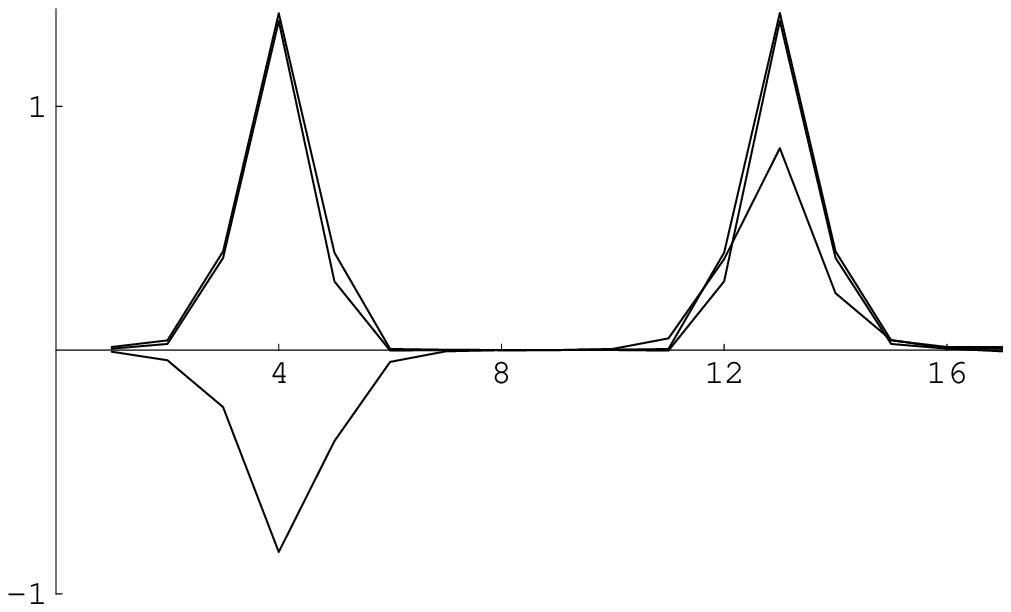}&
\includegraphics[width=0.32\linewidth]
{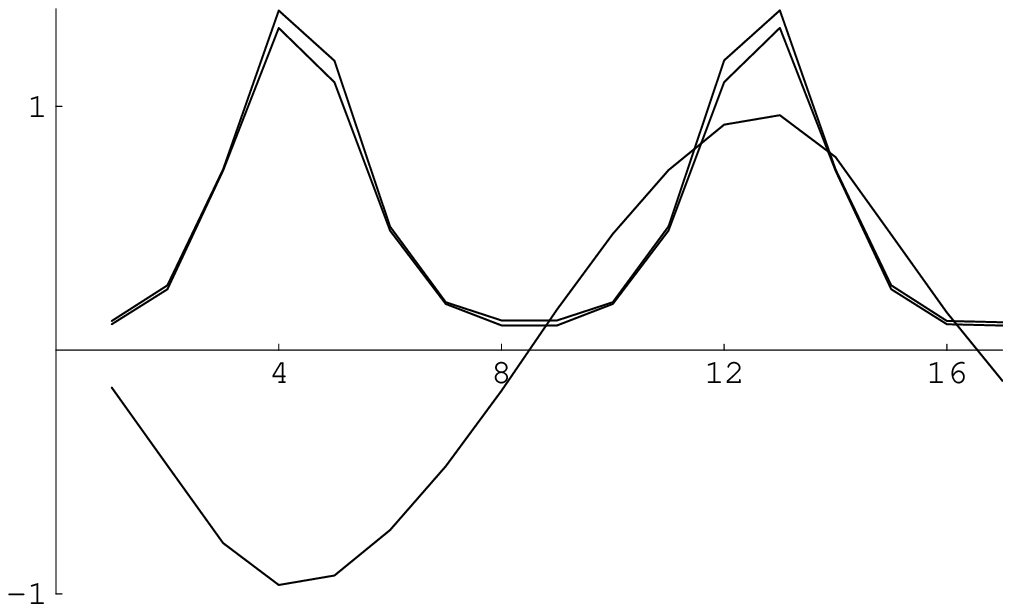}\\
(a)&(b)&(c)
\end{tabular}
\caption{Action density, topological density and Polyakov loop
 computed from filtered links in the background of
the large caloron of Fig.\ \ref{fig_bosonic}. (b) $\z=1/4$ with $N=4$ modes
(action and topological density multiplied by 100); (c) more
modes $N=150$ (scale 400 just as in the original plot Fig.\
\ref{fig_bosonic} (a)). (a) periodic boundary condition $\z=0$ with $N=4$ modes
(scale 30).}
\label{fig_filter_bosonic_caloron}
\end{figure}

This picture stays the same when more modes are taken into
account\footnote{Although, of course, in the limit $N\rightarrow\N$ the Polyakov
loop is the original one for all $\z$.}.
The Polyakov loop for the periodic case develops a stronger
dip then, but is still very close to $\Eins_2$ anywhere else.
For the intermediate
case the average stays close to 0.
Locally it agrees almost
perfectly with the one of the original configuration,
see Fig.\ \ref{fig_filter_bosonic_caloron} (c) for $N=150$ vs.\ Fig.\
\ref{fig_bosonic} (a). To make this more
quantitative we give the values of the Polyakov loop at the action density lumps
(see below) and on average in Tab.\ \ref{tab_obs}.

\begin{table}[h]
\centering
\begin{tabular}{cccccc|c}
$N$&4&10&100&150&200&original\\[0.5ex]\hline\\[-2.8ex]
$x_3$ at lumps& 4.0&3.7&4.1&4.3&3.7&4.4 \\
&13.0&13.3&12.9&12.8&13.3&12.6\\
action density &  0.014& 0.010&0.0023&0.0037&0.0026&0.0032\\          
top. density  &  0.013&0.010&0.0019&0.0035&0.0026&0.0032\\
Pol. loop & $\pm\,$0.84&$\pm\,$0.80&$\pm\,$1.00&$\pm\,$0.99&$\pm\,$0.99&$\pm\,$0.97\\[0.8ex]\hline\\[-2.9ex]
total action&              2.4&2.2&1.62&1.34&1.63&1.07\\
total top. charge&         0.98&0.99&1.00&1.00&1.00&1.00\\
average Pol. loop&   $-2\cdot 10^{-6}$&$-4\cdot 10^{-6}$&$-0.02$&$0.0001$&$-0.0002$&$-0.0003$
\end{tabular}
\caption{Quality of the reconstruction of a large caloron
from the filter with different number
of modes. Both, local quantities (interpolated) and global quantities converge to
the original values within some error margin.}
\label{tab_obs}
\end{table}

The findings for the reconstructed action and topological charge are analogous.
In order to compare the two, we have computed them with the
help of an $O(a^4)$ improved field strength tensor \cite{bilson-thompson:02}. 
The periodic case sees only one monopole\footnote{
Increasing $N$ makes the other monopole visible, but with a height much lower
than that of the first monopole.
Thus this choice of $\z$ reproduces the equal mass
constituents quite badly.}, while the
intermediate case is able to detect both (Fig.\
\ref{fig_filter_bosonic_caloron}).
We conclude that the boundary condition $\z=1/4$ is the appropriate one
when one wants to describe calorons with maximally nontrivial
holonomy through the filter,
as is also clear from the discussion of the contribution
of a single mode in the last section.

\enlargethispage{2\baselineskip}

Remarkably, the filter with the number of modes as low as $N=4$
already has quite some knowledge
about the classical structures. Although
the action density lumps are a bit spiky and their locations are not perfect
as recorded in  Tab.\ \ref{tab_obs}
(notice also the different scales for the
action and topological density used in Fig.\ \ref{fig_filter_bosonic_caloron}
(b) vs.\ (c) and Fig.\ \ref{fig_bosonic} (a)),
they clearly reflect the constituent structure with opposite
Polyakov loops. The two lumps are almost selfdual
(cf.\ third and fourth row in Tab.\ \ref{tab_obs}).
Furthermore, the adopted definition of the
topological charge works quite well for the filtered links
and gives a total topological charge close to 1.
We stress that the filtered
configurations have not undergone further cooling
and their approximate selfduality
is a remnant of the original configuration
passed on by its lowest-lying Laplacian modes.

\begin{figure}[b]
\centering
\begin{tabular}{ccc}
\includegraphics[width=0.35\linewidth]
{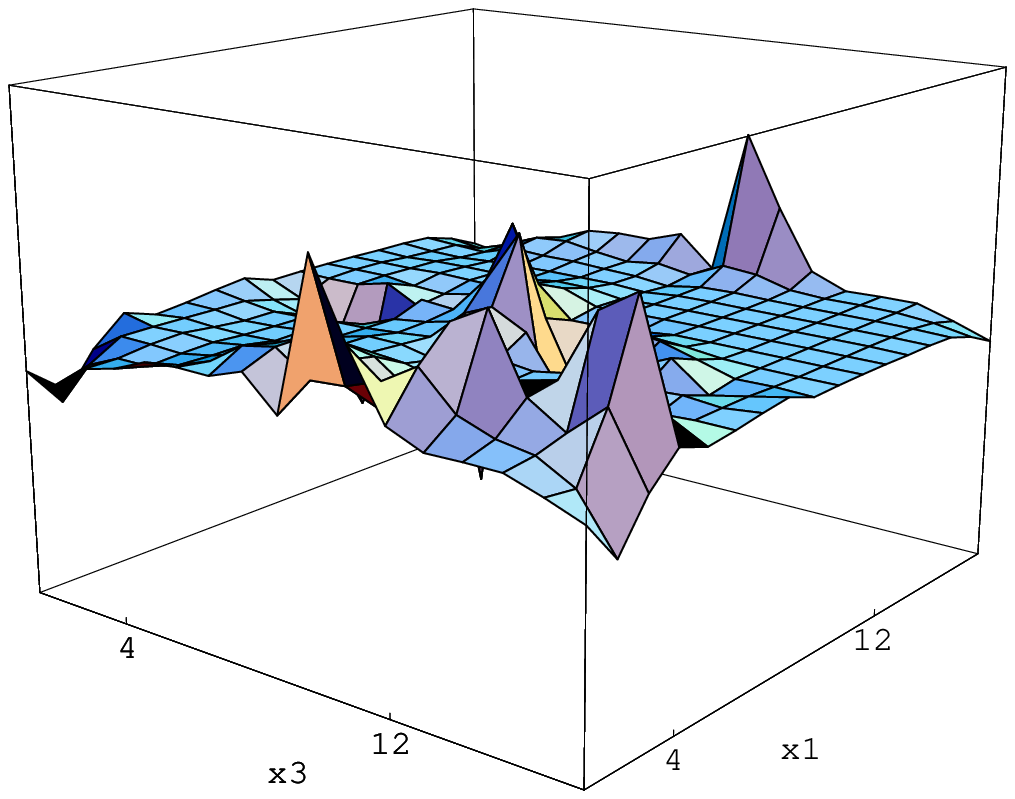}&&
\includegraphics[width=0.35\linewidth]
{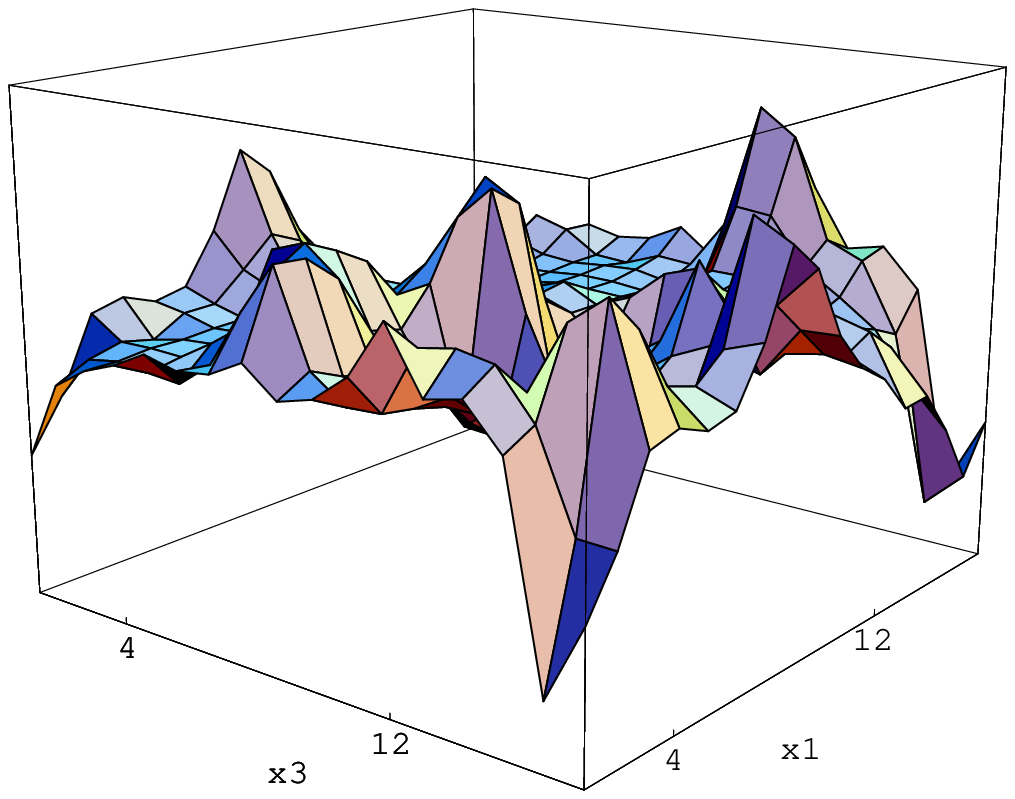}\\
(a)&&(b)\\
\\
\includegraphics[width=0.35\linewidth]
{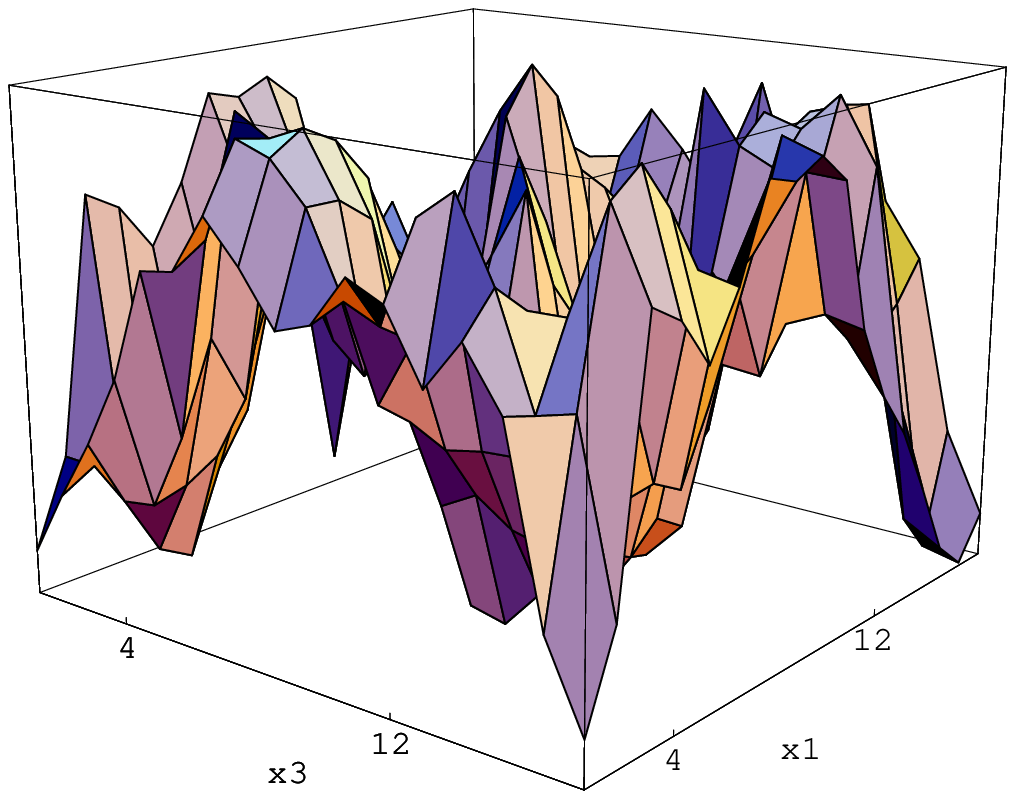}&&
\includegraphics[width=0.35\linewidth]
{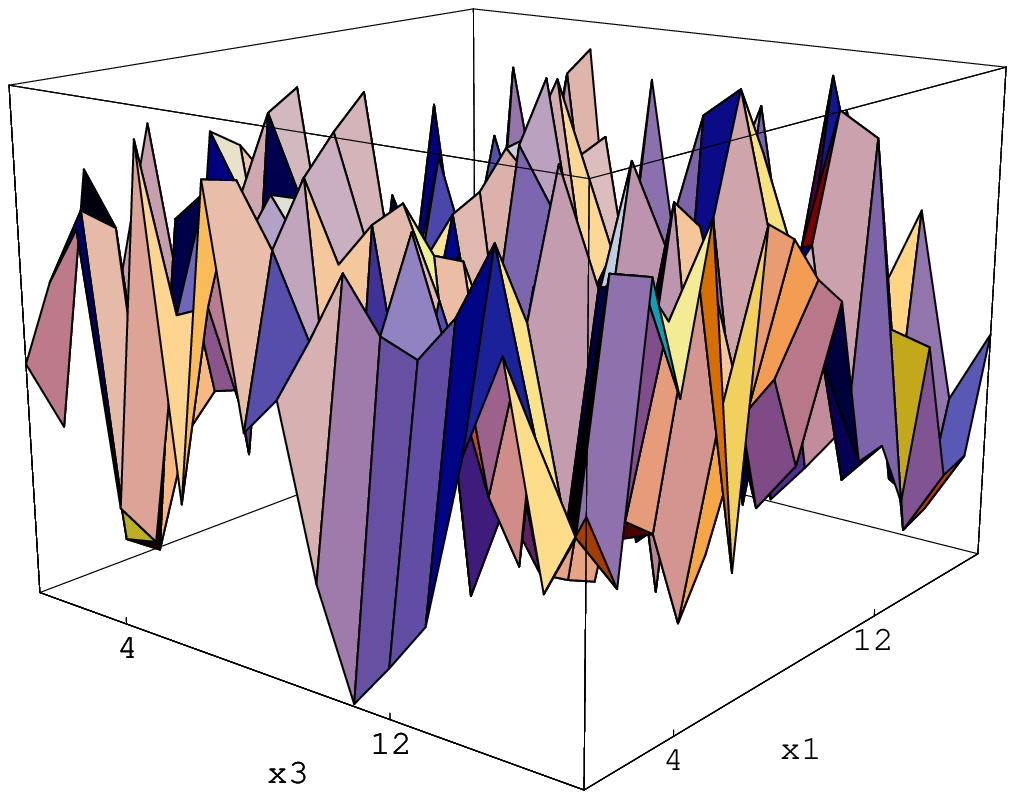}\\
(c)&&(d)
\end{tabular}
\caption{Polyakov loop `evolution' with increasing number $N$ of modes for a
thermalized configuration in a fixed lattice plane
(the one of Fig.\ \ref{fig_struc_filter}).
(a) $N=2$, (b) $N=10$, (c) $N=100$ and (d) the original configuration.
The vertical axes are from $-1$ to $1$.}
\label{fig_struc_filter_polloop}
\end{figure}

The inaccuracies of the $N=4$ case 
are cured by increasing the number $N$ of modes used in the filter,
however, not systematically. The quite
perfect case
$N=150$ used for the plot in Fig.\ \ref{fig_filter_bosonic_caloron} (c) is
contrasted by $N=200$, where some of the observables are further away
from the original configuration. We interpret this as a
limitation on the precision of the filter. Away from the ultimately exact limit
$N=\N$, the filter will reproduce the classical background only within some
error and hence it is not useful to take more than a few
hundred modes into account (at least for this example).
This does not spoil the use of the filter at all,
since its intended application will be mainly
to thermalized configurations (see below).
There it is not the aim to reproduce
the rough background to a high precision,
but to keep only a minimal structure of it
such that it still captures the physical randomness.

We note in passing that the Taubes winding inside caloron configurations
as well as the ring structure for
the charge 2 example are displayed by the filtered configurations, too.

\subsection{Are the filtered fields confining?}

The next two subsections are devoted to the application of the filter to
thermalized configurations. One of the main issues here is whether the filtered
links still give rise to confinement.

To this end we have measured the Polyakov loop correlator on the ensemble of
50 configurations on a $16^3\cdot 4$ lattice created at $\beta=2.2$. 
The Polyakov loop as the deconfinement order parameter has average $0$ and
therefore we choose $\z=1/4$ in the construction of the filter, Eq.\ 
(\ref{eqn_link_filtered_final}).
Indeed, the filtered Polyakov loop
$\langle \tr\,\tilde{\P}/2\rangle$ averaged over the lattice has an expectation
value compatible with 0 (with 50 configurations giving standard
devitations from 0.015 for $N=2$ to 0.083 for $N=100$
whereas the original standard deviation is 0.025). 
This again confirms that the average Polyakov
loop for small $N$ follows the one of the trivial case $N=1$.
In a way, we adapt the Polyakov loop average by a parameter in our filter
and then look at correlations in fluctations on top of it.

Fig.\ \ref{fig_struc_filter_polloop} shows how the Polyakov loop looks locally
(for a fixed configuration and lattice plane). When taking more modes into
account, the Polyakov loop deviates further and more often from the average 0.
Fig.\ \ref{fig_filter_distribution} makes this statement more quantitative. It
displays the distribution of Polyakov loops for one filtered configuration\footnote{ 
We thank A.\ Wipf for suggesting to show the $N$-dependence of the local Polyakov
loop distribution.}.
As is clear from the discussion so far, the distribution for $N=2$ is quite narrow
and broadens with growing $N$.
At $N=100$ it is very close to the one of the original distribution
and the Haar measure $\sqrt{1-(\tr\,\P/2)^2}$.
 
\begin{figure}[h]
\centering
\includegraphics[width=0.6\linewidth]
{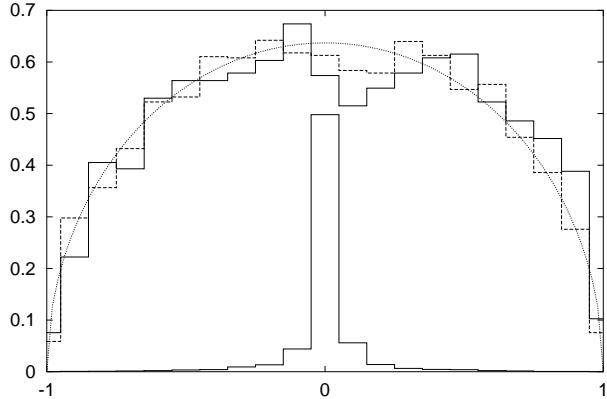}
\caption{
Distribution of Polyakov loops over the lattice sites for one 
configuration when filtered with the intermediate boundary condition 
$\z=1/4$ for 
$N=2$ (narrow distribution around 0, divided by 15 to fit in the same plot)
and $N=100$  (solid line) compared to the orignal configuration (dashed) and the Haar measure (dotted).}
\label{fig_filter_distribution}
\end{figure}

Fig.\ \ref{fig_filter_string} shows the logarithm of
the Polyakov loop correlator,
related to the interquark potential, measured at the
filtered ensembles with $N=2$ to $N=100$ together with the unfiltered
one. There is clear evidence that
{\em the filtered Polyakov loop correlator decays exponentially
with the distance}, 
in roughly the same window as the original one does.
The potential after filtering has no
sign of a Coulomb regime since} the filter has washed out short range
fluctuations as it should.
The filtered curves are also shifted vertically.
The value at zero distance represents the width
of the Polyakov loop distribution. 
It approaches the original one from below,
because the filtered distributions are narrower, 
as was shown in Fig.\ \ref{fig_filter_distribution}.

In order to cast the confining behaviour into numbers
we have performed an exponential
fit $c\,\exp(-\bar{\sigma} r)$ in the range between $3$ and $6$ lattice
spacings\footnote{
Our lattice spacing is $0.21\,fm$ provided $\sigma(0)=(440\,MeV)^2$.}.
The slope $\bar{\sigma}$ is directly proportional to the string tension 
$\sigma=\bar{\sigma}/N_4\,a^2$.
We will use the corresponding string tension of the original
configurations\footnote{
The obtained value of the string tension (for the original configurations, see
Tab.\ \ref{tab_act_vs_clusters}) is
equivalent to $0.78\,\sigma(0)$ whereas a parametrization
$\sigma=\sqrt{1-(T/T_c)^2}\,\sigma(0)$ \cite{pisarski:82}
would predict $0.66\,\sigma(0)$.} as the
reference observable.
The lower end of the fit region was
taken such that the numerical value of $\bar{\sigma}$
for the original configuration
is stable when compared with a fit over 1 to 6 lattice spacings including a
Coulomb part.
Towards large distances the fit was
limited by statistical errors.
The obtained values for the string tension
and estimates of its error are given in Tab.\ 
\ref{tab_act_vs_clusters}.
Although the approach of the filtered string tensions
for the number of modes between $N=2$ and $100$
to the original one is again not monotonic,
the latter is reproduced within 15\%
(for $N=4$ modes the reproduced string tension is almost perfect).
 
\begin{figure}[t]
\centering
\hspace{-0.5cm}
\includegraphics[width=0.6\linewidth]
{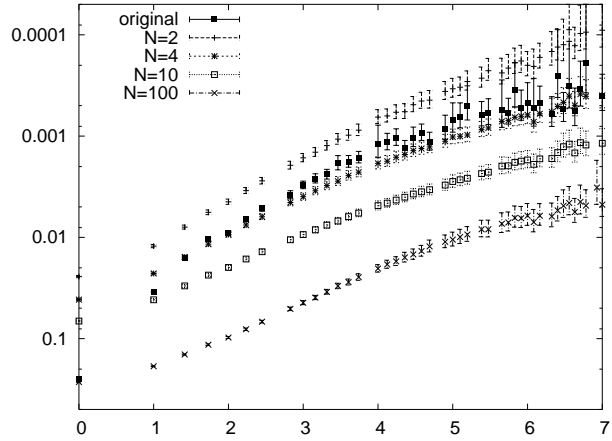}
\caption{
Polyakov loop correlator 
plotted on an inverse logarithmic scale over the distance
for different $N$ compared to the original one.}
\label{fig_filter_string}
\end{figure}

\begin{figure}[b]
\centering
\hspace{-0.5cm}
\includegraphics[width=0.6\linewidth]
{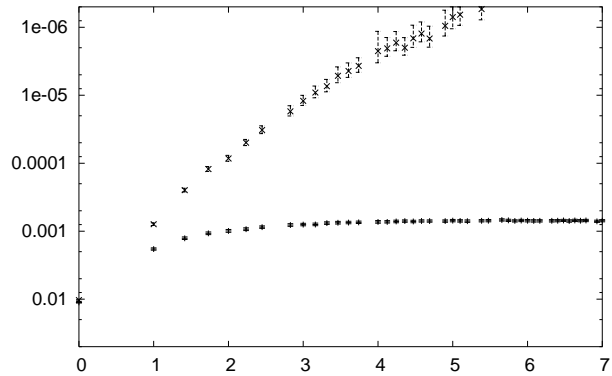}
\caption{
Test whether the original and filtered configurations contain
correlated fluctuations disordering the Polyakov loop. Correlators $C(r)$ (upper
curve) and $C_{\rm ind}(r)$ (lower curve), see text.}
\label{fig_filter_crosscorr}
\end{figure}

A priori it is not obvious whether the Polyakov loop fluctuations in the filtered
and the original configurations are correlated or whether the filter creates
independent fluctuations
(a pointwise scatterplot of the two respective Polyakov loops against each other
shows a correlation setting in not below $N=50$ modes).
A similar question has been investigated
in \cite{deldebbio:98}
concerning two Wilson loops $W_1(C)$ and $W_2(C)$\footnote{
We thank J.\ Greensite for urging us to perform the following test.}.
There the correlation
$\langle W_1(C) W_2(C)\rangle$
(i.e.\ between different Wilson loops along the same curve $C$) was compared to
$\langle W_1(C)\rangle \langle W_2(C)\rangle$
representing the hypothetic independence of
the Wilson loops (and decaying exponentially with the
area of $C$). Adopting this idea to our case we have to compare
\begin{eqnarray}
C(|\vec{x}-\vec{y}|)&=&
\frac{1}{16}\,\langle \tr\, \P(\vec{x})\tr\, \P(\vec{y})\cdot
\tr\, \tilde{\P}_N(\vec{x})\tr\, \tilde{\P}_N(\vec{y})\rangle\qquad {\rm  to}\\  
C_{{\rm ind}}(|\vec{x}-\vec{y}|)&=&
\frac{1}{16}\,\langle \tr\, \P(\vec{x})\tr\, \P(\vec{y})\rangle\langle
\tr\, \tilde{\P}_N(\vec{x})\tr\, \tilde{\P}_N(\vec{y})\rangle\,.   
\end{eqnarray}
We did this for $N=4$ and show in 
Fig.\ \ref{fig_filter_crosscorr} that the correlator $C(r)$ becomes
independent of the distance $r$ rather quickly. $C_{{\rm ind}}(r)$ of
course decays exponentially with the sum of the string tensions
of the original and the filtered case.
We conclude that the fluctuations disordering the
Polyakov loop for the original and the filtered configuration are not
independent but rather correlated.

\subsection{Structures found by the filter}

Finally, we give a first description of the vacuum structures
that appear when the
filter is applied to an individual configuration
from an equilibrium ensemble, representing finite temperature in our case.

\begin{table}[h]
\centering
\begin{tabular}{c|ccccc|c}
$N$&2&4&10&50&100&original\\
\hline
$\sigma/10^5\, MeV^2$&$1.732(31)$&$1.447(25)$&
$1.299(17)$&$1.385(13)$&$1.574(25)$&$1.509(70)$\\
Wilson action& 66 &84&106&177&234&2131\\
max. \# clusters&37&50&70&79&104&342
\end{tabular}
\caption{Observables to characterise the filter
when acting on equilibrium configurations.
The string tension is computed from 50 configurations,
the other two observables are for the example configuration
characterised locally in Fig.\ \ref{fig_struc_filter_polloop}
and \ref{fig_struc_filter}.}
\label{tab_act_vs_clusters}
\end{table}

\begin{figure}
\centering
\begin{tabular}{ccc}
\includegraphics[width=0.35\linewidth]
{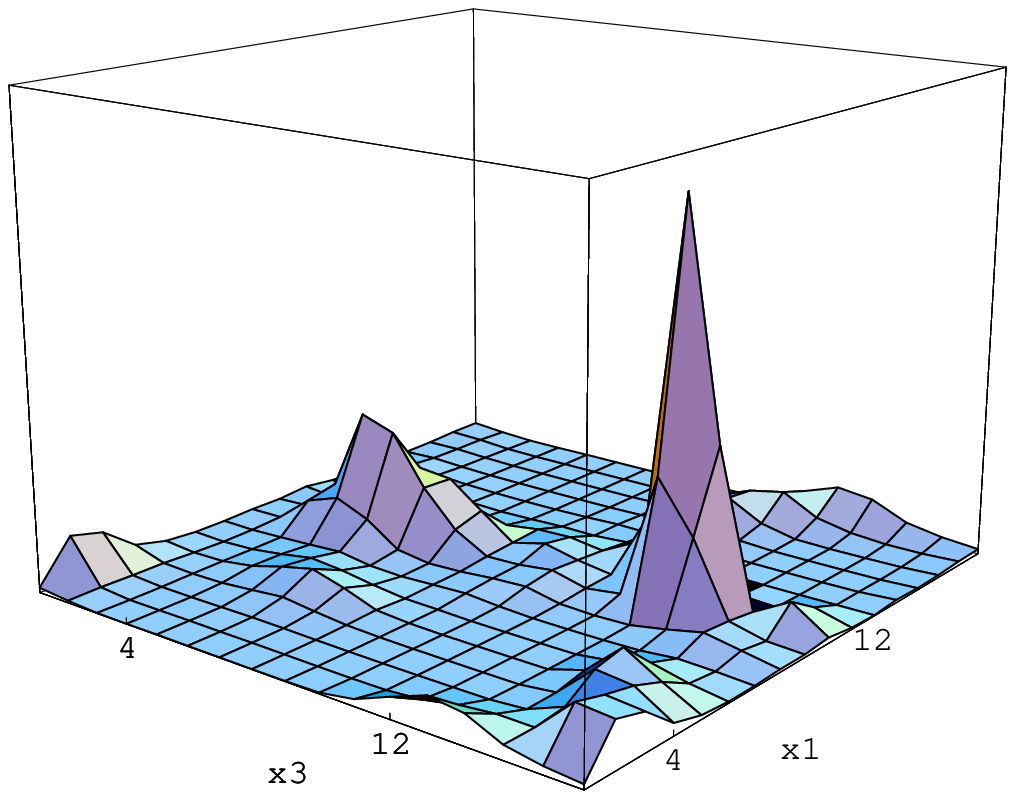}
\hspace{-1cm}(a)&&
\includegraphics[width=0.35\linewidth]
{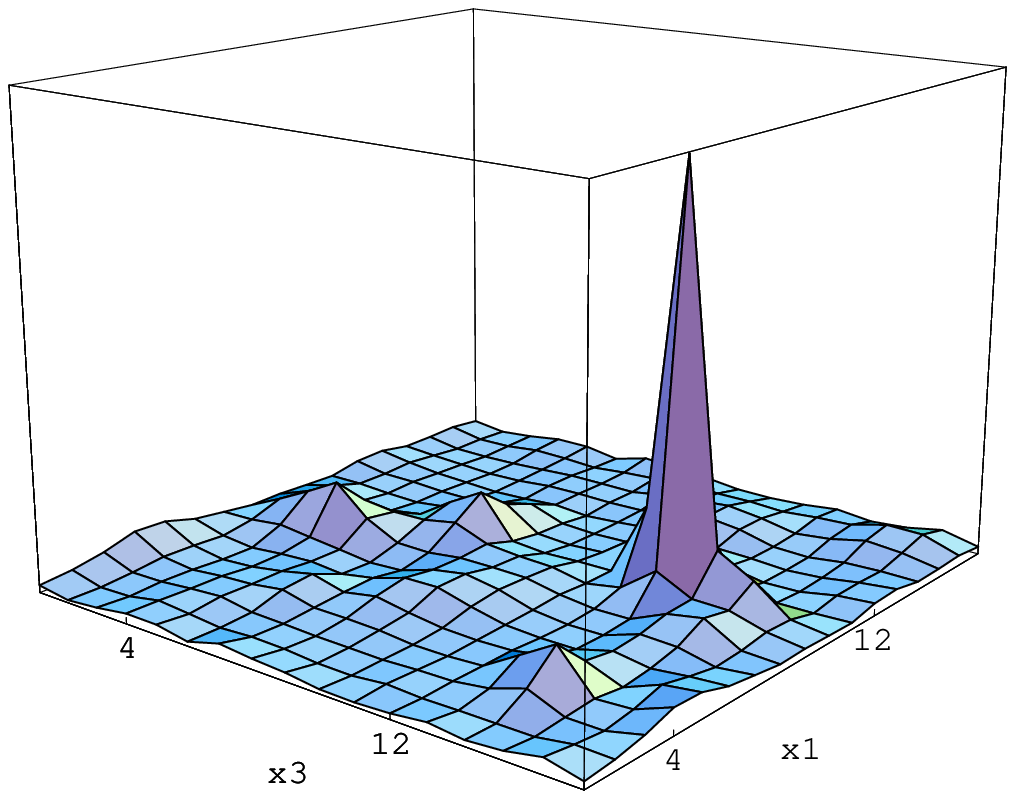}
\hspace{-1cm}(b)\\
\includegraphics[width=0.35\linewidth]
{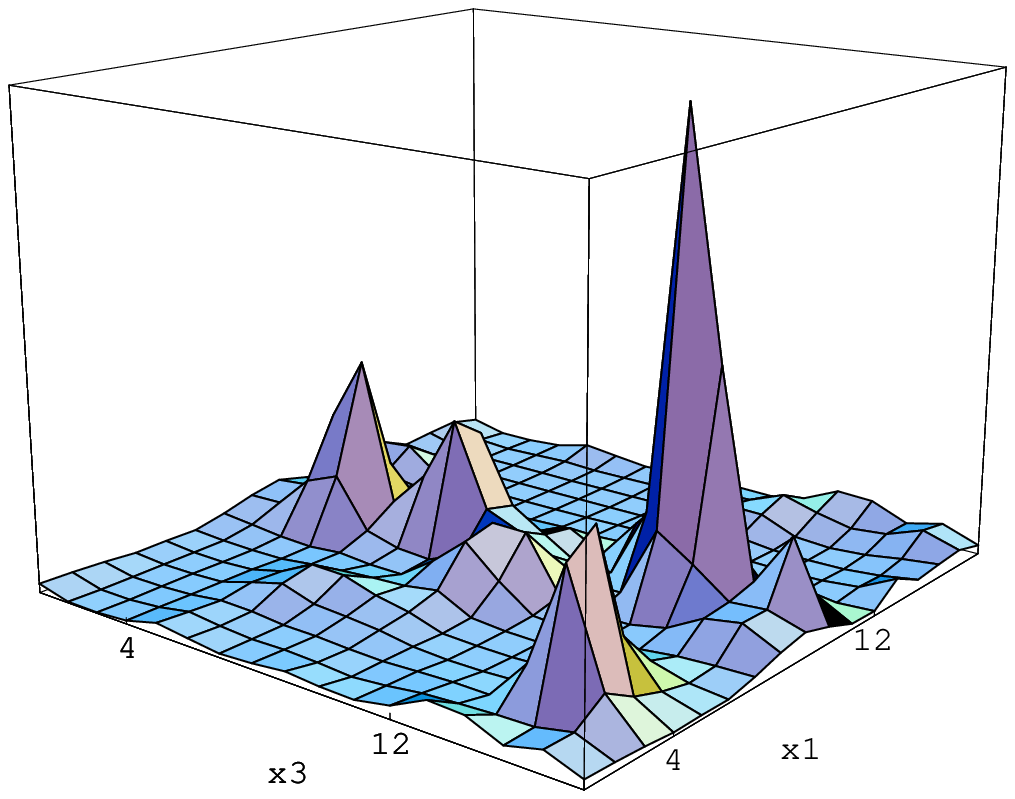}
\hspace{-1cm}(c)&&
\includegraphics[width=0.35\linewidth]
{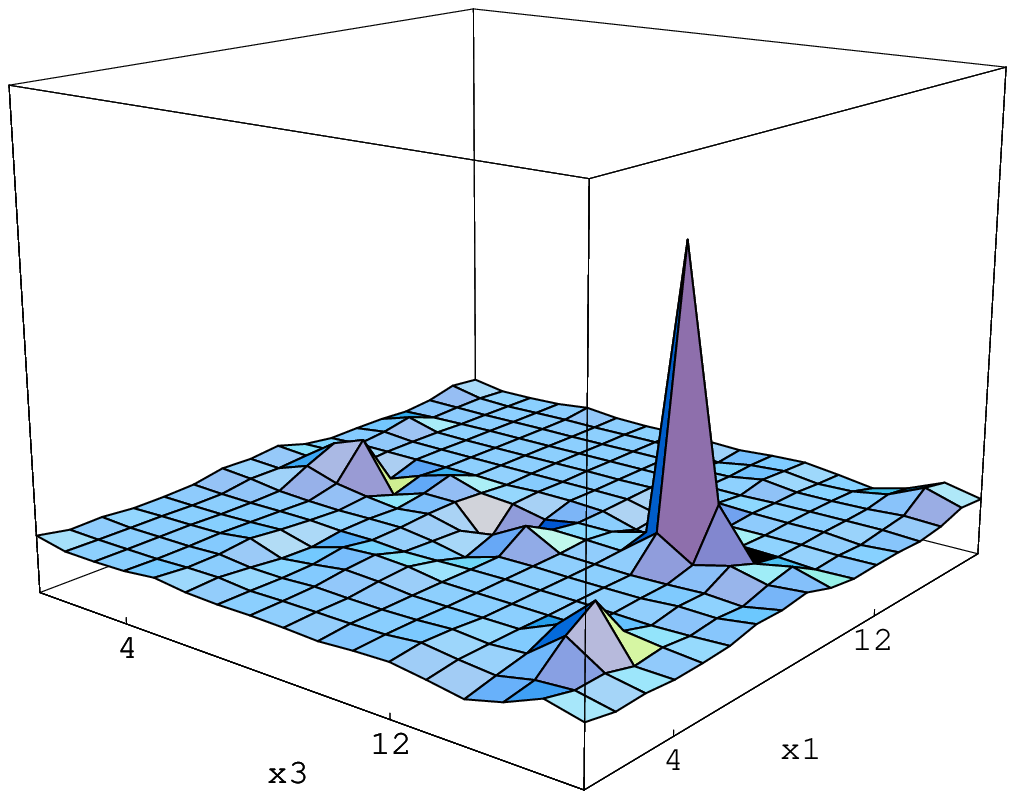}
\hspace{-1cm}(d)\\
\includegraphics[width=0.35\linewidth]
{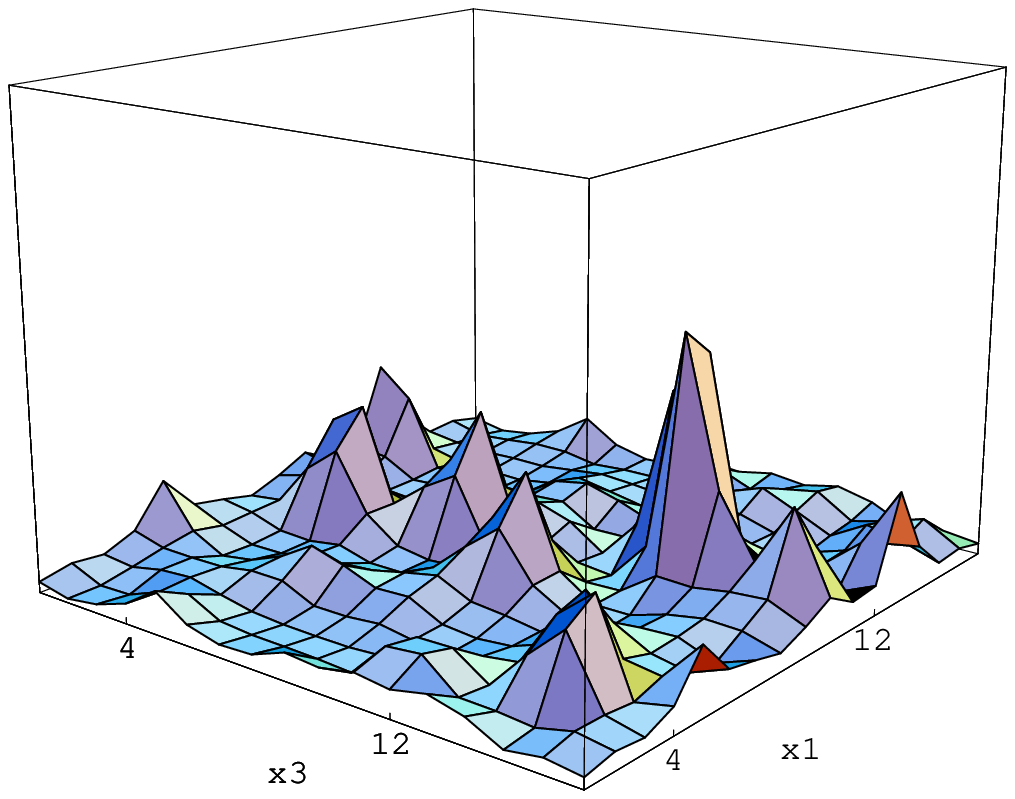}
\hspace{-1cm}(e)&&
\includegraphics[width=0.35\linewidth]
{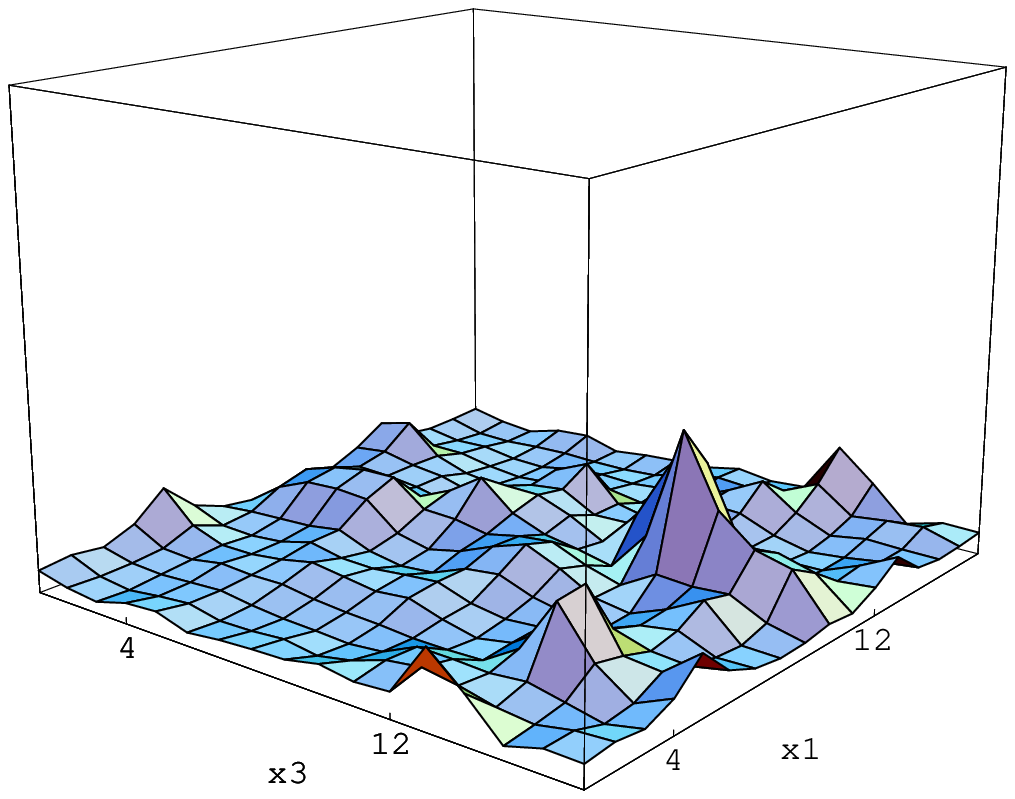}
\hspace{-1cm}(f)\\
\includegraphics[width=0.35\linewidth]
{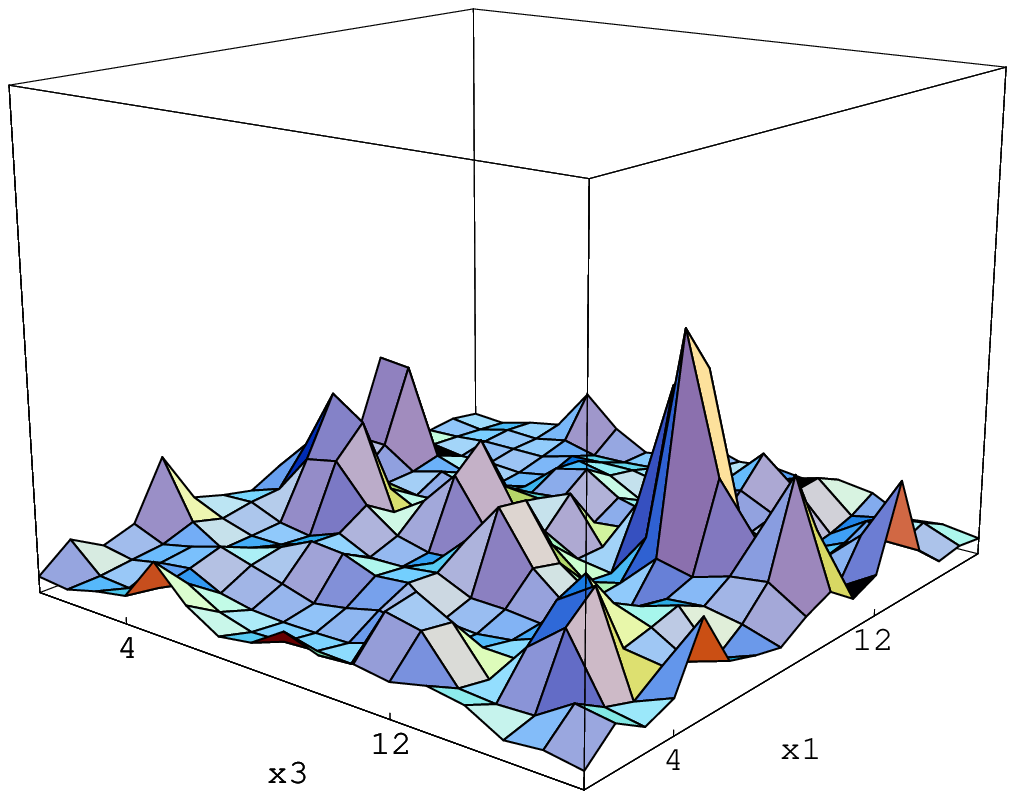}
\hspace{-1cm}(g)&&
\includegraphics[width=0.35\linewidth]
{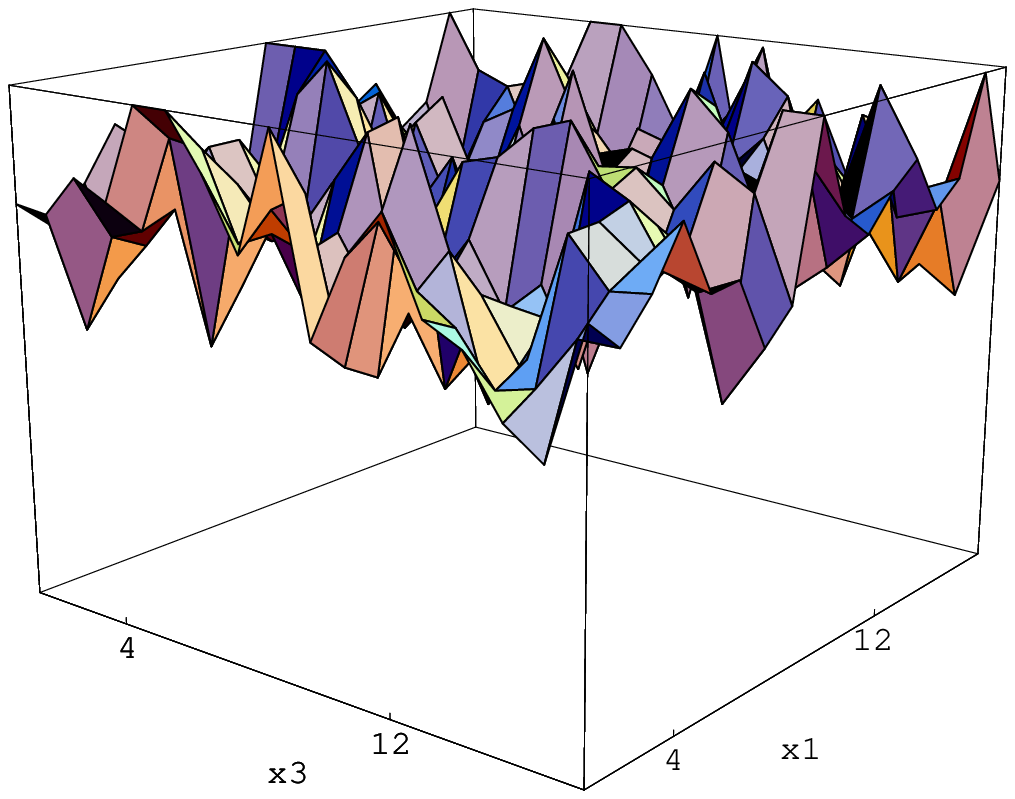}
\hspace{-1cm}(h)
\end{tabular}
\caption{
Left column: `Evolution' of the action density with
growing number of modes in the filter, (a) $N=4$,
(c) $N=10$, (e) $N=50$ and (g) $N=100$.
Chosen is the lattice plane $(x_2,x_4)=(6,4)$
which contains the global maximum for $N=10$ sitting
at $(x_1,x_3)=(10,13)$.   
For comparison the action densities for the same configuration after 5
steps of smearing (b), after 2 steps of cooling (f)
and for the original configuration
(h) are presented with uniform scale.
In the second row a comparison is made
between the action density (c) and the topological density (d),
both after filtering with $N=10$.}
\label{fig_struc_filter}
\end{figure}

The behaviour of the Polyakov loop was shown in the previous section (Fig.\
\ref{fig_struc_filter_polloop}). The total action of the filtered configuration
behaves like $N^{1/3}$ in a range from $N=2$ to $N=100$ eigenmodes.
The interpretation of this particular dependence is not obvious,
but the overall picture is clear: the more modes are included in the filter,
the more fluctuations occur and contribute to the total action.
In the example we use (cf.\ Fig.\ \ref{fig_struc_filter_polloop}
and \ref{fig_struc_filter})
the action in the first nontrivial case $N=2$ is 66
in instanton units of $2\pi^2$, compared
to a total plaquette action of 2131 for the original configuration.
The first number
should be read as the minimal content that survives the
filter. The only way to achieve an `even smoother configuration' would be to
apply the filter to the already filtered configuration.

The global maximum of the action density is only weakly varying with $N$,
meaning that there are peaks of about the
same height at every stage of the filter.
From Fig.\ \ref{fig_struc_filter}, showing the action density in a
fixed lattice plane for different $N$, it is evident that these peaks are quite
narrow. This effect is counterintuitive, but has also been observed for caloron
backgrounds in the last section.
The peaks look the same in a space-time plane, too, i.e.\ they are not static. The
figure also shows that at least some of the peaks are stable w.r.t.\ $N$, that is
they stay as local maxima. Their width is not changing much either.

In order to more quantitatively describe the filtered structure we have
performed a cluster analysis.
Lowering the threshold we have
recorded the number of clusters and their respective
volume, size and accumulated action.
The number of clusters
first raises and after reaching a plateau decreases again due to cluster
mergings, while the peaks remain.

\begin{figure}[b]
\centering
\includegraphics[width=0.6\linewidth]
{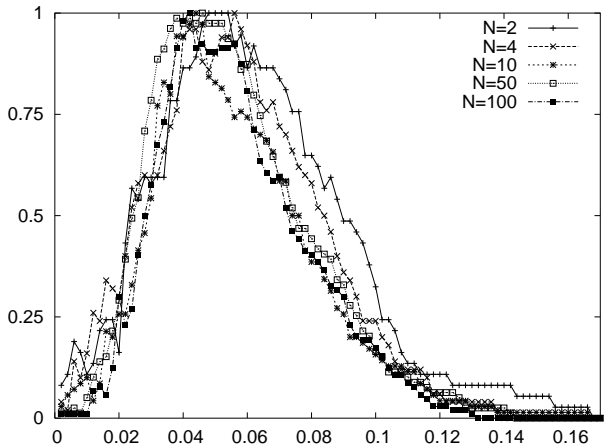}
\caption{Relative number of action density clusters as a function of the lower
threshold for a filtered configuration at different $N$.}
\label{fig_struc_filter_cluster}
\end{figure}

As shown in Fig.\ \ref{fig_struc_filter_cluster},
the corresponding curves fall almost on top of each other,
when the number of clusters is taken relative to its
maximum at that $N$ (see Tab.\ \ref{tab_act_vs_clusters}).
That suggests that the difference in the filtered action for
different $N$ is due to the different {\em number} of clusters.
Otherwise they are
rather similar. Other indications for that are the volume, the size  and the
action per cluster,
which -- plotted as a function of the threshold --
fall on top of each other as well.

In Tab.\ \ref{tab_act_vs_clusters} we give for each $N$ the maximal number
of clusters. The latter appears at a threshold of roughly a third of the global
maximum, where all
clusters together
have accumulated a few percent of the lattice volume and roughly 15\%
of the action.
It is interesting to notice that the maximal number of action density clusters
roughly agrees with half the total action (in instanton units,
see Tab.\ \ref{tab_act_vs_clusters} third and second row).
A lower bound for the estimated action per peak is thus 0.3. More realistic
is to assign part of the remaining action below the threshold to the peaks
(sitting in their tails). From Fig.\ \ref{fig_struc_filter_cluster} we read off
that at a threshold of 0.01 (around $1/15$ of the global maximum) the number of
clusters starts to fluctuate because the region of action density noise has been
reached. At that threshold the accumulated action of the clusters is roughly 60\%
of the full one. This results in an average action per peak of 1.2 in instanton units.
Hence, from an action density point of view, this estimate is compatible with
an interpretation of the peaks as (possibly fractional) instanton lumps.

Indeed, the local maxima of the action density are often accompanied by local
extrema in the topological charge density. This is visualized for $N=10$ in
Fig.\ \ref{fig_struc_filter} (d). However, the topological charge density does not
equal the action density at the peaks.
For the global maximum at $N=10$ the ratio of both quantities is 0.87,
but for other local maxima no signature of (anti)selfduality was found.
Moreover, the total topological charge of generic configurations expressed in
terms of the filtered links is not close to an integer.
In this respect, the configuration, although it is
filtered, is not smooth enough to make the $O(a^4)$ improved
field strength definition of the
topological charge work. Fermionic definitions via the index
should be used to determine whether
the filter preserves the total topological charge. 

The peaks seem to have nothing to do with the maxima of the modulus
of the Laplacian mode in that background,
but with the determinant used as the normalization
factor in Eq. (\ref{eqn_link_filtered_final}).
Obviously this determinant is given in terms of
eigenmodes (and eigenvalues) of the Laplacian, too, thus it also contains information
about the gauge background.

Another interesting question is whether the new filter is related to
smearing and cooling.
In Fig.\ \ref{fig_struc_filter} we compare the corresponding action
densities to the filtered ones, where the number $N$ of modes was chosen such
that the total actions are comparable.
For instance, 2 cooling steps result in
an action of 172, which therefore compares to $N=50$
(see Fig.\ \ref{tab_act_vs_clusters} second row) and 5 smearing steps result in
84 which compares to $N=4$. As the third and the first row of
Fig.\ \ref{fig_struc_filter} show,
the corresponding peaks seem to agree locally, which is a nontrivial correlation,
because both methods lower the total action by a smoothing procedure, but in
completely different ways.
In contrast, 5 cooling steps lead to an action of 56,
which seems to be comparable to $N=2$.
The action density, however, looks very different.
It is clear that for more cooling steps
the correlation to the filter has to break down,
since cooling typically drives towards
action plateaus of a few instanton units and finally removes the string
tension. This is avoided in the filter method. 

More work has to be done to better understand
the ($N$-dependent) spiky structures induced by the filter.

\section{Discussion and outlook}

We have investigated Laplacian eigenmodes in both classical and thermalized gauge
configurations with respect to their capability to analyze certain
properties of the background.
Our localization studies for classical backgrounds lead to the conclusion that
there is an analogy to fermionic (near zero) modes.
In caloron backgrounds they detect the monopoles by a minimum and a maximum
and hop between these constituents upon changing the
boundary condition angle $\z$. 
We conclude that such a localization is merely due to information about the
gauge background in the covariant derivative, which is modified by the angle
$\z$ in the same way, and does not depend much on the
spin of the analyzing field.

Yet, there is some difference between fermionic and Laplacian
modes. This concerns for instance the behaviour under intermediate boundary conditions.
For the caloron fermionic zero mode one lump grows at the expense of the
other, while the Laplacian ground state develops a valley.
Also the precision of the localization is lower than for the
fermions: the maximum of the modulus
is less pronounced and static even for a time-dependent
action density of the background. 
For the minimum these problems do not occur,
but the locations of both, minimum and maximum, deviate from the constituents.
Moreover, the modulus of the mode
approaches the average value away from those
structures and hence the IPR's are rather small.
That is why the Laplacian modes resemble modified waves rather
than exponentially localized discrete states (in continuum language). 

The wave character also applies to the Laplacian modes in the adjoint
representation, where the underlying structures (constituent monopoles or
instantons) are detected by minima only\footnote{
For adjoint {\em fermions}
there are more zero modes,
with maxima at the constituents \cite{gonzalez-arroyo:05}.}.
We expect them to become zeros in the continuum,
which is natural from the point of view of the Laplacian Abelian projection.
For antiperiodic adjoint modes these minima form even two-dimensional sheets
between the constituent monopoles.
We have illustrated this for a semiclassical background, too.
Signatures of the Taubes winding can be found in both representations.


In thermalized backgrounds, there is clear evidence that the lowest mode 
changes its global maximum with the boundary conditions.
Different intervals in the boundary condition angle $\z$ emerge, where different
local maxima take over the role of the global maximum.
We have observed up to 4 jumps per configuration and the corresponding
lattice locations seem to be close to randomly distributed.
These locations are also visited by excited Laplacian modes and those in the
adjoint representation.

The corresponding global minima are not stable under the boundary condition (and the
number of local minima is large). Therefore they could not be used as a
practical tool for localization.
Thus, a straightforward interpretation of Laplacian modes on
thermalized backgrounds in terms of
classical objects is not possible.
In this context it would be interesting to study the effect of
quantum fluctuations on the Laplacian modes by heating a classical background.

Actually, the IPR is basically insensitive to minima.
Another observation that questions the use of the IPR is that, in most cases
we studied, it is proportional to the value at the global maximum. This large
value apparently dominates the sum in the IPR definition and therefore the latter
gives no new information.
In addition to the IPR measurements we have performed a cluster investigation.
Somewhat unexpectedly we have found that the fundamental Laplacian mode can
even be characterised as a global structure (depending on $\z$).

Apart from the weaker localization and their more static nature, the Laplacian modes
behave again similar to fermionic modes. A natural next step is to clarify whether
Laplacian and fermionic modes on the same configuration see the same locations.
In order to eventually go beyond a purely empirical description it
is desirable to sort out and measure the relevant gluonic features in the gauge field
backgrounds, for instance the topological charge density.

Smoothing techniques like smearing could be used for this purpose, although
they have the disadvantage to modify the gauge background. Then the Laplacian
modes become similar to those in classical backgrounds: they level off thereby
lowering (and joining or slightly moving) the lumps and smoothing the minima.

We have pointed out an interesting relation of the unsmeared Laplacian mode to a
gluonic observable after smearing: the smeared Polyakov loop provides pinning
centres, some of which the Laplacian mode occupies.
Such a finding in itself is in accordance with both hypotheses,
calorons underlying the thermalized configuration
and Anderson localization.
More work has to be done to better assess the real mechanism. 

Another interesting question is whether the Laplacian modes reveal signatures of
the deconfinement phase transition.
As a first signal we have seen a qualitative difference in the $\z$-dependence
of the spectrum.
Of course the Laplacian modes should also be investigated at zero
temperature, to see to what extent
the properties described in this paper remain.\\

Laplacian eigenmodes are the natural ingredients for a Fourier-like filter. We have
described a new method to reconstruct the link variables by truncating a sum over
Laplacian modes. Because of the use of eigenmodes one might view this
technique as a nonlocal smoothing. It is actually not too expensive, the
computation of $N=10$ modes on a $16^3\cdot 4$ lattice including the
reconstruction of the links takes a minute on a 1.7 GHz PC.
It should be stressed that the filtered links allow for the
measurement of any quantity and that the procedure is not biased towards any
particular degree of freedom in the QCD vacuum.

The striking properties of the filter are that it
reproduces classical structures, in particular selfduality,
and preserves the string tension (within 15\%)
when applied to equilibrium configurations.
The number of Laplacian modes can be kept remarkably low,
typically\footnote{This number, referring to $\beta=2.2$ on a $16^3\cdot 4$ lattice,
might depend on the lattice volume and the coupling constant.} 4
out of more than $10^5$ eigenmodes start
to reproduce the mentioned features qualitatively.
Because the filter keeps the relevant long-range disorder,
we claim a `low mode dominance' in the confining properties
of lattice gauge theory.

The filter has a lower limit, namely $N=2$ modes.
It might well be that the content
of action density and Polyakov loop fluctuations for this case is minimally
required to keep the long-range physics.
For the same reason it is clear that the observed similarity of the filter
with smearing and cooling is limited, for instance to early stages of cooling.

Apart from the number of modes the only filter parameter is the phase in the
boundary condition. We have shown that it influences the quality of the filter
and that in the confined phase (at finite temperature) it is best set to the
intermediate value $\z=1/4$.

Concerning the tomography of the filtered configurations
we have not yet reached a final understanding.
The filtered Polyakov loop is narrowly distributed around 0 and approaches the
original distribution for $N\simeq 100$.
In the emerging action density isolated peaks appear, both for
caloron and equilibrium configurations.
This seems counterintuitive, taking into account that the filter
uses the {\em lowest} Laplacian modes.
One should keep in mind, though, that the contribution of
the high end of the spectrum is almost the same as for the low end
(due to the staggered symmetry).

It seems that the peaks are correlated to the normalization factor
(inverse square root of the determinant)
that has to be applied to project the filtered link back to $SU(2)$.
This mechanism as well as a possible
relation to the structures found by Fourier-filtering in Landau gauge
\cite{gutbrod:04} and to singular gauge fields \cite{zakharov:04}
has to be clarified.

In this context it might be helpful to improve the filter procedure. 
First of all, the lattice Laplace operator could be replaced by an
improved version. The second opportunity is to interpolate the eigenmodes. This
leads to filtered links on finer lattices (similar to inverse blocking
\cite{degrand:96a,degrand:96b}), which can be subject to blocking.
A prerequisite for this idea is that not only the modulus but also the
components of the lowest-lying eigenmodes are fairly smooth. This is actually the case
in Laplacian gauge.

There are some possibilities to generalize our method or to apply it in another
physical context. The obvious applications are to
zero temperature and to the deconfined phase, respectively. For the latter 
we have checked that filtering with $\z=1/4$ results in no string
tension, however, one might be forced to fix $\z$ in a different way. The
quality of the filter in this phase could be checked by virtue of the spatial string
tension.

The generalization to higher gauge
groups is not completely trivial.
We have used the charge conjugation symmetry (relying on
the pseudoreal nature of $SU(2)$) in the derivation of the filtered links
becoming unitary.
For gauge groups $SU(N)$ the singular value decomposition
seems to be the only alternative.

Fermionic modes could also be used to reconstruct the gauge field, provided one
is able to project out the spin indices to arrive at an exact formula for the
links. This variant of the filter will be more expensive, but might be advantageous
concerning topological properties of the gauge field background.

\section*{Acknowledgements}

We are grateful to Pierre van Baal to make our collaboration possible. We would
like to thank him as well as Philippe de Forcrand, Jeff Greensite, Hans Joos,
Kurt Langfeld, Michael M{\"u}ller-Preussker, Daniel Nogradi, Stefan Olejnik, Ion-Olimpiu Stamatescu
and Andreas Wipf for helpful discussions.
EMI gratefully appreciates the hospitality at the Instituut Lorentz
of Leiden University and FB that of the Theoretical Particle Physics group of
Humboldt University Berlin.
This work was supported by FOM and FB by DFG (No.\ BR 2872/2-1).

\pagebreak


\providecommand{\href}[2]{#2}\begingroup\raggedright\endgroup

\end{document}